%% file: Token-Operations-Oriented-Inference-Optimization-Techniques-for-Large-Models.tex
\documentclass{article}

\usepackage{arxiv}
\usepackage[utf8]{inputenc}
\usepackage[T1]{fontenc}
\usepackage{hyperref}
\usepackage{xurl}
\usepackage{booktabs}
\usepackage{amsmath,amsfonts,amssymb}
\usepackage{microtype}
\usepackage{graphicx}
\usepackage{subcaption}
\usepackage{float}
\usepackage[numbers,sort&compress]{natbib}

\graphicspath{{figures/}}

\title{Token-Operations-Oriented Inference Optimization Techniques for Large Models}

\makeatletter
\long\gdef\@author{
\begin{minipage}{0.94\textwidth}
\centering
{\small\bfseries\linespread{0.98}\selectfont \mbox{Shiguo Lian$^{1,2,*}$},\allowbreak{} \mbox{Kai Wang$^{1,2}$},\allowbreak{} \mbox{Zhaoxiang Liu$^{1,2}$},\allowbreak{} \mbox{Wen Liu$^{1,2}$},\allowbreak{} \mbox{Minjie Hua$^{1,2}$},\allowbreak{} \mbox{Yutong Liu$^{1,2}$},\allowbreak{} \mbox{Jiangze Yan$^{1,2}$},\allowbreak{} \mbox{Xin Wang$^{1,2}$},\allowbreak{} \mbox{Cong Wang$^{1,2}$},\allowbreak{} \mbox{Yilin Zhang$^{1,2}$},\allowbreak{} \mbox{Yi Shen$^{1,2}$},\allowbreak{} \mbox{Jieyun Huang$^{1,2}$},\allowbreak{} \mbox{Fang Zhao$^{1,2}$},\allowbreak{} \mbox{Huanlin Gao$^{1,2}$},\allowbreak{} \mbox{Ping Chen$^{1,2}$},\allowbreak{} \mbox{Xinyu Yang$^{1,2}$},\allowbreak{} \mbox{Kaikai Zhao$^{1,2,3}$},\allowbreak{} \mbox{Yantao Li$^{1,2,4}$},\allowbreak{} \mbox{Yao Zhao$^5$},\allowbreak{} \mbox{Xinggang Wang$^6$},\allowbreak{} \mbox{Huishuai Zhang$^7$},\allowbreak{} \mbox{Dongyan Zhao$^7$},\allowbreak{} \mbox{Junping Du$^8$},\allowbreak{} \mbox{Tao Chen$^9$},\allowbreak{} \mbox{Xiang Gao$^{10}$},\allowbreak{} and \mbox{Qinghuai Ma$^{11}$}\par}
\vspace{0.25em}
{\footnotesize\normalfont\linespread{1.00}\selectfont $^1$ Unicom Data Intelligence, China Unicom\\
$^2$ Data Science and Artificial Intelligence Research Institute, China Unicom\\
$^3$ Tsinghua University\\
$^4$ Nanjing University\\
$^5$ Beijing Jiaotong University\\
$^6$ Huazhong University of Science and Technology\\
$^7$ Peking University\\
$^8$ Beijing University of Posts and Telecommunications\\
$^9$ Fudan University\\
$^{10}$ Zhejiang Lab\\
$^{11}$ Hygon Information Technology Co., Ltd.\\[0.25em]
$^*$ Corresponding author: \texttt{liansg@chinaunicom.cn}\par}
\end{minipage}
}
\makeatother

\date{}

\begin{document}
\maketitle

\begin{abstract}
Large model inference optimization serves as a key foundation for supporting the scalable, low-cost, and highly stable operation of large model services. Centered on token-oriented inference optimization technology, this paper proposes for the first time a four-layer technical architecture consisting of "Multi-model Fusion, Model Optimization, Compute-Model Fusion, and Compute-Network-Model Fusion." It systematically reviews the key technologies and current industry status across these four levels and analyzes the application value of related technologies in real-world business scenarios. This paper provides a practical technical path for reducing token production costs, improving token service efficiency, ensuring the stability of token supply, and driving the transition of large model services from being merely "callable" to being "operable."
\end{abstract}


\vfill
\begin{figure}[H]
\centering
\includegraphics[width=0.95\linewidth]{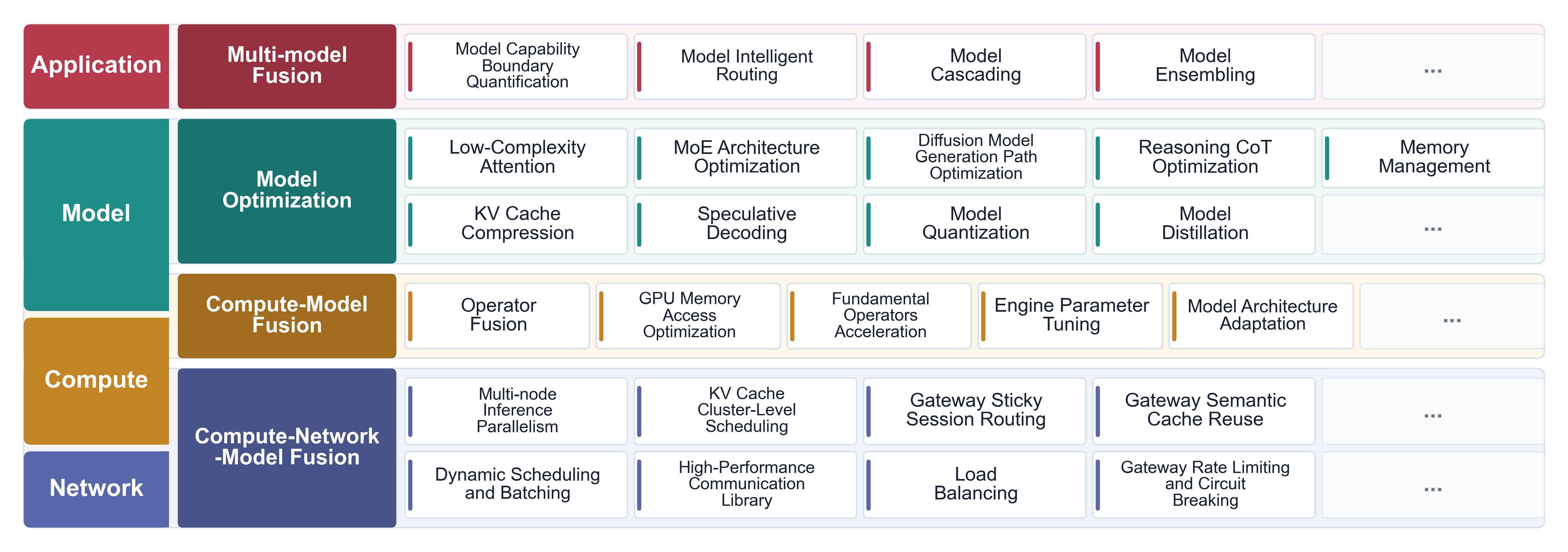}
\caption{Overview of large model inference optimization technologies.}
\label{fig:0_1}
\end{figure}

\nocite{*}

\input{sections/01_body}

\bibliographystyle{unsrtnat}
\bibliography{references}

\end{document}

%% file: sections/01_body.tex
\section{Introduction}

Artificial intelligence is moving rapidly from technological breakthroughs to industrial-scale deployment. Supported by China's policy environment, expanding computing infrastructure, and growing application demand, China's large AI model industry has entered a stage of accelerated growth. In this process, the token has become a core operational unit for large model services. As the minimum unit of information processed by large models, a token is measurable, billable, and tradable, making token volume an important indicator of service activity, resource consumption, and industrial value. According to data released by China's National Data Administration, China's average daily token processing volume had exceeded 140 trillion by March 2026.

The rapid growth of token volume is pushing large model services into a new stage of large-scale token operations. At this stage, the competitiveness of a token platform depends not only on model access and computing capacity, but also on the ability to deliver low-cost, low-latency, high-quality, and governance-ready services under production workloads. These workloads often involve massive requests, high-concurrency access, long-context interaction, multi-model collaboration, and complex agent workflows. Meanwhile, large-scale token operations introduce significant system-level challenges. Stronger models are better suited to complex tasks, but they also bring higher per-token cost, larger GPU memory consumption, and longer inference latency. In contrast, production traffic contains many lightweight or repetitive requests, such as short-form Q\&A, structured extraction, long-document processing, multi-turn dialogue, and tool invocation. Processing all requests with the strongest model would lead to inefficient resource use, higher tail latency, and reduced service stability. Fine-grained scheduling, cache reuse, inference acceleration, and service governance are therefore essential. These challenges become more pronounced as agent applications evolve from conversational interaction to decision-making and task execution. The inference pipeline is no longer a single model call. Instead, end-to-end service efficiency is jointly affected by Prefill, Decode, KV cache management, model routing, tool invocation, gateway scheduling, and fallback control. Large model inference optimization has therefore become a key technical foundation for token operations. Its objective is not merely to accelerate a single model, but to achieve a systematic balance among quality, cost, latency, throughput, stability, and security.

This paper reviews inference optimization technologies for token operations across four complementary dimensions: Multi-model Fusion, Model Optimization, Compute-Model Fusion, and Compute-Network-Model Fusion. Multi-model fusion focuses on request-level coordination among different models. It includes model capability boundary quantification, intelligent routing, model cascading, and model ensembling, enabling each request to be assigned to an appropriate model or model combination according to task complexity, quality requirements, and cost constraints. Model optimization targets the intrinsic cost of token generation by improving attention mechanisms, MoE architectures, generation paths, chain-of-thought processes, memory management, KV cache compression, speculative decoding, model quantization, and model distillation. Compute-model fusion improves the execution efficiency of model workloads on hardware through operator fusion, memory access optimization, foundational operator acceleration, engine parameter tuning, and model architecture adaptation. Compute-network-model fusion further addresses large-scale serving efficiency by coordinating multi-node inference parallelism, KV cache cluster scheduling, sticky-session routing, semantic cache reuse, dynamic batching, high-performance communication libraries, load balancing, rate limiting, and graceful degradation, thereby supporting stable token service delivery under high-concurrency and multi-tenant workloads.

Overall, inference optimization for token operations is a foundational engineering capability for large-scale model deployment. Effective optimization requires coordinated design across models, computing resources, networks, gateways, and business traffic. Such system-level coordination is essential for reducing token production costs, improving service efficiency, ensuring stable token supply, and enabling large model services to evolve from basic API accessibility to sustainable operational capability.

\section{Multi-model Fusion}

Multi-model fusion is a key technical direction for large model services to evolve from "a single model handling all requests" to "optimal allocation through multi-model collaboration". Its core lies in achieving a systematic balance among quality, cost, and latency through mechanisms such as model capability evaluation, adaptive request dispatching, sequential verification, and parallel fusion. This section focuses on four aspects: model capability boundary quantification, intelligent routing, model cascading, and model ensembling. Among these, model capability boundary quantification is the prerequisite for fusion decisions. It characterizes the applicable boundaries of each model across dimensions such as knowledge, reasoning, coding, and long contexts through standardized evaluation and runtime observation, providing a comparable evidence base for subsequent scheduling. Intelligent routing makes a one-time dispatching decision at the request entry point based on task characteristics and model profiles, aiming for high cost-effectiveness (small models handling simple requests) or high performance (dispatching tasks to the most suitable model). Model cascading adopts a ``try-then-judge'' sequential structure: a lightweight model first provides an answer, a validator dynamically evaluates its quality, and only when the quality is insufficient does the system escalate to a stronger model, thereby balancing cost and accuracy under uncertain request distributions. Model ensembling invokes multiple complementary models in parallel and integrates their strengths through ranking, voting, or generative fusion, breaking through the capability ceiling of a single model. The four technologies undertake capability profiling, request distribution, quality validation, and result fusion, respectively, collectively supporting fine-grained token operations for Model-as-a-Service (MaaS) platforms under a multi-model matrix.

\subsection{Model Capability Boundary Quantification}

\subsubsection{Technical Definition}

Model capability boundary quantification refers to the quantitative characterization of a large model's applicable scope across dimensions such as knowledge, reasoning, mathematics, coding, long contexts, tool use, and multimodality, based on standardized evaluation, scenario-specific samples, runtime observation, and statistical analysis. It emphasizes answering ``what the model can do, under which conditions it performs stably, and on which tasks or input distributions it tends to fail'', rather than merely describing model capability using a single leaderboard rank or a vague notion of strength. From the perspective of token operations, the core value of capability boundary quantification lies in transforming model capabilities from experience-based judgments into a comparable, repeatable, and traceable evidence system. Its results can serve as input for model selection, version admission, critical business stress testing, risk grading, and model routing strategies.

\begin{figure}[htbp]
\centering
\includegraphics[width=0.95\linewidth]{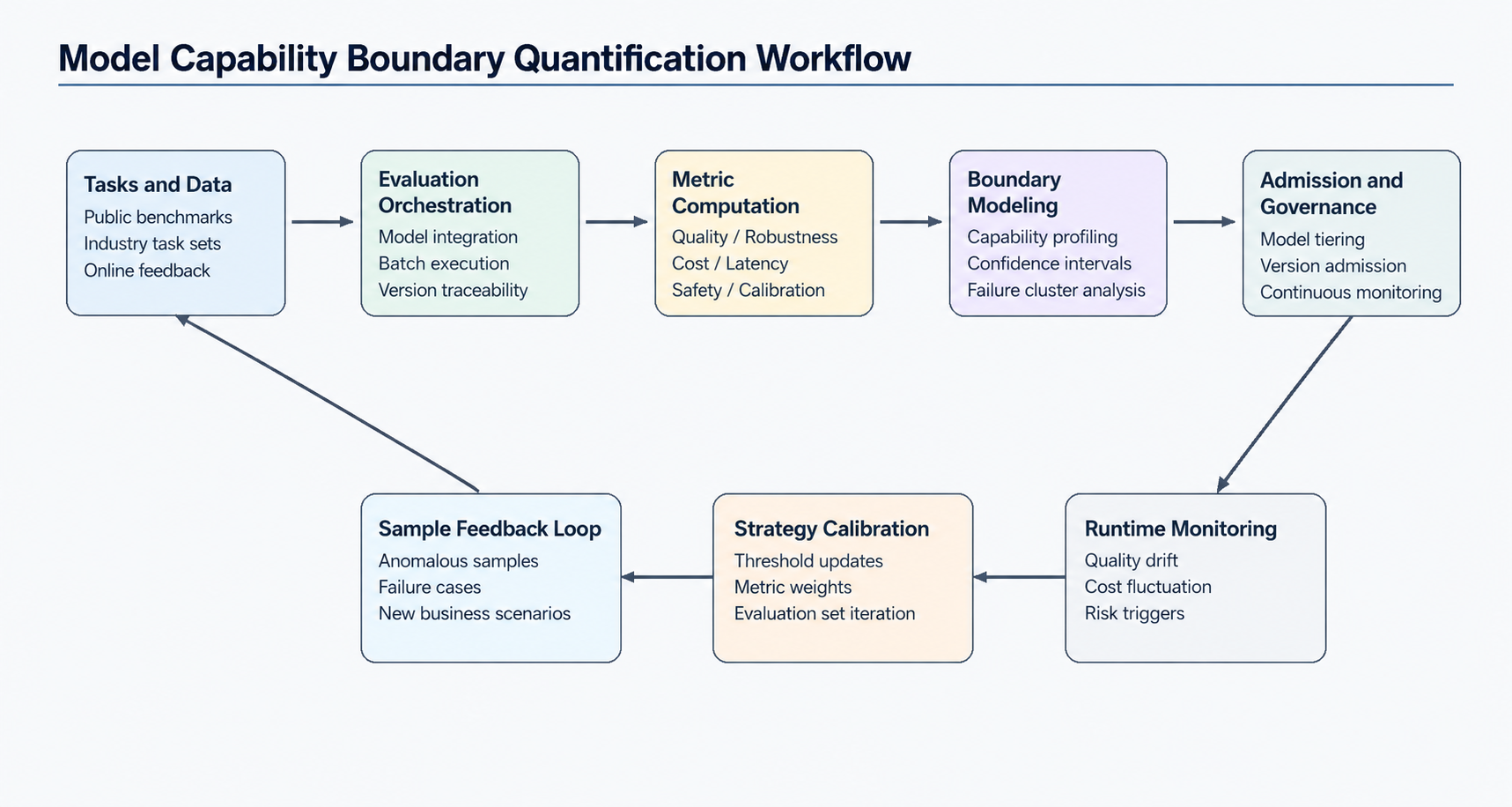}
\caption{Model capability boundary quantification workflow.}
\label{fig:1_1}
\end{figure}

\subsubsection{Industry Trends and Developments}

Industry research on model capability boundaries is moving from single-benchmark rankings toward systematic evaluation methodologies. One survey categorizes evaluation tasks into general capabilities, domain capabilities, and targeted capabilities, and points out that issues such as data contamination, cultural and linguistic bias, and insufficient dynamic environment evaluation can affect evaluation conclusions \cite{ref001}. An earlier comprehensive survey on large language model evaluation also emphasizes that evaluation data, prompt templates, evaluators, statistical significance, and reproducibility conditions can all alter model capability judgments \cite{ref002}. This indicates that ``capability boundaries'' are not fixed labels but observable ranges determined jointly by task distribution, input constraints, evaluation metrics, and runtime environments.

\begin{figure}[htbp]
\centering
\includegraphics[width=0.95\linewidth]{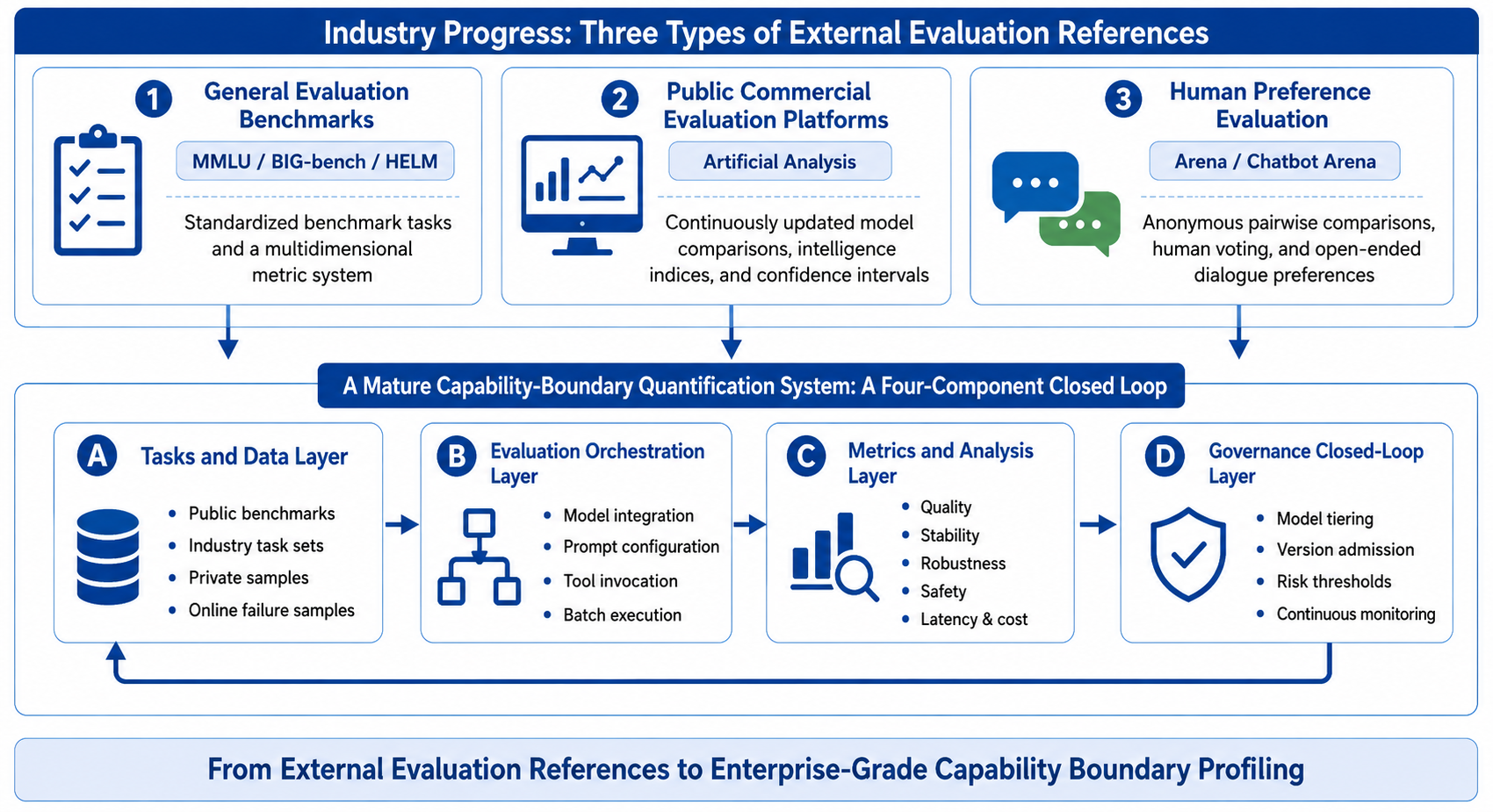}
\caption{Industry progress and system architecture for model capability boundary quantification.}
\label{fig:1_2}
\end{figure}

General evaluation benchmarks remain the foundation for quantifying capability boundaries. MMLU measures model knowledge and problem-solving ability through a multi-subject, multiple-choice question set \cite{ref003}, while BIG-bench observes the extrapolation trends of language models in reasoning, memory, symbolic operations, etc., using a larger and more open set of tasks \cite{ref004}. HELM represents a more comprehensive evaluation paradigm, comparing models under uniform conditions while simultaneously incorporating metrics such as accuracy, calibration, robustness, fairness, bias, toxicity, and efficiency \cite{ref005}. These benchmarks have facilitated cross-model horizontal comparisons, but they also reveal the limitations of aggregate scoring: a model that leads in comprehensive metrics may still exhibit weaknesses in domain-specific question answering, long-document understanding, structured output, or safety scenarios.

Public commercial evaluation platforms further enable continuous observation of capability boundaries. Artificial Analysis provides continuously updated model evaluations and comparisons for frontier models, with a key metric called the ``Intelligence Index'' that is composed of multiple public benchmarks covering agent tasks, coding, general ability, scientific reasoning, and other dimensions, while also disclosing methodological details such as test composition, weights, and confidence intervals \cite{ref006}. The value of such platforms lies in providing a continuously trackable horizontal reference, allowing model version updates, capability drift, and differences among suppliers to be sensed more quickly; their limitation is that public benchmarks may still suffer from data contamination, task bias, and mismatch with business scenarios, and cannot directly replace internal enterprise-specific evaluations.

Human preference benchmarks represented by Arena supplement traditional automated evaluations. Arena uses anonymous pairwise comparison and human voting to measure open-ended conversational experiences, covering aspects of expressive quality, instruction-following ability, and interaction preferences that are difficult for automatic scoring to evaluate \cite{ref007}. The Arena leaderboard extends public arena-style evaluation into dynamic rankings for capabilities including text, code, image, and video, providing an important reference for observing a model's relative performance in real user interactions \cite{ref008}. However, human preference evaluations are also influenced by voter population, prompt distribution, language environment, and temporal changes, making them more suitable as one piece of evidence for capability boundary profiling rather than the sole criterion.

From an engineering perspective, a mature capability boundary quantification system typically includes four components. The first is the task and data layer, covering public benchmarks, industry task sets, private business samples, red-team samples, and online failure samples. The second is the evaluation orchestration layer, responsible for model integration, prompt configuration, tool invocation, multi-turn interactions, version logging, and batch execution. The third is the metric and analysis layer, which computes quality, stability, calibration, robustness, safety, latency, and resource consumption metrics, and identifies failure clusters, boundary samples, and out-of-distribution samples. The fourth is the governance and closed-loop layer, which translates evaluation results into model grading, version admission, risk thresholds, critical scenario regression testing, and continuous monitoring mechanisms. Domestic industrial practice is advancing along a ``capability--scenario'' bidirectional driver. For example, China Unicom, based on the A-Eval benchmark, categorizes typical application scenarios into multi-level capability classes and constructs a mapping graph from ``model $\rightarrow$ capability class $\rightarrow$ capability level $\rightarrow$ scenario'' \cite{ref009,ref010,ref011}; in the OpenClaw agent scenario, it further extends capability boundary quantification to real task sets, task metadata, and traceable decision bases, shifting evaluation from general horizontal comparison to task-specific evaluation oriented to business pipelines \cite{ref012}. Such closed-loop processes upgrade capability boundary quantification from a one-time horizontal assessment into a foundational capability supporting long-term token operations.

\subsection{Model Intelligent Routing}

\subsubsection{Technical Definition}

\begin{figure}[htbp]
\centering
\includegraphics[width=0.95\linewidth]{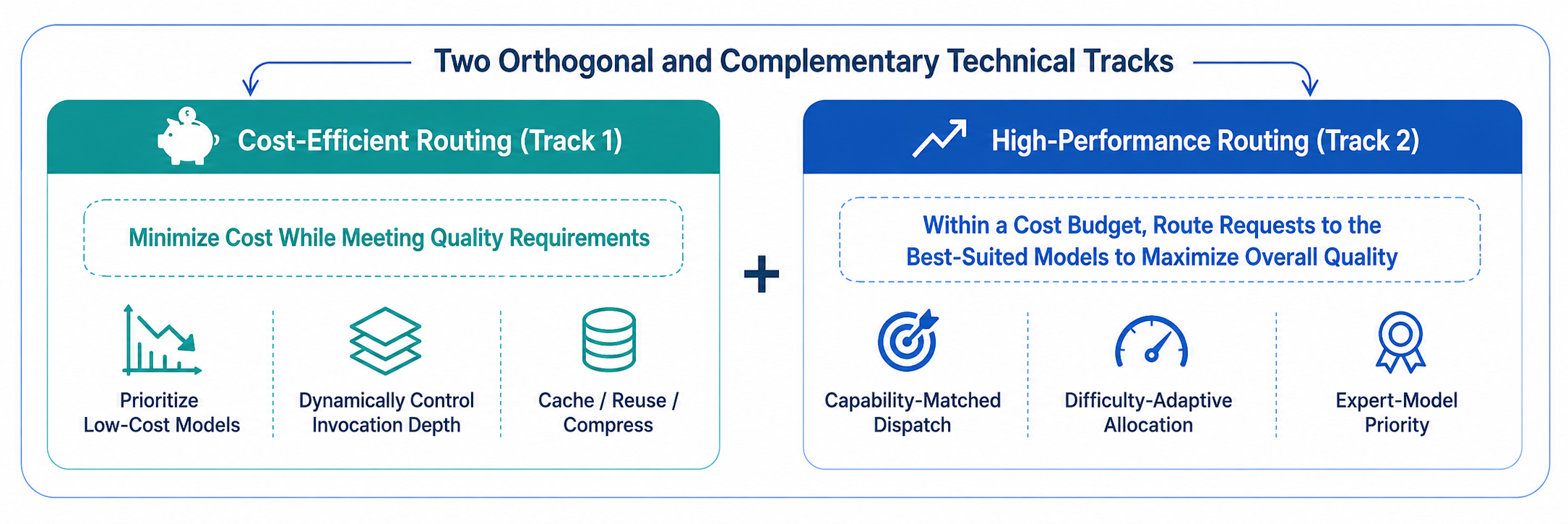}
\caption{Cost-effective and high-performance routing.}
\label{fig:1_3}
\end{figure}

Intelligent routing is the core capability of the MaaS platform for multi-model collaborative scheduling. It means that when a request enters the inference gateway, a lightweight decision-making module comprehensively evaluates the request characteristics, model capability profiles, and cost-quality requirements to automatically select the most suitable model backend for inference. The problem stems from the cost-quality contradiction in single large model services: the primary model offers higher quality for complex tasks but at a significantly higher cost, while real traffic includes a large number of low-complexity requests such as short Q\&A, format conversion, and structured extraction, which can be well handled by smaller parameter models. If all requests are routed to the primary model, it leads to unnecessary token costs and latency pressure. By goal, intelligent routing is divided into two orthogonal yet combinable technical lines: first, cost-effective routing, which minimizes inference costs under quality constraints, typically implemented as a dual-model routing where a small auxiliary model handles requests deemed simple, while the primary model serves as the quality baseline and fallback; second, high-performance routing, which, under a given cost budget, distributes requests based on task type, domain characteristics, and modality to the most proficient model, maximizing overall output quality and breaking through the capability ceiling of a single model.

\subsubsection{Industry Trends and Developments}

\begin{figure}[htbp]
\centering
\includegraphics[width=0.95\linewidth]{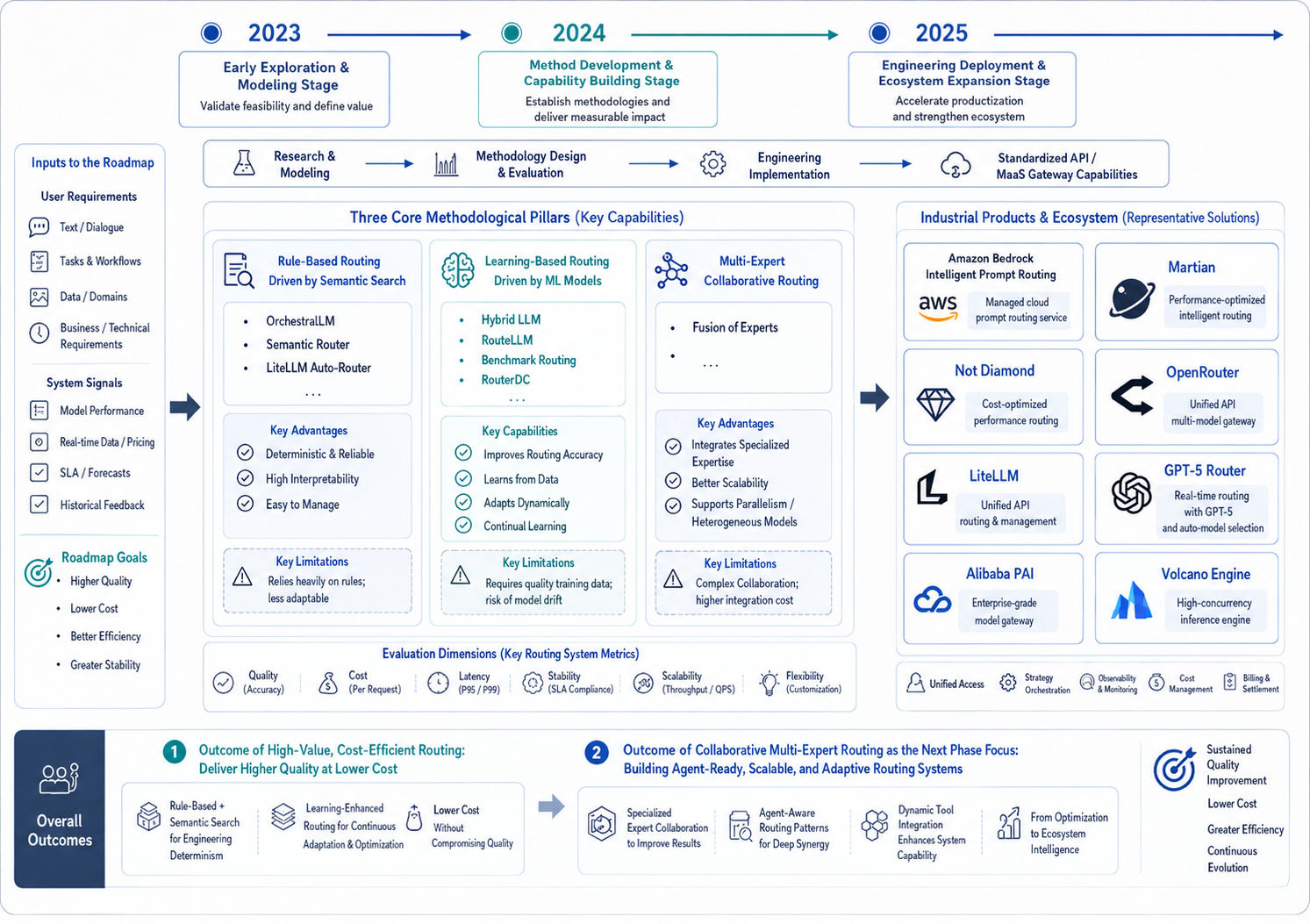}
\caption{Industry progress in intelligent routing.}
\label{fig:1_4}
\end{figure}

Since 2023, academia and industry have developed a relatively complete methodological spectrum around intelligent routing. Related work can be broadly categorized into three types: routing based on semantic retrieval and rules, routing based on learning-driven discriminators, and high-performance routing for multi-model complementarity. Building on this, intelligent routing capabilities have gradually been productized by cloud vendors, model platforms, and independent startups, becoming one of the key capabilities of large model API gateways and MaaS platforms.

The first category is routing based on semantic retrieval and rules. These methods typically transform the question of "whether to invoke a small model" or "which expert model to call" into a similarity matching problem within a vector space. For dialogue state tracking tasks, OrchestraLLM constructs separate expert libraries for small and large models, and based on the assumption that "semantically similar samples have similar difficulty," it dynamically decides which expert model to call through nearest neighbor retrieval \cite{ref013}. The open-source project Semantic Router abstracts a similar idea into a general-purpose routing tool, where developers maintain several Routes and their corresponding responses, and the system performs millisecond-level routing decisions using the cosine similarity between request embeddings and Routes \cite{ref014}. This has been integrated as an Auto-Router option by gateways such as LiteLLM \cite{ref015}. The advantage of this approach lies in low cold-start costs, strong interpretability, and minimal inference overhead; its drawback is that its effectiveness heavily depends on the coverage of the response library, and it has limited generalization capability when encountering new requests that are semantically distant.

The second category is based on learning discriminators or preference routing. This approach models routing as classification, regression, or ranking problems, with supervision signals sourced from human-annotated data, Judge-LLM evaluations, human preference data, or public benchmarks. Hybrid LLM trains a query router to predict quality differences between small and large models for queries, allowing users to dynamically balance cost and quality through thresholds. The paper reports a reduction of up to 40\% in large model invocations without compromising response quality \cite{ref016}. RouteLLM trains its router using human preference data from Chatbot Arena, comparing implementations such as similarity-weighted Elo, matrix factorization, BERT classifiers, and Causal-LLM classifiers. Released as an open-source framework, the authors report cost reductions of over 2x in some scenarios without sacrificing response quality \cite{ref017}. Shnitzer et al. further reduce the problem of "selecting the best LLM for a new task" to a set of binary classification tasks, proposing a method that relies solely on existing public benchmarks to train the router, avoiding the need for re-annotation of target tasks \cite{ref018}. RouterDC addresses the convergence issue of traditional routers when multiple models are suitable for a given query by introducing contrastive learning objectives, achieving superior results on both in-distribution and out-of-distribution test sets compared to existing methods \cite{ref019}.

The third category is high-performance routing for multi-model complementarity. When the goal shifts from "saving costs" to "enhancing capability limits," the research focus of routing transitions from "whether to downgrade" to "which expert model is best suited to handle the request." Fusion of Experts formalizes this problem by organizing multiple models with complementary domain expertise into an expert pool, where a learned fusion mechanism either selects a specific expert for each request or weights their outputs, enabling overall performance on cross-domain mixed distributions to surpass any single expert \cite{ref020}. This approach is also applicable in a routing deployment mode: rather than actually invoking all experts in parallel, the request is dispatched only to the expert whose profile best matches.

At the industrial product level, intelligent routing has evolved from research prototypes to standardized APIs and gateway capabilities. Amazon Bedrock Intelligent Prompt Routing supports request-level routing within the same model family based on predicted response quality, with the official claim of reducing costs by up to 30\% without sacrificing accuracy \cite{ref021}. Martian Model Router offers drop-in replacement capabilities via an OpenAI-compatible interface, estimating the performance of different models on specific requests through predictive Model Mapping, and exposing explicit cost-quality trade-off parameters \cite{ref022}. Not Diamond is positioned as an agent-oriented model router, focusing on training custom routers based on client evaluation data, achieving over 50\% cost savings and over 10\% accuracy improvement \cite{ref023}. OpenRouter serves as a model aggregation layer, providing unified API and billing capabilities, and has integrated over 400 active models from more than 60 service providers, routing based on cost, speed, and availability in the backend \cite{ref024}. LiteLLM leans more towards enterprise self-hosted gateways, offering unified OpenAI protocol, virtual keys, quota management, observability, retries, degradation, and load balancing capabilities, and has introduced an Auto-Router option driven by Semantic Router \cite{ref015}.

Intelligent routing has also begun to enter the ecosystem of leading model platforms and cloud providers. OpenAI introduced built-in real-time routing in GPT-5, released in August 2025, combining a fast general-purpose model with a deep reasoning model into a unified system. The router dynamically decides which sub-model to call based on conversation type, complexity, tool requirements, and user intent, and continuously trains using real signals such as user model-switching behavior, response preference rates, and accuracy measurements \cite{ref025}. Domestic cloud providers have also launched relevant capabilities: Alibaba Cloud PAI's "LLM Intelligent Routing" primarily focuses on load balancing and scheduling among multiple inference instances, optimizing based on metrics such as prompt length and generate length \cite{ref026}; Volcengine Ark's "Intelligent Model Routing" selects models on a per-request basis, dynamically matching the most suitable model according to the prompt of each request, and supports two strategies: "cost-priority" and "effect-priority" \cite{ref027}.

In the practice of MaaS platforms for large model services, the industry has gradually reached a consensus on high-cost-performance intelligent routing: building an iterative closed loop around "evaluation---capability profiling---routing---gateway---feedback," first establishing benchmarks in scenarios such as Q\&A, reasoning, code, mathematics, and tool invocation, characterizing each model's capabilities from the dimensions of quality, speed, cost, and stability, and then scheduling accordingly. Routing strategies have generally evolved from rule-based and semantic matching to intelligent scheduling based on difficulty discrimination---low-complexity requests are diverted to lightweight auxiliary models to reduce costs, while difficult-to-judge or anomalous requests are automatically reverted to the main model to ensure a baseline. For high-stability scenarios such as agents and tool invocation, a conservative strategy prioritizing the main model is adopted. At the system level, a pluggable proxy gateway is typically used to uniformly handle distribution and logging, while offline evaluations and online data feedback continuously drive iteration. Multi-party experience shows that in natural language Q\&A and even some tool invocation scenarios, auxiliary models can achieve results comparable to the main model, significantly reducing costs without substantial quality loss. A natural extension is to upgrade from a binary downgrade to fine-grained scheduling across multiple models and scenarios, and to expand into complex agent, multi-turn, and multimodal scenarios.

Overall, the current industry shows two clear trends in intelligent routing. First, cost-effective routing remains the most mature and mainstream direction, with industrial products generally taking "saving money without compromising quality" as their core selling point, corresponding to the academic strong-weak model binary routing and learning discriminator approach. Second, high-performance routing and multi-model complementarity are becoming the next key focus. The GPT-5 real-time router evolves routing from a gateway-side bypass decision into an endogenous module of the model system, indicating that intelligent routing capabilities will further deeply integrate with the base model itself and the Agent framework.

\subsection{Model Cascading}

\subsubsection{Technical Definition}

\begin{figure}[htbp]
\centering
\includegraphics[width=0.95\linewidth]{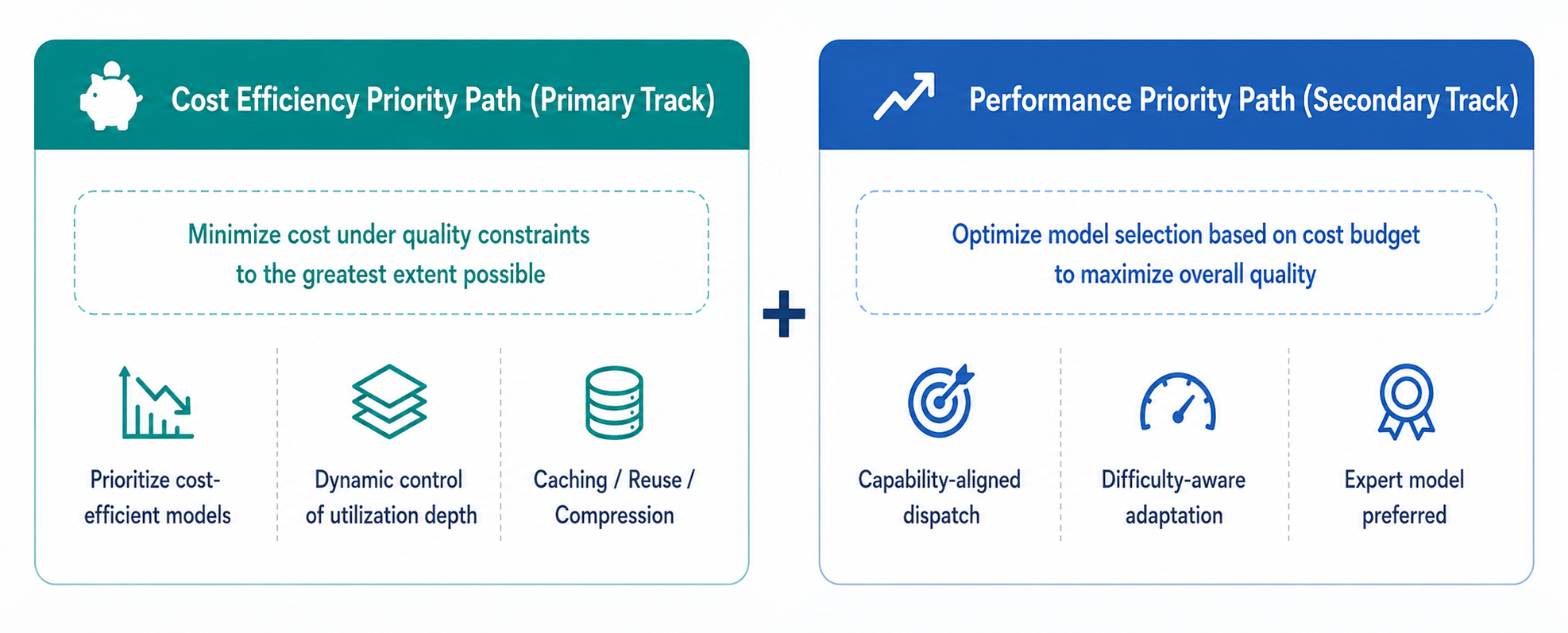}
\caption{Scenario-oriented objectives of model cascading.}
\label{fig:1_5}
\end{figure}

Model cascading refers to the MaaS platform processing a request by sequentially feeding it to multiple candidate models in order of increasing cost. At each step, a validator or scoring function determines whether the current response meets an acceptable quality threshold. If it does, the process terminates immediately, and the response is returned. Otherwise, it escalates to a stronger, more expensive model until a final response is given. The motivation for this approach lies in the unreliability of static difficulty prediction in intelligent routing for tasks such as complex reasoning, code generation, and domain-specific Q\&A. Relying solely on the request text to gauge difficulty in advance is often insufficient. Cascading transforms difficulty assessment from static prediction at the request side to dynamic evaluation of the actual model output through a "try first, judge later" serial structure, thereby balancing cost and quality across a broader distribution of requests. In terms of objectives, cascading is divided into two main lines: one focuses on cost-saving cascades, where as many requests as possible are handled by the smaller models at the front of the chain, primarily suited for tasks like knowledge Q\&A and text classification that can be measured by one-shot responses. The other aims at quality improvement, leveraging a small model for an initial attempt and a stronger model for the final decision, thereby enhancing overall accuracy on difficult problems beyond any single model. This is mainly applicable to tasks requiring high correctness, such as mathematical reasoning, code generation, and tool invocation.

\subsubsection{Industry Trends and Developments}

\begin{figure}[htbp]
\centering
\includegraphics[width=0.95\linewidth]{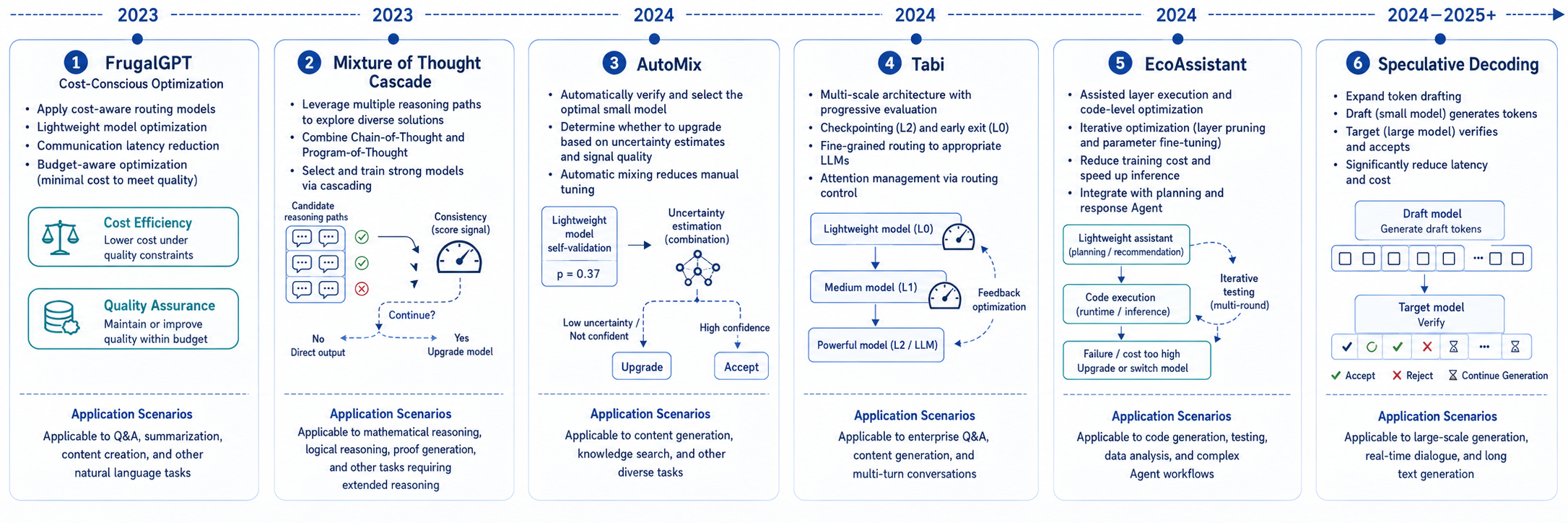}
\caption{Evolution of model cascading.}
\label{fig:1_6}
\end{figure}

Model cascading is one of the most systematically studied directions in large model cost optimization, having established a clear methodological genealogy since 2023, and gradually expanding to complex tasks involving agents, reasoning, and tool invocation.

FrugalGPT is a foundational work in this direction. Chen, Zaharia, and Zou proposed the LLM Cascade concept in their paper, which calls candidate models in ascending order of cost. A lightweight, independently trained scoring function assigns a reliability score between 0 and 1 to each step's answer. If the score reaches a threshold, the process terminates early; otherwise, it proceeds to query the next model. The authors model the selection of the candidate model chain and the threshold as an optimization problem with budget constraints. They report that on multiple datasets, FrugalGPT can match the performance of strong models like GPT-4 while reducing costs by up to 98\%, or improve accuracy by 4 percentage points at the same cost \cite{ref028}. This work explicitly defines two typical usage modes: "maintaining quality while significantly cutting costs" and "enhancing quality at the same cost." It remains the most frequently cited design paradigm for implementing model cascades in industry.

Mixture of Thought Cascade extends the cascading idea to reasoning tasks. Yue et al. observed that relying solely on single-response reliability scores is unstable in complex mathematical and symbolic reasoning, and thus proposed using the consistency of weak model answers as a difficulty signal: for the same problem, the weak model samples multiple reasoning paths, statistically evaluates the consistency of answers, and combines Chain-of-Thought and Program-of-Thought representations for voting or verification. Only problems with consistency below a threshold are escalated to the strong model. The paper demonstrates on six mathematical and symbolic reasoning datasets that this cascade can achieve accuracy close to that of using GPT-4 alone, but at only about 40\% of its cost \cite{ref029}. The significance of this work lies in extending the verifier from an "independent small model" to a "statistical criterion based on the weak model's own outputs," avoiding the cost of training an additional verifier and making it more suitable for deployment in reasoning tasks.

AutoMix further integrates cascade with few-shot self-verification. Madaan et al. propose generating answers and self-verification probabilities with a small model, then a meta-verifier based on a partially observable Markov decision process decides whether to escalate the problem to a stronger model; the paper reports that across multiple language models and datasets, this approach can reduce computational costs by over 50\% while maintaining comparable performance \cite{ref030}. At the system level, HKUST's Tabi work targets discriminative NLP services, using calibrated confidence to first respond to low-risk requests quickly with a small model, escalating low-confidence requests to an LLM, and introducing attention pruning and weighted ensemble on the escalation path to compensate for system overhead, demonstrating from an engineering perspective the feasibility of cascade architecture in production inference services \cite{ref031}.

Model cascading has also been validated in complex agent tasks for tool invocation and code generation. EcoAssistant introduces the cascading concept into code-driven question-answering agents, building a hierarchy of assistants from "cheap to expensive": it first engages the cheapest assistant in dialogue with the code executor to iteratively generate and debug code, only upgrading to a more expensive assistant when the current dialogue fails to solve the problem, and uses historical successful question-answer pairs as examples to benefit the low-cost assistant. The authors report that this approach achieves over 10 percentage points higher success rates than a GPT-4 assistant on code-driven question-answering tasks in areas such as weather, stocks, and locations, while costing less than half of GPT-4 \cite{ref032}. This result is particularly instructive for agent workloads with heavy tool invocation: it suggests that in complex agent-oriented tasks, cascading not only saves money but may also be more robust than a single direct call to a strong model due to the iterative debugging mechanism.

In recent years, model cascading has further integrated with inference acceleration techniques at the system level. Speculative decoding can be seen as the ultimate embodiment of the cascading idea at the token level: a small draft model generates multiple candidate tokens in parallel, and then the target model verifies which tokens are acceptable in a single forward pass, significantly improving decoding throughput while ensuring the output distribution remains unchanged \cite{ref033}. Mainstream inference frameworks such as vLLM and SGLang have already adopted speculative decoding as one of their default capabilities, extending the cascading mechanism from the "request level" to the "token level," thereby further reducing the cost per token.

At the level of industrial products, cascading and fallback capabilities have become standard features of mainstream LLM gateways. The Databricks Unity AI Gateway provides a fallback mechanism that automatically switches to backup models in a preset order when the primary model returns rate-limiting or 5xx errors, logging all calls to a unified usage and load log \cite{ref034}. OpenRouter also offers automatic fallback capabilities at the underlying level, based on cost, speed, and availability \cite{ref024}. These product-level solutions typically start with "failure fallback" as an entry point, but their underlying calling patterns are consistent with score-based cascading, serving as an engineering baseline for cost-saving cascading.

In the practice of MaaS platforms oriented towards large model services, the industry generally adopts intelligent routing as the main form of multi-model collaboration, while simultaneously planning an evolution toward model cascading: routing addresses the "difficulty of pre-execution prediction," while cascading handles the "in-execution evaluation of results," with the two coordinating to serve requests of varying quality sensitivity and cost budgets. At the model level, a multi-parameter tier of self-developed language models, combined with external strong models, can organize a cascading structure of ``lightweight self-developed model $\rightarrow$ medium-scale self-developed model $\rightarrow$ main model or external strong model.'' At the evaluation level, multi-dimensional benchmarks and offline evaluation processes established around routing can simultaneously train request-level difficulty classifiers and answer-level reliability scorers, reusing existing data loops. At the system level, a pluggable proxy gateway supports extending a single decision into multiple serial calls, recording the responses, costs, latency, and final source of each model level through a unified logging and observability system, making the cascading effect quantifiable and auditable. Subsequent focus typically covers three types of scenarios: tasks with verifiable consistency, such as mathematical reasoning and code generation, leveraging Mixture of Thought Cascade and AutoMix, using the consistency of weak model answers as a difficulty signal; agent and tool-call-intensive tasks, leveraging EcoAssistant, using tool execution results as scoring signals; and high-concurrency dialogue and discriminative tasks, combining cascading with speculative decoding, extending cost optimization from the request level to the token level. This, together with intelligent routing, forms a two-layer cost optimization mechanism of "pre-execution prediction + in-execution verification."

Overall, research and industrial practice in model cascading exhibit two clear evolutionary paths. First, the validation signals have shifted from "independently trained scoring functions" to "consistency, self-verification, and execution feedback of the model output itself," making cascading easier to deploy on new tasks lacking annotations. Second, the granularity of cascading has extended from the request level to the token level, and from single-turn Q\&A to complex agent workflows, deeply integrating cascading with inference acceleration and agent orchestration.

\subsection{Model Ensembling}

\subsubsection{Technical Definition}

\begin{figure}[htbp]
\centering
\includegraphics[width=0.95\linewidth]{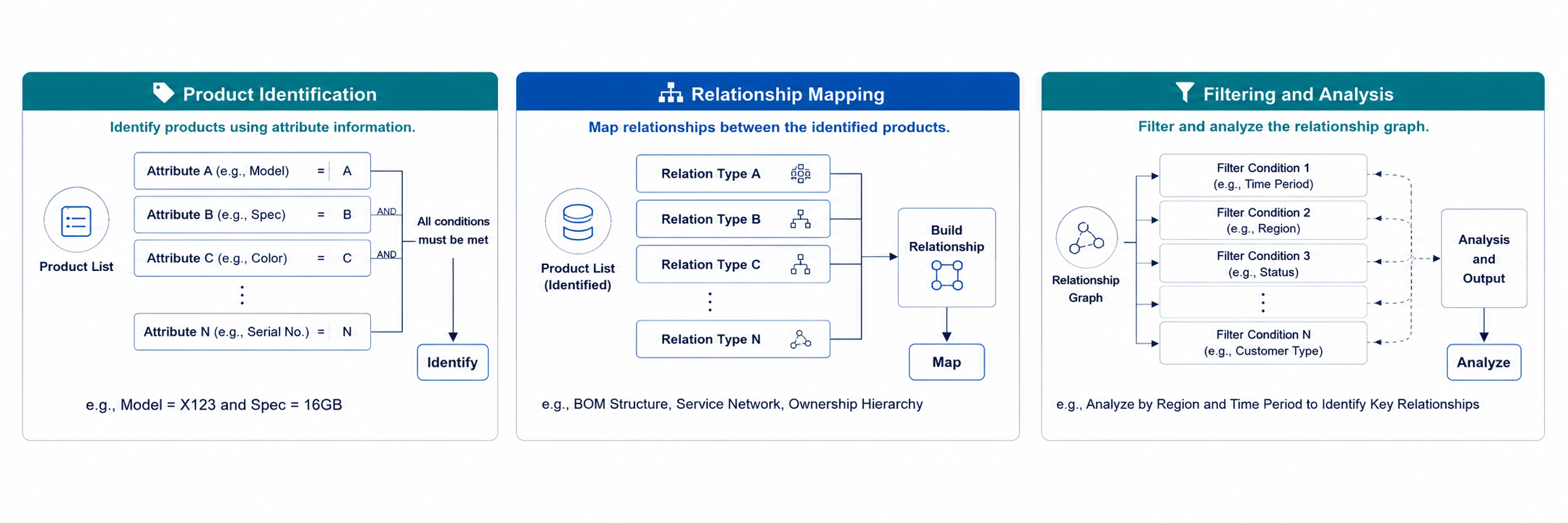}
\caption{Three forms of model ensembling.}
\label{fig:1_7}
\end{figure}

Model ensembling refers to the MaaS platform's parallel invocation of multiple candidate models for the same request, followed by evaluation, weighting, or generative merging of the candidate outputs by a sorter, aggregator, or fusion module. This ultimately produces a response that integrates the strengths of multiple models, reducing the variance and bias of any single model through statistical averaging, voting, or learning-based fusion, so that the overall output surpasses the individual performance of any participating model. The motivation for this approach lies in the inherent capability ceilings and domain blind spots of single models in complex tasks. Different models exhibit significant variations in strengths across task subsets such as mathematical reasoning, code generation, and long-form text writing, necessitating parallel invocation and explicit fusion mechanisms to leverage each model's strengths and compensate for their weaknesses. Ensembling primarily serves high-performance scenarios and takes three typical forms: first, self-ensembling of the same model, where multiple invocations are made using different sampling or thinking paths, and the most stable answer is selected through majority voting or self-consistency to reduce randomness errors and improve reasoning accuracy; second, heterogeneous model ensembling, where complementary models with different structures and training data are invoked in parallel, and their outputs are integrated by a sorter or fusion module to combine the domain strengths of different models; third, multi-agent collaboration, where multiple models are organized into an agent structure, with the front-end model generating candidates and the back-end model aggregating and refining them, used for complex creation, debate, and decision-making tasks.

\subsubsection{Industry Trends and Developments}

\begin{figure}[htbp]
\centering
\includegraphics[width=0.95\linewidth]{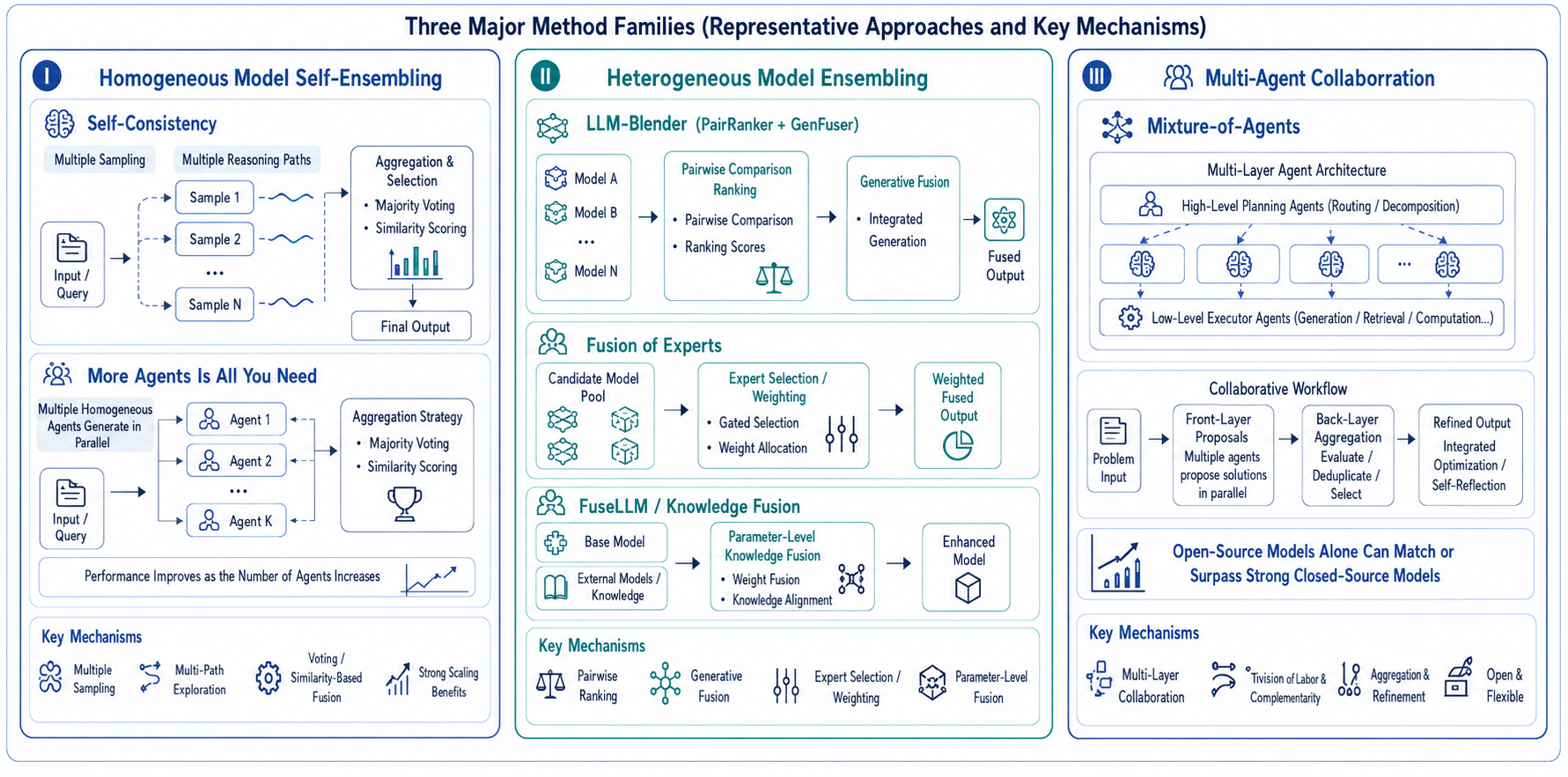}
\caption{Three major industry approaches to model ensembling.}
\label{fig:1_8}
\end{figure}

Around model ensembling, academia and industry have developed a complete spectrum of methods encompassing self-ensembling of identical models, heterogeneous model ensembling, and multi-agent collaboration, with significant quality improvements verified in tasks such as mathematical reasoning, code generation, and open-ended writing.

A representative work of self-ensembling with the same model is self-consistency. Wang et al. proposed the Self-Consistency method based on Chain-of-Thought prompting: for the same problem, the model generates multiple reasoning paths under temperature-based sampling, then performs majority voting on the final answers, replacing single greedy decoding with marginalization over multiple reasoning paths, thereby significantly improving accuracy in mathematical and symbolic reasoning tasks \cite{ref035}. This method is one of the most influential implementations of model ensembling in the era of large models and remains a common baseline for various reasoning benchmarks. The Tencent team's work "More Agents Is All You Need" further extends the self-consistency idea into a more general sampling-voting framework: instantiating multiple independent agents for the same LLM, each answering the same problem, and then aggregating through majority voting or similarity scoring. The paper reports that the performance of this method steadily improves with the number of agents and is orthogonal to existing prompt engineering, CoT, debate frameworks, etc., allowing for combined use \cite{ref036}. This approach offers controllable costs and easy engineering implementation, making it the most direct form of self-ensembling with the same model deployed on MaaS platforms.

A representative work in heterogeneous model ensembling is LLM-Blender. Jiang, Ren, and Lin proposed PairRanker for pairwise comparisons among outputs from multiple candidate LLMs, followed by GenFuser to generate fused outputs. The authors reported overall performance significantly surpassing any single model and various baseline methods, and constructed the MixInstruct dataset to support training based on pairwise comparisons \cite{ref037}. Fusion of Experts formalized a similar problem by combining multiple expert models with complementary domain expertise, using a fusion module trained via supervised learning to either select a single expert or weight multiple expert outputs per request. It demonstrated performance exceeding any single expert on cross-domain mixed distributions, and explicitly controlled the number of expert invocations in a frugal setting \cite{ref020}. Knowledge Fusion advances fusion from the output level to the parameter level. The FuseLLM work distills knowledge from the generation distributions of source models, integrating the capabilities of multiple open-source LLMs with different architectures and training data into a target LLM, validating that the fused model outperforms any source model on reasoning, commonsense, and code generation tasks \cite{ref038}. These works provide a complete methodology for heterogeneous model ensembling, ranging from shallow output fusion to deep parameter fusion.

A representative work in multi-agent collaboration is Mixture-of-Agents. Wang et al. proposed organizing multiple open-source LLMs into a multi-layer agent structure, where the front-layer models propose multiple candidate answers, and the back-layer models aggregate and refine the outputs from the front layers, progressively improving answer quality. The paper reports that MoA, using only open-source models, surpasses GPT-4 Omni on benchmarks such as AlpacaEval 2.0 \cite{ref039}. This result suggests that, through reasonable parallel invocation and layer-by-layer aggregation, ensembling heterogeneous open-source models can potentially approach or even exceed the capability ceiling of closed-source single strong models, providing an important reference for high-performance ensembling.

The concept of model ensembling also appears in a more fine-grained form in the field of inference acceleration. Speculative decoding simultaneously utilizes both the draft model and the target model at each step of the decoding process, with the target model performing parallel verification on multiple candidate tokens from the draft model. In a sense, this is a token-level lightweight ensemble that significantly improves throughput while ensuring the output distribution remains unchanged \cite{ref033}. This approach is philosophically consistent with request-level ensembling, both of which achieve improvements in quality or efficiency by simultaneously invoking multiple models and aggregating outputs through a trusted verification mechanism.

At the industrial product level, model ensembling naturally incurs higher costs, so it currently appears more as an optional capability or a targeted solution for high-value scenarios. Databricks Unity AI Gateway offers fallback orchestration capabilities while also supporting multi-model A/B testing and gradual traffic shifting, laying the groundwork for ensembling based on parallel multi-model validation \cite{ref034}; mechanisms such as multi-model review and multi-model voting for agent workflows are also beginning to appear in enterprise-level agent platforms. Overall, the industry's pace of adopting ensembling is significantly slower than that of routing and cascading. On one hand, ensembling yields stable results in verifiable tasks like mathematical reasoning and code generation, but it is difficult to find a universal, low-cost combiner for open-ended generation. On the other hand, the cost structure of ensembling makes it suitable only for high-value tasks, requiring routing capabilities to jointly determine when ensembling is worthwhile.

In the practice of MaaS platforms oriented towards large model services, model ensembling is typically positioned as a core method for high-performance scenarios, aiming to precisely dispatch complex tasks to the strongest models, leverage the complementary advantages of multiple models, and break through the capability ceiling of a single model. At the model level, a self-developed language ranging from lightweight to main models, multimodal models, and external strong models form a multimodal symbiotic matrix. Combined with capability profiles derived from multi-dimensional evaluations, ensembling combinations can be organized based on complementarity rather than similarity. At the evaluation level, a two-stage offline evaluation process established around intelligent routing also serves ensembling: the first stage labels are used to train rankers and fusers, and the second stage evaluation extends to ensembled outputs, making ensembling comparable and observable under the same metric system as single models. At the system level, a pluggable proxy gateway uniformly handles authentication, parallel invocation, logging, and timeout control, recording candidate outputs, weights, and final results during the fusion stage. Streaming and asynchronous capabilities support long-generation and multi-turn scenarios. Future focus typically involves three directions: first, verifiable tasks such as mathematical reasoning and code generation, introducing Self-Consistency sampling and voting; second, long-text and cross-modal creation, leveraging the fuser to integrate advantages of multiple models; third, complex agents and multi-step decision-making, adopting a Mixture-of-Agents structure with propose-criticize-aggregate layering. Considering that ensembling inherently incurs multiplied inference costs, platforms generally do not enable ensembling for all requests. Instead, intelligent routing first filters high-value requests, which are then handled by ensembling, ultimately forming a three-layer collaborative system of "routing identifies value, cascading controls depth, and ensembling breaks through limits."

Overall, model ensembling is the approach within the multi-model collaboration paradigm that can most directly break through the capability ceiling of a single model, yet it is also the most computationally intensive. Its optimal form in a MaaS platform involves combining it with routing and cascading: routing determines which requests are worth using ensembling for, while cascading controls the depth and cost of the ensemble, ultimately yielding output quality significantly higher than that of a single model on high-value complex tasks at a controllable cost premium.

\section{Model Optimization}

Model optimization is a core technical direction for reducing per-token generation cost and improving generation efficiency in large model inference. Its essence lies in systematically optimizing around model structural characteristics, inference process control, intermediate state compression, and computational precision reduction, thereby achieving equivalent or higher quality output with less computational overhead. This section analyzes nine aspects: low-complexity attention mechanism optimization, MoE architecture optimization, diffusion model generation path optimization, chain-of-thought optimization, memory management, key-value cache compression, speculative decoding, model quantization, and model distillation. Specifically, low-complexity attention mechanisms reduce the computational, caching, and memory access complexity of self-attention through routes such as key-value representation compression, sparse/mixed attention, and linearized state recurrence. MoE architecture optimization leverages sparse activation and expert parallelism to control per-token computation while scaling up model capacity. Diffusion model generation path optimization targets diffusion/flow-matching models, accelerating multi-step generation through training-time trajectory compression or training-free cache reuse. Chain-of-thought optimization reduces ineffective reasoning tokens under the premise of maintaining inference quality via budget control, difficulty adaptation, and redundancy suppression. Memory management structurally stores, retrieves, and updates historical interactions and experiences, reducing redundant computation for long-context and multi-turn dialogues. Key-value cache compression cuts down memory usage and memory access pressure in long-context inference through importance eviction, low-bit quantization, and prefix reuse. Speculative decoding introduces a lightweight draft model to generate candidate tokens in parallel, which are then verified by the target model in one pass, producing the same output with fewer model invocations. Model quantization converts weights and activations to low-bit representations, compressing memory and accelerating matrix operations. Model distillation transfers the knowledge and reasoning capabilities of a large model to a smaller model, enabling low-cost, high-efficiency deployment. Together, these techniques form a multi-level optimization system spanning the structural layer (attention, MoE, diffusion path), the process layer (chain-of-thought, memory management), and the compression layer (key-value cache, speculative decoding, quantization, distillation), providing an essential foundation for scaling large models from ``capable of reasoning'' to ``low-cost, high-efficiency'' real-world deployment.

\subsection{Low-Complexity Attention}

\subsubsection{Technical Definition}

Low-complexity attention mechanisms refer to improvements made to the algorithmic structure of the Transformer self-attention layer in terms of how tokens exchange information, how the Q/K/V (query/key/value) representations are organised, the scope of context visibility, and the paths for long-range memory. The goal is to reduce the computational complexity, KV cache footprint, and memory access pressure during decoding for long-sequence modelling, while preserving the model's ability to capture global semantics, local details, cross-segment dependencies, and complex reasoning chains.

From a technical paradigm perspective, current low-complexity attention mechanisms mainly fall into three categories. The first category consists of full-attention variants with KV compression, where MHA (Multi-Head Attention) typically serves as the standard full-attention baseline, while MQA (Multi-Query Attention), GQA (Grouped-Query Attention), MLA (Multi-Head Latent Attention), etc., reduce the number of KV heads or compress KV representations, thereby lowering KV cache and memory access costs while retaining global modelling capability. The second category is sparse / hybrid attention, which uses mechanisms such as sliding windows, global attention, dynamic sparse selection, KV compression, and attention pooling to reduce ineffective attention connections in long contexts. The third category is linear or more general efficient attention, which alleviates the quadratic growth of standard Softmax attention with context length by linearising computations, using state recursion, or combining attention with state-space structures. In the context of token-oriented operations, low-complexity attention mechanisms directly affect long-context carrying capacity, per-token inference cost, KV cache growth rate, stability in multi-turn tasks, and quality in long-document / long-code tasks.

\begin{figure}[htbp]
\centering
\includegraphics[width=0.95\linewidth]{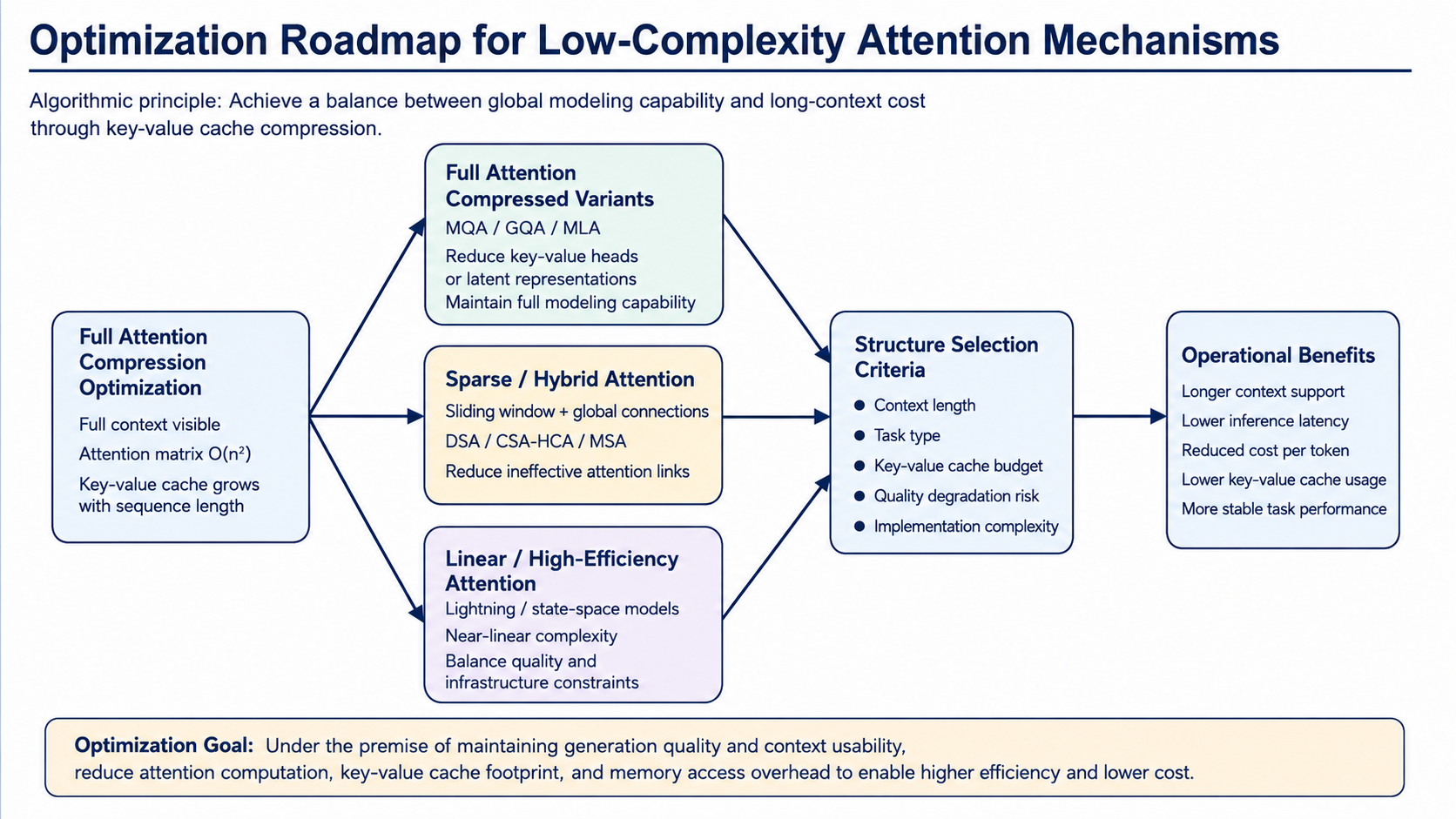}
\caption{Optimization pathways for low-complexity attention mechanisms.}
\label{fig:2_1}
\end{figure}

\subsubsection{Industry Trends and Developments}

In recent years, research on low-complexity attention has shifted from ``accelerating standard attention implementations'' to ``re-engineering long-context information flow''. Relevant surveys point out that long-context large models must systematically trade off global modelling capability, computational complexity, cache cost, and the risk of quality degradation \cite{ref040}. This means that reducing complexity is not the only goal: mainstream models do not simply pursue the lowest complexity order. Instead, under different model scales and task scenarios, they combine full-attention compression, sparse/hybrid attention, and linear/efficient attention to strike a balance among complexity, capability, and cost.

\begin{figure}[htbp]
\centering
\includegraphics[width=0.95\linewidth]{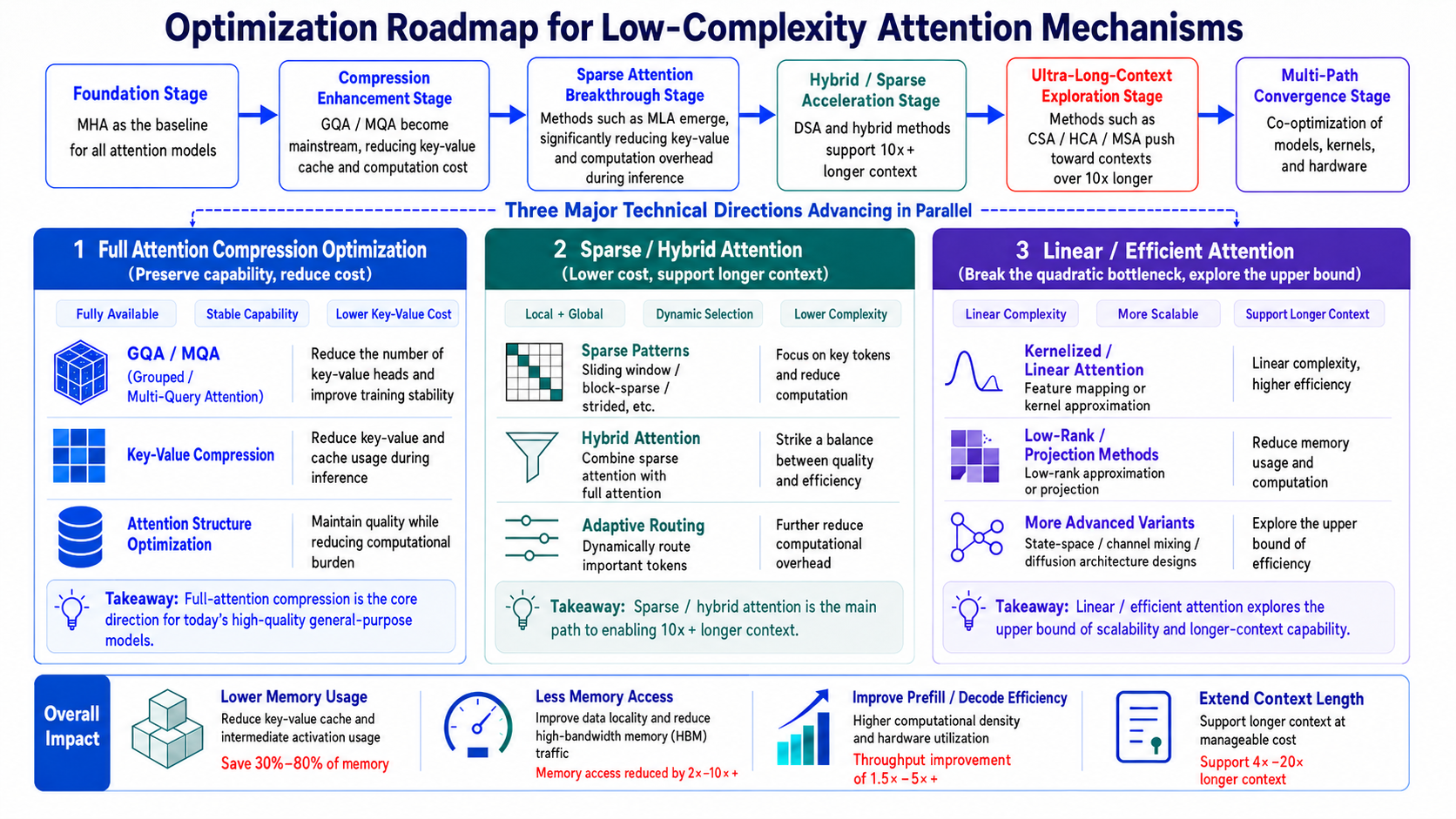}
\caption{Industry progress in low-complexity attention mechanisms.}
\label{fig:2_2}
\end{figure}

The first category is full attention and its KV-compressed variants. MHA remains the most robust standard full-attention structure, but it incurs high KV cache and memory access costs during long-context decoding. GQA reduces the cache size by cutting the number of KV heads. Qwen3 shows that both dense and MoE models employ GQA, combined with query-key normalisation to enhance training stability \cite{ref041}. DeepSeek-V3 adopts MLA, compressing KV into a latent representation space to reduce the KV cache cost during inference \cite{ref042}. MiniMax-M2 uses full-context self-attention in all layers with a GQA configuration for its Q/K/V heads. Its paper explicitly states that although linear and sparse attention offer efficiency advantages, quality loss assessment and infrastructure maturity remain critical constraints for production tasks such as complex reasoning, coding, and agent applications \cite{ref043}. This indicates that KV-compressed full attention remains an important low-complexity baseline for high-quality general-purpose models.

The second category is sparse/hybrid attention. This line no longer requires full pairwise visibility among all tokens; instead, it reduces long-context computational cost through local windows, global connections, dynamic sparse selection, or KV compression. Qwen3.5-Omni employs a hybrid attention MoE framework to support 256K long contexts and multi-modal long-sequence reasoning \cite{ref050}. DeepSeek-V3.2 proposes DeepSeek Sparse Attention (DSA) to lower the attention computation complexity for long contexts \cite{ref051}; DeepSeek-V4 further adopts a hybrid attention architecture combining Compressed Sparse Attention (CSA) and Heavily Compressed Attention (HCA), reducing per-token computation and KV cache footprint in million-token scenarios \cite{ref046}. GLM-5 and GLM-5.1 also reflect the use of the DSA approach for long-context efficiency optimisation \cite{ref047}. MiMo-V2.5 inherits the hybrid sliding-window attention language backbone from MiMo-V2-Flash and supports up to one million tokens of context \cite{ref213}.

The third category is linear or more general efficient attention. This direction attempts to break through the quadratic complexity bottleneck of the standard attention mechanism, but in production-grade large models it typically needs to simultaneously address quality stability, low-precision representation, prefix caching, speculative decoding, and inference framework adaptation. A representative work in this line is the Mamba architecture based on the State Space Model (SSM), which models long sequences with linear time complexity through selective state recursion, compressing the entire history into a fixed-size hidden state and fundamentally eliminating the linear growth of KV cache with context length \cite{ref214}. Notable examples include Falcon Mamba, which trained a 7B base model without any attention layers on 5.8 trillion tokens, demonstrating that a pure Mamba architecture can perform on par with or even better than mainstream Transformer models at that scale, while maintaining constant memory footprint during long-sequence generation \cite{ref215}. Jamba was the first to interleave Transformer attention layers and Mamba layers in a staggered ratio and combine them with MoE, reducing KV cache footprint by approximately an order of magnitude while supporting 256K long contexts, validating the feasibility of ``attention + state recursion'' hybrid architectures for scaled deployment \cite{ref216}. In addition, the MiniMax family provides a clear evolution case: MiniMax-Text-01 used a hybrid design of Lightning linear attention interleaved with full attention; MiniMax-M2 reverted to the full-attention GQA route for quality reliability reasons; and MiniMax-M3 proposed MiniMax Sparse Attention (MSA) to support ultra-long contexts up to one million tokens through more precise KV chunking and sparse selection \cite{ref043}\cite{ref049}. This shows that efficient attention is not simply about pursuing low complexity, but about continuously balancing model capability, infrastructure maturity, and ultra-long-context cost.

Overall, low-complexity attention mechanisms are forming a three-pronged technical landscape: the KV-compressed full-attention route emphasises stable capability and deployment predictability; the sparse/hybrid attention route focuses on cost control at million-token contexts; and the linear/efficient attention route continues to explore even longer contexts and lower complexity. In the future, large models will not be dominated by a single mechanism. Instead, depending on model scale, task type, context length, KV cache budget, and inference cost targets, they will adopt hybrid designs combining multiple low-complexity structures.

\subsection{MoE Architecture Optimization}

\subsubsection{Technical Definition}

MoE (Mixture of Experts) architecture optimization refers to replacing a portion of the dense feed-forward networks in large models with multiple expert networks, where a router dynamically selects a small number of experts for each token to participate in computation. This approach expands the total parameter count of the model while controlling the amount of activated computation per token. Unlike dense models where all parameters participate in computation for every input, sparse MoE allows different tokens to access different expert combinations, theoretically increasing model capacity and task coverage while concentrating computation on the few experts selected by the routing mechanism.

For inference optimization, MoE presents both advantages and challenges. It provides larger total parameter capacity under similar activated computation, but inference systems must address issues such as expert weight residency, uneven routing distribution, cross-GPU/cross-node All-to-All communication, expert load fluctuations within a batch, and expert parallel scheduling. From the perspective of token-oriented operations, the core of MoE architecture optimization is not simply pursuing larger parameter counts, but striking a balance among model quality, activated computation, communication overhead, memory footprint, and service stability.

\begin{figure}[htbp]
\centering
\includegraphics[width=0.95\linewidth]{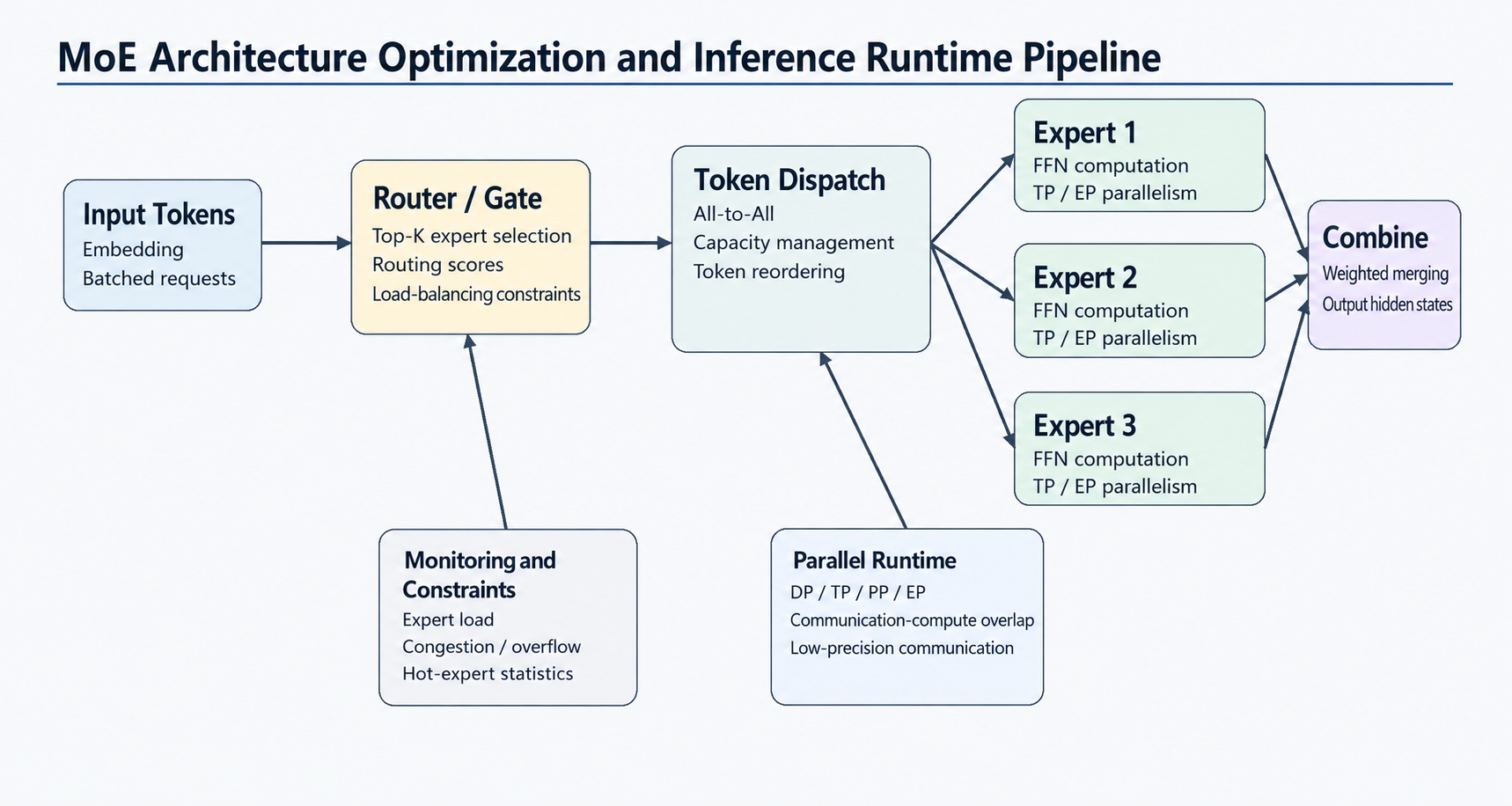}
\caption{MoE architecture optimization and inference runtime pipeline.}
\label{fig:2_3}
\end{figure}

\subsubsection{Industry Trends and Developments}

A 2025 survey on MoE systematically examines the mixture-of-experts architecture in large language models across dimensions including algorithm design, system design, and application scenarios. It points out that the advantage of MoE lies in scaling model capacity with low activation computation, but deployment benefits depend on routing stability, expert load balancing, and system parallelism capabilities \cite{ref052}. This conclusion aligns with practical inference system experience: whether MoE reduces token cost cannot be judged solely by total parameter count or activated parameter count; one must also consider communication, batching patterns, and expert load distribution.

\begin{figure}[htbp]
\centering
\includegraphics[width=0.95\linewidth]{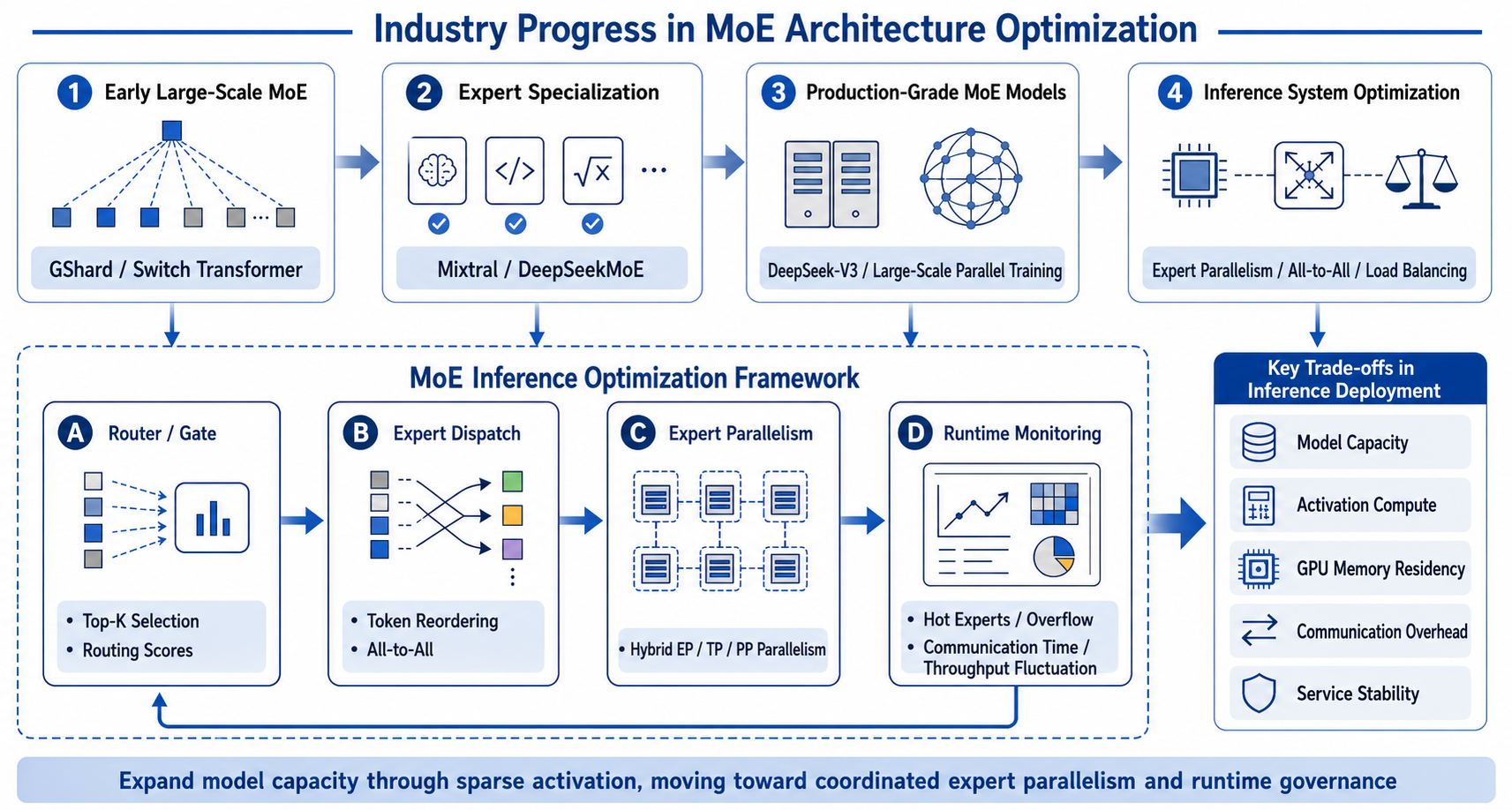}
\caption{Industry progress in MoE architecture optimization.}
\label{fig:2_4}
\end{figure}

The scaling path of MoE can be traced back to GShard and Switch Transformer. GShard combines conditional computation with automatic sharding to scale model capacity using expert layers \cite{ref053}. Switch Transformer simplifies routing to Top-1 expert selection and mitigates expert congestion and training instability using mechanisms such as capacity factor and load-balancing loss \cite{ref054}. These works established the basic MoE paradigm: select experts via a router, control per-token computation through sparse activation, and avoid over-heating of a few experts using load-balancing mechanisms.

In the era of large language models, MoE places greater emphasis on expert specialization and inference usability. Mixtral 8x7B uses 8 experts per MoE layer and selects 2 experts per token, achieving strong overall capability with low activated parameter count \cite{ref055}. DeepSeekMoE proposes fine-grained expert segmentation and shared expert isolation to reduce expert redundancy and enhance expert specialization \cite{ref056}. Subsequent models such as DeepSeek-V3 further integrate MoE with large-scale parallel training, load balancing, and inference deployment, demonstrating that MoE has progressively moved from a research architecture to production-grade large model systems \cite{ref042}.

A major part of MoE architecture optimization challenges lies at the system level. Expert routing distributes different tokens to experts on different GPUs or nodes, making All-to-All communication and intra-batch load imbalance major bottlenecks. Inference frameworks such as TensorRT-LLM support expert parallelism, tensor parallelism, and hybrid approaches to distribute expert weights across multiple GPUs \cite{ref058}. Addressing the straggler effect caused by imbalanced expert load, Capacity-Aware Inference treats the global waiting induced by overloaded experts as a key source of MoE inference latency, and proposes capacity-aware token drop and expanded drop mechanisms to improve expert utilization and end-to-end inference efficiency \cite{ref059}. Looking ahead, MoE research is moving from homogeneous experts and single-sparse-activation mechanisms toward heterogeneous experts, grouped experts, and resource-aware routing design. To address the issue that traditional MoE experts are relatively uniform in scale and struggle to match varying token complexity, heterogeneous grouped expert architectures have been proposed. These architectures use a two-level routing mechanism to achieve more flexible resource-aware expert combinations, and introduce expert group constraints and expert assignment strategies that account for GPU load imbalance and parameter utilization efficiency. This allows tokens of different complexity to be directed to more suitable expert groups while balancing computational distribution across multiple GPUs \cite{ref060}. Training and inference ecosystems such as DeepSpeed-MoE, Megatron-Core, and Tutel continue to optimize around expert parallelism, communication reordering, low-precision communication, computation-communication overlap, and dynamic scheduling. The industry consensus is that MoE deployment efficiency cannot be judged by activated parameter count alone; one must simultaneously consider routing distribution, expert communication, batch shapes, load balancing, and kernel support.

From the token operations perspective, MoE architecture introduces new optimization dimensions. On one hand, sparse activation helps increase model capacity under limited computational budgets. On the other hand, the per-token cost of MoE models is more heavily influenced by expert distribution, parallel topology, and runtime scheduling. Practical deployment requires monitoring expert load, overflow, token drop, hot experts, cross-node communication time, and throughput fluctuations under different batch sizes, and forming a closed loop among model training, deployment configuration, and scheduling policies. Only when routing stability, expert specialization, and system parallelism capabilities mature together can MoE reliably translate into inference cost benefits.

\subsection{Diffusion Model Generation Path Optimization}

\subsubsection{Technical Definition}

Generation path optimization for diffusion models refers to a class of techniques designed for the generation mechanism of diffusion models and flow matching models, which progressively denoise and gradually approximate the target distribution. From the perspective of denoising trajectories or generation paths, these techniques optimize key components that affect generation efficiency and quality, with the goal of accelerating high-quality generation, reducing computational cost, and improving generation stability.

The core focus of this direction is not merely to ``reduce computation,'' but to restructure how the generation process advances across time steps, updates intermediate states, and reuses information across steps, while preserving generation quality, semantic consistency, and content stability as much as possible. In this way, path optimization improves the overall inference efficiency of generative models.

\begin{figure}[htbp]
\centering
\includegraphics[width=0.95\linewidth]{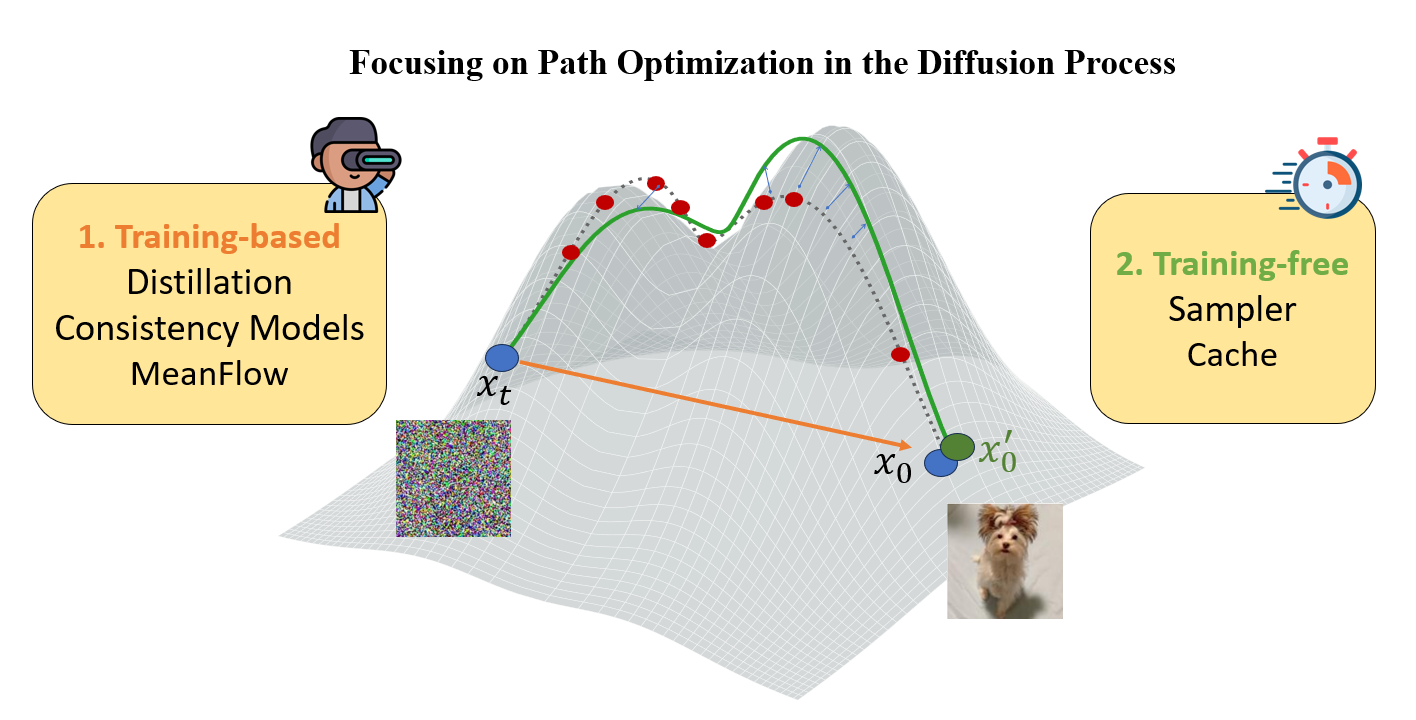}
\caption{Illustration of generation path optimization for diffusion models.}
\label{fig:2_5}
\end{figure}

\subsubsection{Industry Trends and Developments}

Generation path optimization for diffusion models has evolved from an early focus on ``reducing sampling steps'' into a broader technical system in which training-based optimization and training-free optimization develop in parallel. The former compresses generation paths or reconstructs the generation mechanism through additional training, while the latter improves the existing inference pipeline as much as possible without modifying the original model parameters, thereby enhancing generation efficiency and engineering adaptability.

In the direction of training-based optimization, representative methods include distillation and consistency models. These methods usually learn a shorter generation path, compressing the originally multi-step iterative diffusion generation process into few-step or even one-step generation. They show strong potential for extreme latency reduction and are well suited for generation scenarios with stringent real-time requirements. Their advantages lie in high compression capability and a high theoretical acceleration ceiling. However, training-based optimization often depends on additional training data, training resources, and model adaptation procedures, resulting in relatively high R\&D costs and longer migration cycles. For rapidly evolving open-source models, commercial models, and diverse task scenarios, their generalization ability and deployment flexibility may also be constrained.

\begin{figure}[htbp]
\centering
\includegraphics[width=0.95\linewidth]{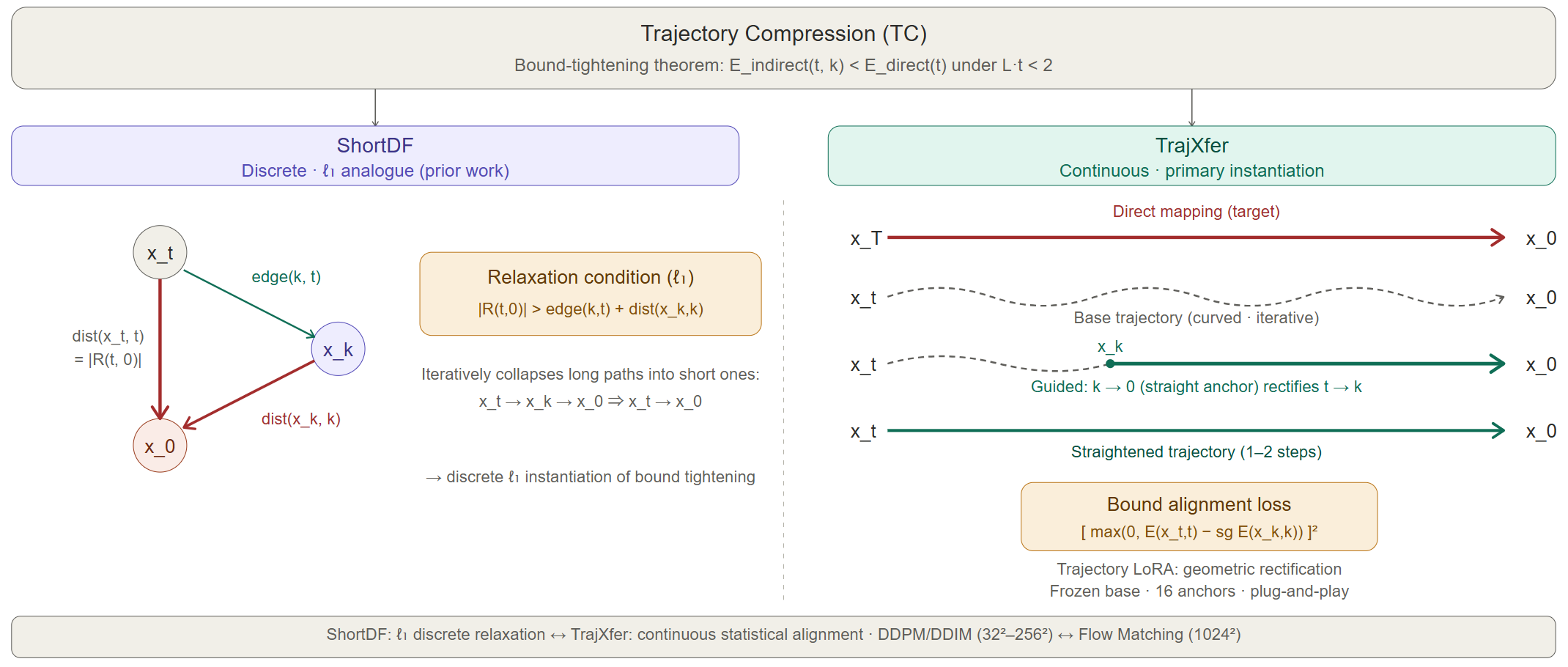}
\caption{Two complementary instances under the unified TC framework---ShortDF and TrajXfer.}
\label{fig:2_6}
\end{figure}

A representative approach in the direction of training optimization is the ``Trajectory Compression (TC)'' idea, which compresses multi-step decoding trajectories into fewer steps through training. Based on a unified TC framework, the China Unicom Yuanjing team characterizes both diffusion and flow matching generative models from the perspective of probability flow ODE, and proves via a bound-tightening theorem that, under mild stability conditions, the global truncation error bound of stepwise trajectories is strictly smaller than that of direct single-step generation. This provides a formal explanation for ``trading off generation quality with finer-grained trajectories'' and lays a theoretical foundation for a unified acceleration scheme from low-resolution diffusion to high-resolution flow matching foundation models. Along the same theoretical axis, the TC framework further materializes into two complementary methods under discrete and continuous mathematical formulations. On the discrete side and under l\_1-norm constraints, ShortDF models reverse diffusion as a shortest path problem on a weighted graph, compressing redundant paths iteratively via inter-step reverse graph construction and residual relaxation loss. It reduces the number of sampling steps from 10 to 2 on CIFAR-10 while improving FID by 18.5\%, and achieves state-of-the-art results on benchmarks such as CelebA and LSUN Churches. This work has been accepted as a CVPR 2025 Highlight \cite{ref061}. On the continuous side, TrajXfer targets large-scale text-to-image foundation models that are difficult to retrain fully, translates the bound-tightening theorem into a few-shot geometric alignment objective, and uses a low-rank adapter (Trajectory LoRA) as a pluggable trajectory correction module. With only 16 domain-specific prompts, it compresses the inference steps of FLUX.1-dev from 20 to 2, achieving about 15x acceleration, and can naturally combine with community style LoRAs without breaking semantic expression.

At the same time, there is also a class of training-free generation path optimization methods. Compared with training-based optimization, training-free optimization emphasizes low-intrusion modification of existing models. This route usually does not change the original model parameters. Instead, it improves inference efficiency by optimizing the sampling process or reusing intermediate computation. Representative directions include sampler optimization and cache reuse. Sampler optimization shortens the generation path by improving the denoising solver, reducing ineffective time steps, or enhancing the efficiency of single-step updates, while keeping the main model architecture unchanged. Cache reuse further exploits redundancy among adjacent time steps, different network layers, tokens or patches, and cross-step states. By selectively reusing intermediate results, cache-based methods reduce repeated forward computation and intermediate state read/write overhead. Related surveys usually classify this route under diffusion caching or cache-based acceleration \cite{ref062}.

Among training-free methods, cache reuse has become one of the most active research directions in recent years. An early representative work is DeepCache, whose core idea is to exploit feature similarity between adjacent diffusion time steps and cache and reuse part of the intermediate features \cite{ref063}. Methods at this stage mainly relied on fixed intervals and static reuse. They demonstrated that temporal redundancy in diffusion generation can be systematically exploited and laid the foundation for subsequent research on path optimization.

With the development of DiT, video generation, and larger-scale generative models, path optimization has gradually shifted from static caching to dynamic scheduling. Methods represented by TeaCache and DiCache introduce time-step-related change estimation mechanisms, deciding whether to refresh the cache based on state variation, thereby improving the trade-off between speed and quality \cite{ref064}. Meanwhile, methods such as DuCa further push the caching granularity down to the token level \cite{ref065}. This indicates that the focus of the field has shifted from ``whether computation can be reused'' to ``which parts are suitable for reuse and which parts must be precisely updated.\cite{ref066}''

\begin{figure}[htbp]
\centering
\includegraphics[width=0.95\linewidth]{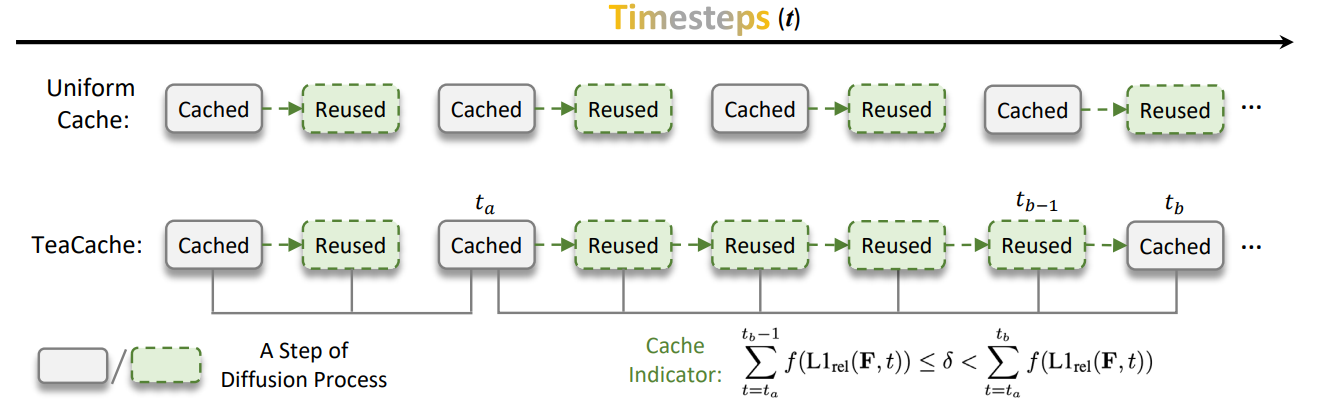}
\caption{Trajectory Optimization Rules of TeaCache; adapted from Reference~\cite{ref064}.}
\label{fig:2_7}
\end{figure}

As AIGC evolves from diffusion models toward flow matching models, cache reuse techniques for flow matching models have also attracted substantial attention. Flow matching models learn the velocity field of the generative trajectory, and the perspective of cache reuse has expanded from caching intermediate states in the diffusion process to reusing early velocity fields. However, aggressive velocity reuse may cause drift in the decoding trajectory. To address this issue, the China Unicom Yuanjing large model team proposed MeanCache, accepted by ICLR 2026. From the perspective of average velocity, MeanCache uses Jacobian-Vector Product to construct interval-wise average velocity and combines it with a trajectory-stability-oriented scheduling strategy. This reduces local bias accumulation and trajectory drift under high-ratio cache reuse \cite{ref067}.

\begin{figure}[htbp]
\centering
\begin{subfigure}{0.48\linewidth}
\centering
\includegraphics[width=\linewidth]{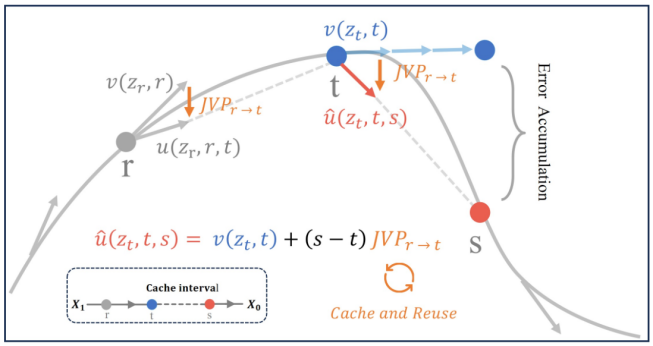}
\end{subfigure}
\hfill
\begin{subfigure}{0.48\linewidth}
\centering
\includegraphics[width=\linewidth]{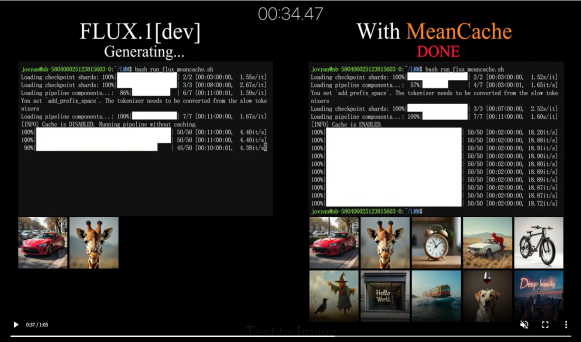}
\end{subfigure}
\caption{MeanCache Optimizes Generation Trajectories from the Perspective of Average Velocity.}
\label{fig:2_8}
\end{figure}

Related surveys further point out that the overall evolution of cache reuse is moving from ``static reuse'' toward ``dynamic prediction,'' and gradually from empirical local rules toward more fine-grained and globally informed path scheduling frameworks. Overall, training-based path optimization is more suitable for scenarios that pursue extreme compression, while training-free path optimization is more suitable for rapid deployment and continuous optimization of existing models. Among training-free approaches, cache-based path optimization is becoming an important direction for accelerating inference in generative diffusion models, due to its low-intrusion nature, transferability, and potential for continuous evolution. In particular, in scenarios such as text-to-video generation, high-resolution image generation, and next-generation flow matching models, how to conduct cache scheduling, error control, and stability optimization around the complete denoising path has become a key competitive issue in the industry.

\subsection{Reasoning CoT Optimization}

\subsubsection{Technical Definition}

Reasoning Chain-of-Thought Optimization refers to a family of techniques for generating, utilizing, and controlling the intermediate reasoning processes of large language models when solving complex problems. Traditional large language models typically produce answers through a direct "input-to-output" paradigm. In contrast, Chain-of-Thought (CoT) methods explicitly generate intermediate reasoning steps, enabling models to decompose complex tasks into smaller, more manageable subproblems, thereby improving performance on mathematical reasoning, code generation, commonsense reasoning, symbolic reasoning, and related tasks. In modern reasoning-oriented models, a reasoning chain is no longer limited to human-readable text; it may also take the form of internal reasoning tokens, scratchpads, tool-calling plans, search-tree nodes, or compressed latent states. Consequently, the core objective of reasoning chain optimization is not to make models "think indefinitely," but rather to achieve an optimal balance among accuracy, latency, cost, interpretability, and safety.

From the perspective of technical objectives, reasoning chain optimization primarily addresses four categories of challenges. The first is reasoning quality optimization, which seeks to improve answer correctness through better decomposition, verification, reflection, and path selection. The second is reasoning efficiency optimization, which aims to reduce redundant, repetitive, or ineffective intermediate steps, thereby lowering token consumption, inference latency, and deployment costs. The third is reasoning structure optimization, which extends linear Chain-of-Thought reasoning into more sophisticated paradigms such as divide-and-conquer strategies, tree-based search, programmatic reasoning, or interleaved "reasoning-and-acting" frameworks. The fourth is reasoning controllability optimization, which dynamically allocates a thinking budget according to task difficulty, allowing simple tasks to be answered quickly while providing sufficient reasoning resources for more complex tasks. Recent work on efficient reasoning often evaluates models using metrics such as correctness per token, token-to-correct ratio, and the budget--accuracy Pareto frontier.

From an implementation perspective, reasoning chain optimization can be categorized into five major approaches. The first is prompt-level optimization, including methods such as Few-Shot CoT, Zero-Shot CoT, and Least-to-Most prompting, which decompose complex tasks into smaller solvable steps. The second is decoding-level optimization, exemplified by Self-Consistency, which improves robustness by sampling multiple reasoning paths and aggregating them through voting, albeit at the cost of increased inference overhead. The third is search and planning optimization, represented by approaches such as Tree of Thoughts, which formulate reasoning as a search process over thought nodes and support backtracking and self-evaluation. The fourth is training-level optimization, including methods such as STaR, which iteratively trains models on self-generated correct reasoning trajectories, and DeepSeek-R1, which demonstrates how reinforcement learning can be used to elicit advanced reasoning capabilities. The fifth is budget and compression optimization, which employs techniques such as token budgeting, length penalties, concise drafts, and reflection suppression to reduce reasoning-token usage while maintaining performance.

At its core, reasoning chain optimization concerns how a model should reason, how long it should reason, what reasoning structure it should adopt, and which parts of the reasoning process should be preserved or compressed. In practical deployments, the complete raw reasoning chain is not necessarily exposed to users. Instead, it is often represented through internal reasoning tokens, summarized rationales, controllable drafts, or final explanations. This design helps reduce information leakage and safety risks while also preventing users from mistaking partially faithful intermediate text for the model's true causal reasoning process. Recent developments such as Anthropic's extended thinking, Google's thinking budget in Gemini, and Qwen3's thinking/non-thinking modes all reflect a broader trend toward treating reasoning as a controllable and allocatable computational resource.

\subsubsection{Industry Trends and Developments}

\begin{figure}[htbp]
\centering
\includegraphics[width=0.95\linewidth]{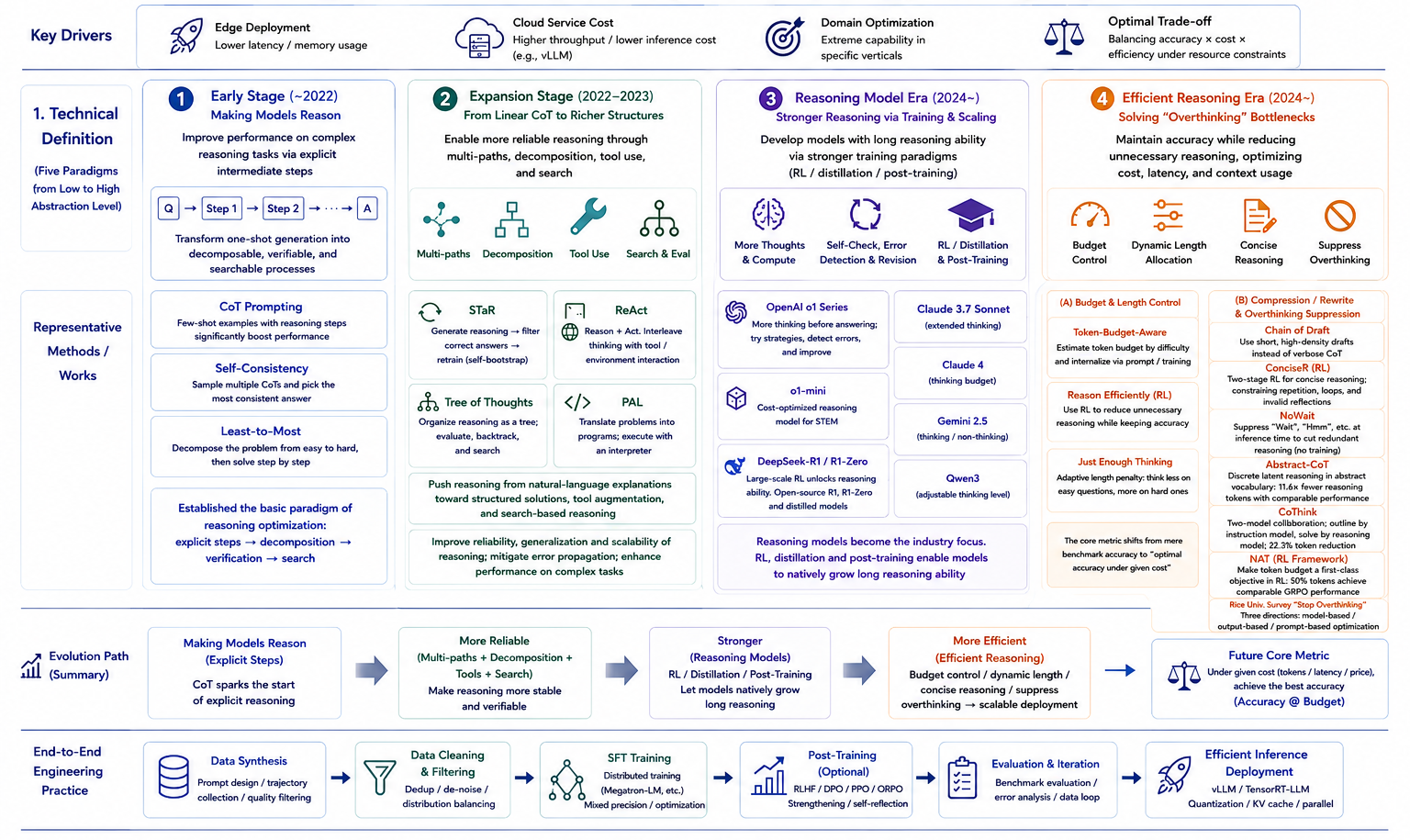}
\caption{Industry development path of reasoning chain-of-thought optimization.}
\label{fig:2_9}
\end{figure}

In the early stage of development, research primarily focused on a fundamental question: how to enable language models to reason. Chain-of-Thought (CoT) Prompting demonstrated that providing a sufficiently large model with a small number of examples containing explicit reasoning steps can substantially improve performance on complex reasoning tasks \cite{ref068}. Self-Consistency further showed that complex problems often admit multiple valid reasoning paths leading to the same correct answer; therefore, sampling multiple CoT trajectories and selecting the most consistent answer can further improve performance \cite{ref035}. Least-to-Most Prompting decomposes a problem into a sequence of simpler subproblems and solves them incrementally, thereby enhancing generalization from easy examples to more challenging tasks \cite{ref069}. Together, these methods established the foundational paradigm of reasoning optimization: transforming one-shot generation into a decomposable, verifiable, and searchable process through explicit intermediate reasoning steps.

Subsequently, research attention expanded from linear chains of thought to more sophisticated reasoning structures. STaR introduced a self-bootstrapping framework in which models iteratively generate reasoning traces, filter for correct answers, and retrain on the resulting data \cite{ref070}. ReAct interleaves reasoning traces with actions, enabling models to think while simultaneously interacting with external tools or environments, thereby mitigating hallucinations and error propagation in factual reasoning tasks \cite{ref071}. Tree of Thoughts organizes reasoning into a tree structure, allowing models to evaluate, backtrack, and search among multiple candidate reasoning paths \cite{ref072}. PAL transforms natural language problems into intermediate program representations and delegates computation to an interpreter, significantly improving reliability on arithmetic and symbolic reasoning tasks \cite{ref073}. These representative works advanced the notion of a "chain of thought" from textual explanations toward structured problem solving, tool-augmented reasoning, and search-based inference.

After 2024, reasoning-centric foundation models became a major focus of industry competition. The OpenAI o1 series emphasized allocating more computation to reasoning before generating an answer and employed training strategies that enabled models to explore alternative solution paths, identify mistakes, and refine their reasoning \cite{ref074}. The o1-mini variant further optimized costs for STEM-oriented applications, highlighting the emergence of specialized compact reasoning models \cite{ref075}. DeepSeek-R1 demonstrated the feasibility of eliciting strong reasoning capabilities through large-scale reinforcement learning and released open-source versions including R1, R1-Zero, and distilled models, facilitating the diffusion of advanced reasoning capabilities from proprietary frontier systems into the open-source ecosystem \cite{ref076}. Meanwhile, products such as Anthropic Claude 3.7 Sonnet, Claude 4, Google Gemini 2.5, and Qwen3 increasingly treated reasoning intensity as a controllable resource through mechanisms such as extended thinking, thinking budgets, and thinking/non-thinking modes \cite{ref077}\cite{ref078}\cite{ref079}.

As reasoning models entered real-world deployment, a new bottleneck emerged: overthinking. For example, DeepSeek-R1 may generate hundreds of reasoning tokens even for a simple arithmetic problem such as "100 + 200 - 300 = ?", resulting in substantial resource waste and increased latency. Recent research on efficient reasoning has therefore focused on a central challenge: how to reduce reasoning costs while preserving reasoning capability. Token-Budget-Aware LLM Reasoning proposes dynamically estimating a reasoning-token budget according to problem complexity and internalizing budget awareness through prompting or post-training, enabling models to determine in advance how many tokens should be spent on reasoning \cite{ref080}. Training Language Models to Reason Efficiently employs reinforcement learning to directly optimize token efficiency by incorporating length penalties into the reward function, encouraging models to minimize unnecessary reasoning while maintaining accuracy \cite{ref081}. Just Enough Thinking further introduces adaptive length penalties that vary according to problem difficulty, encouraging models to spend less computation on simple tasks while reserving greater computational resources for harder ones \cite{ref082}. Collectively, these studies indicate a shift in evaluation priorities: the key objective is no longer benchmark accuracy alone, but rather achieving the highest possible accuracy under a fixed computational budget.

Another line of efficient reasoning research focuses on compressing or rewriting reasoning chains. Chain of Draft proposes replacing lengthy CoT trajectories with shorter, higher-information-density drafts to reduce latency and token costs \cite{ref083}. ConciseR adopts a two-stage reinforcement learning framework to encourage concise reasoning and explicitly suppress repetitive, circular, or unproductive reflections commonly found in long CoT traces \cite{ref084}. NoWait observes that explicit self-reflection tokens such as "Wait" and "Hmm" in some R1-style models may trigger excessive reasoning; it therefore suppresses such tokens during inference to reduce redundant reasoning without requiring retraining \cite{ref085}. Abstract-CoT takes a different approach by constructing a discrete latent reasoning mechanism in which models reason using abstract symbols rather than natural language, reducing reasoning-token usage by 11.6x while maintaining comparable performance \cite{ref086}. CoThink adopts a dual-model collaboration paradigm, in which an instruction-following model first generates a high-level solution outline and a reasoning model subsequently expands it into a detailed solution, achieving a 22.3\% reduction in token consumption across three major benchmark datasets \cite{ref087}. The NAT framework incorporates token budget as a primary optimization objective during reinforcement learning and achieves performance comparable to full-sequence GRPO training while using only 50\% of the tokens \cite{ref088}. The survey Stop Overthinking, published by Rice University, provides a systematic review of this emerging field and categorizes existing approaches into three major directions: model-based optimization, reasoning-output-based optimization, and input-prompt-based optimization \cite{ref089}.

However, most existing efficient reasoning methods adopt a one-size-fits-all compression strategy, uniformly reducing reasoning tokens or truncating reasoning chains regardless of problem difficulty. While such approaches may perform adequately on simple tasks, they can impair reasoning performance on more complex problems. To address this limitation, China Unicom proposed a difficulty-adaptive optimization paradigm, representing a notable industrial effort to move efficient reasoning from uniform compression toward differentiated regulation. In January 2025, Unicom Data Intelligence Co., Ltd. released the Yuanjing Chain-of-Thought Large Model, the first open-source general-purpose reasoning model developed by a Chinese central state-owned enterprise. The model reportedly outperformed Qwen-QwQ and achieved performance comparable to OpenAI o1 across multidisciplinary and multi-scenario reasoning tasks. The core innovation of Yuanjing is its Adaptive Slow Thinking framework, which consists of both task-adaptive and difficulty-adaptive mechanisms. Task adaptation is achieved through mixed fine-tuning that carefully balances general instruction-following data with long-chain reasoning data, enabling the model to employ extended reasoning for reasoning-intensive tasks while generating concise responses for routine tasks. Empirical results show that Yuanjing produces substantially shorter responses than Qwen-QwQ on non-reasoning tasks while maintaining comparable accuracy. Difficulty adaptation is implemented through the DAST (Difficulty-Adaptive Slow-Thinking) framework \cite{ref090}. DAST introduces the Token Length Budget (TLB), a metric that combines sampling accuracy with the average length of correct responses to establish a mapping between problem difficulty and target response length. When sampling accuracy is high, the TLB approaches the average length of correct responses, indicating that simple problems should receive shorter answers. When sampling accuracy is low, the TLB approaches the maximum generation length, encouraging deeper reasoning for more difficult problems. Based on TLB, DAST designs a reward calibration mechanism: correct answers exceeding the allocated budget receive reduced rewards (with stronger penalties for easier questions), whereas incorrect answers that remain below budget are encouraged to perform additional reasoning. Furthermore, DAST constructs two types of preference data---Dual Correct Pairs (DCP) and Dual Incorrect Pairs (DICP)---for SimPO optimization. In practical deployment, China Unicom has applied adaptive slow-thinking techniques to scenarios such as telecommunications supply-chain question answering and safety operations for port-oriented large models, reportedly reducing reasoning computation by approximately 30\% while preserving complex reasoning capabilities.

Overall, reasoning chain optimization is evolving along a clear trajectory. The first stage was CoT prompting, which encouraged models to explicitly generate reasoning steps. The second stage introduced multi-path reasoning, decomposition, tool use, and search mechanisms to improve reasoning reliability. The third stage saw the emergence of reasoning-centric models that acquired intrinsic long-horizon reasoning capabilities through reinforcement learning, distillation, and post-training techniques. The fourth stage is characterized by efficient reasoning, which emphasizes scalable deployment through budget control, dynamic length allocation, concise reasoning, and suppression of excessive reflection. The difficulty-adaptive paradigm represented by China Unicom's DAST is further advancing this stage from uniform compression toward differentiated regulation, enabling reasoning resources to be allocated in proportion to problem complexity.

\subsection{Memory Management}

\subsubsection{Technical Definition}

LLM memory management refers to a systematic mechanism for selectively writing, organizing, compressing, retrieving, updating, and forgetting information during large model inference. Such information may include historical interactions, task states, external knowledge, execution experience, user preferences, and intermediate reasoning traces. Its core objective is not merely to ``extend the context window,'' but rather to provide the model with continuous, controllable, and low-noise information support under constrained context windows, limited inference budgets, and dynamic task environments. In doing so, memory management enhances long-horizon reasoning, personalized interaction, multi-turn task execution, and cross-task transfer capabilities.

From the perspective of inference optimization, memory management typically consists of three components: memory formation, memory evolution, and memory retrieval. Memory formation is responsible for transforming raw interactions, tool invocation results, environmental feedback, code execution traces, planning processes, or self-reflection outputs into reusable memories, rather than preserving the entire context verbatim. Common techniques include summary-based compression, experience extraction, knowledge distillation, graph-structured construction, vectorized encoding, and parametric internalization. Memory evolution is responsible for maintaining the quality of the memory repository, including deduplication and merging, conflict resolution, elimination of low-value information, removal of outdated information, and abstraction of recurring experiences into higher-level strategies. Memory retrieval is responsible for recalling relevant memories before or during inference based on the current task, user intent, and contextual state, and then injecting them into the model context after reranking, filtering, aggregation, or compression. This lifecycle can be summarized as a closed loop of ``write, organize, recall, use, and update.'' Relevant definitions can be found in the uploaded reference material's descriptions of memory states, memory formation, memory evolution, and memory retrieval operators.

By representation form, LLM memory can be divided into three categories. The first category is token-level explicit memory, where information is stored as discrete units such as text snippets, dialogue records, task trajectories, code fragments, user profiles, knowledge graph nodes, or multimodal entries. This type of memory is typically implemented through vector databases, key-value stores, databases, graph databases, or hierarchical indexes. It is transparent, editable, and auditable, making it the dominant form in current engineering practice. The second category is parametric memory, where knowledge and experience are written into model weights through fine-tuning, LoRA, model editing, or continual learning, allowing them to take effect implicitly in subsequent inference. Its advantages include low invocation cost and no need for explicit retrieval, while its challenges lie in update risk, forgetting control, and limited interpretability. The third category is latent memory, where historical information is encoded into hidden states, memory vectors, long-term KV caches, or dedicated memory tokens, allowing the model to continuously access such information during inference. This form is well suited for low-latency, long-sequence, and multi-turn state retention scenarios.

By function, memory management primarily serves three types of inference requirements. Factual memory records user preferences, environmental states, external knowledge, and long-term facts, addressing the question of ``what the model should know.'' Experiential memory records successful and failed cases, strategies, tool usage patterns, and reusable skills from historical tasks, addressing the question of ``how the model has solved problems before.'' Working memory maintains plans, temporary variables, intermediate conclusions, and tool outputs within the current task, addressing the question of ``what the model is currently processing.'' For inference optimization, working memory reduces state loss in long-chain tasks, factual memory improves personalization and consistency, and experiential memory enhances cross-task self-improvement.

\subsubsection{Industry Trends and Developments}

\begin{figure}[htbp]
\centering
\includegraphics[width=0.95\linewidth]{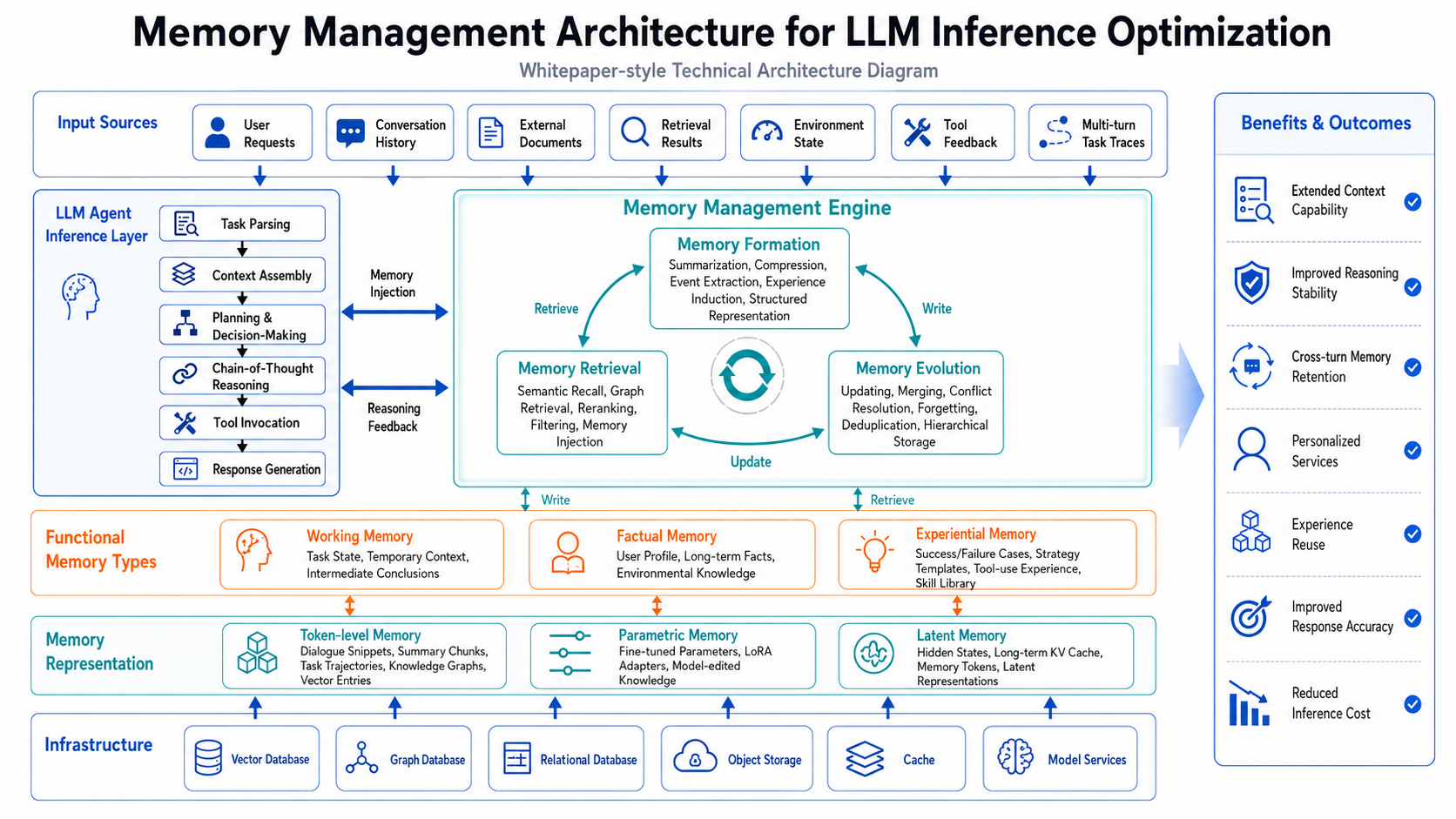}
\caption{Memory management architecture for LLM inference optimization.}
\label{fig:2_10}
\end{figure}

Early memory management primarily focused on long-form dialogue and personalization. MemoryBank enables conversational agents to maintain cross-session consistency and personalized responses by persistently storing user profiles, emotional states, and dialogue histories \cite{ref091}. MemGPT further introduces an operating-system-inspired virtual memory paradigm, treating the limited context window as ``main memory'' and external storage as ``disk,'' while allowing the model to autonomously decide when to page, retrieve, and update memories \cite{ref092}. These methods established an engineering paradigm for explicit memory management, upgrading the context window from a passive input region into a schedulable runtime resource.

Subsequently, memory management gradually expanded from ``preserving facts'' to ``consolidating experiences.'' Reflexion writes verbal feedback from task failures into memory and uses it to guide behavioral correction in subsequent attempts \cite{ref093}. Voyager distills successful exploration experiences in the Minecraft environment into a reusable skill code library, enabling continual learning in open-ended tasks \cite{ref094}. ExpeL induces generalizable experiences and few-shot exemplars from historical trajectories to improve performance on subsequent reasoning tasks \cite{ref095}. These works demonstrate that memory can not only preserve user-specific facts, but also serve as a key mechanism for accumulating strategies, reusing skills, and reducing repeated trial-and-error during inference.

In the past two years, industry and research efforts have increasingly shifted toward structured, automated, and evolvable memory. Mem0 extracts dialogue histories into updatable long-term memories and supports fine-grained user preference management, with an emphasis on low latency, high retrieval hit rates, and controllable updates \cite{ref096}. Engineering systems such as Zep introduce temporally aware and graph-structured memory, supporting long-term dialogue and complex retrieval through entity, event, and relation modeling \cite{ref097}. Recent works such as A-MEM and G-Memory further adopt networked notes, hierarchical graphs, and multi-level summaries to organize fragmented memories into relational structures, thereby improving multi-hop reasoning, context aggregation, and conflict resolution \cite{ref098}\cite{ref099}. This marks a transition of memory systems from ``vector-based retrieval'' toward an organized, reasoning-capable, and maintainable knowledge and experience substrate.

In the direction of inference optimization, the integration of reinforcement learning with memory management has become a representative recent trend. MemAgent, MEM1, and Memory-R1 attempt to enable models to learn when to write, what to write, when to retrieve, and how to compress memories, transforming memory scheduling from rule- and prompt-driven procedures into trainable policies \cite{ref100}\cite{ref101}\cite{ref102}. The value of these methods lies in the fact that they no longer treat memory as a fixed external component; instead, they incorporate memory operations into the model's own reasoning behavior, allowing the model to actively trade off information utility, context cost, and noise risk during inference.

Another important line of work combines memory management with long-context processing, KV caching, and latent states. Traditional long-context methods mainly rely on expanding the context window or compressing tokens, but their cost increases substantially with sequence length. In recent years, techniques such as KV compression, KV reuse, long-term memory tokens, and latent vector memories have sought to preserve key information without fully unfolding historical text. Representative works such as MemoryLLM, M+, and MemGen explore the transformation of historical information into continuously accessible latent memory representations, thereby reducing the cost of explicit text retrieval and context concatenation \cite{ref103}\cite{ref104}\cite{ref105}. In high-frequency interaction, real-time agent, and long-horizon task execution scenarios, such mechanisms are expected to become important inference acceleration techniques beyond explicit memory.

Overall, LLM memory management is evolving from a ``context patch'' into a core infrastructure component of inference systems. In practical engineering deployments, hybrid architectures are commonly adopted: short-term working memory maintains the current task state; vector and graph databases support the retrieval of long-term facts and experiences; summarization and forgetting mechanisms control cost and noise; and parametric or latent memory carries high-frequency, stable capabilities. Future key directions include automated memory scheduling, multimodal memory, verifiable memory, privacy and permission control, shared memory across agents, and inference-cost-oriented memory compression and hierarchical caching.

\subsection{KV Cache Compression}

\subsubsection{Technical Definition}

Key-value cache compression refers to more efficient storage, reuse, quantization, eviction, or hierarchical management of historical tokens' key-value states during the autoregressive inference process of large language models, aiming to reduce GPU memory footprint and memory bandwidth pressure. The role of the key-value cache is to avoid recomputing the key-value states of the entire historical context for each new generated token. However, its size grows linearly with the number of layers, number of attention heads, hidden dimension, batch size, and sequence length. In scenarios such as long context, high concurrency, and multi-turn dialogues, the key-value cache can become the primary resource bottleneck limiting throughput, context length, and per-token cost.

From the perspective of long-context inference optimization, key-value cache compression should be considered together with broader context compression. Context compression refers to transforming the original context into a shorter, cheaper, more reusable, or more computable representation before the model processes long documents, multi-turn history, RAG (Retrieval-Augmented Generation) retrieval results, or code repository contexts. Its targets can be input tokens, soft tokens, summary vectors, memory states, retrieval embeddings, sparse attention indices, or compressed key-value states. Unlike simply expanding the context window, context compression emphasizes reducing prefilling costs, attention computation overhead, key-value cache footprint, and end-to-end latency while preserving task-relevant information as much as possible.

For token-oriented operations, key-value cache compression and context compression jointly affect the achievable number of concurrent requests, long-context capacity, time-to-first-token (TTFT), per-token cost, and long-document task quality. The former primarily acts on inference runtime states, addressing issues of cache GPU memory, memory bandwidth, and cache reuse. The latter mainly acts on the input side and prefilling stage, reducing the scale of context or attention connections that the model has to process directly. For practical deployment in long-document QA, multi-turn dialogues, meeting summaries, enterprise knowledge bases, and RAG systems, a closed loop of ``input-side compression $\rightarrow$ prefix/context reuse $\rightarrow$ KV cache management $\rightarrow$ quality evaluation $\rightarrow$ graceful fallback'' is typically required, rather than using any single compression algorithm in isolation.

\begin{figure}[htbp]
\centering
\includegraphics[width=0.95\linewidth]{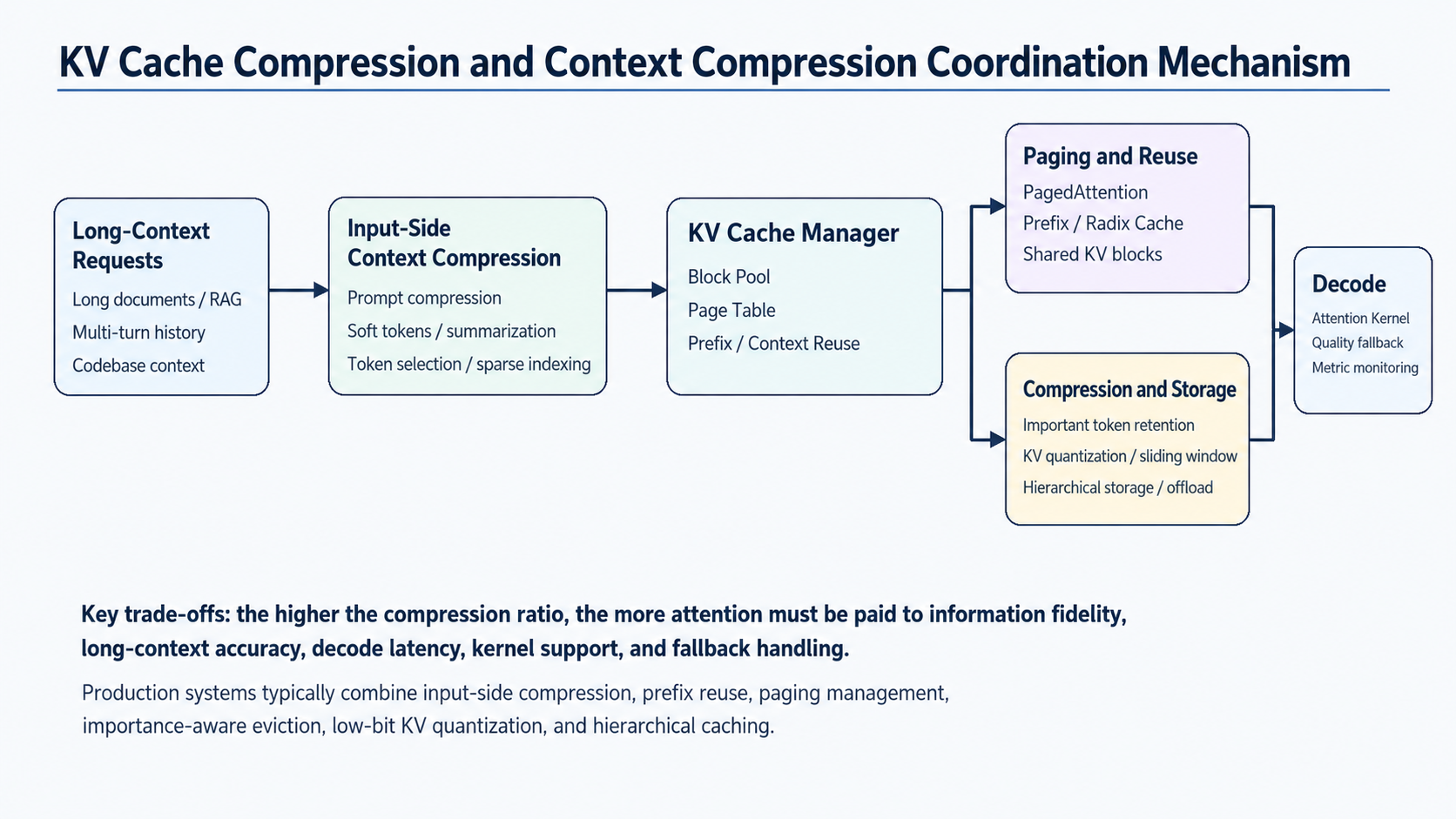}
\caption{Collaborative mechanism between KV cache compression and context compression.}
\label{fig:2_11}
\end{figure}

\subsubsection{Industry Trends and Developments}

A systematic study of key-value cache compression points out that as context length increases, attention computation and KV cache storage jointly amplify inference costs. It classifies existing methods based on their underlying principles and implementation mechanisms, emphasizing that there are significant trade-offs among compression ratio, inference latency, and long-context quality\cite{ref106}. This indicates that KV cache compression cannot be measured solely by the percentage of GPU memory saved; decoding speed, cache hit rate, task accuracy, and graceful fallback must also be evaluated.

\begin{figure}[htbp]
\centering
\includegraphics[width=0.95\linewidth]{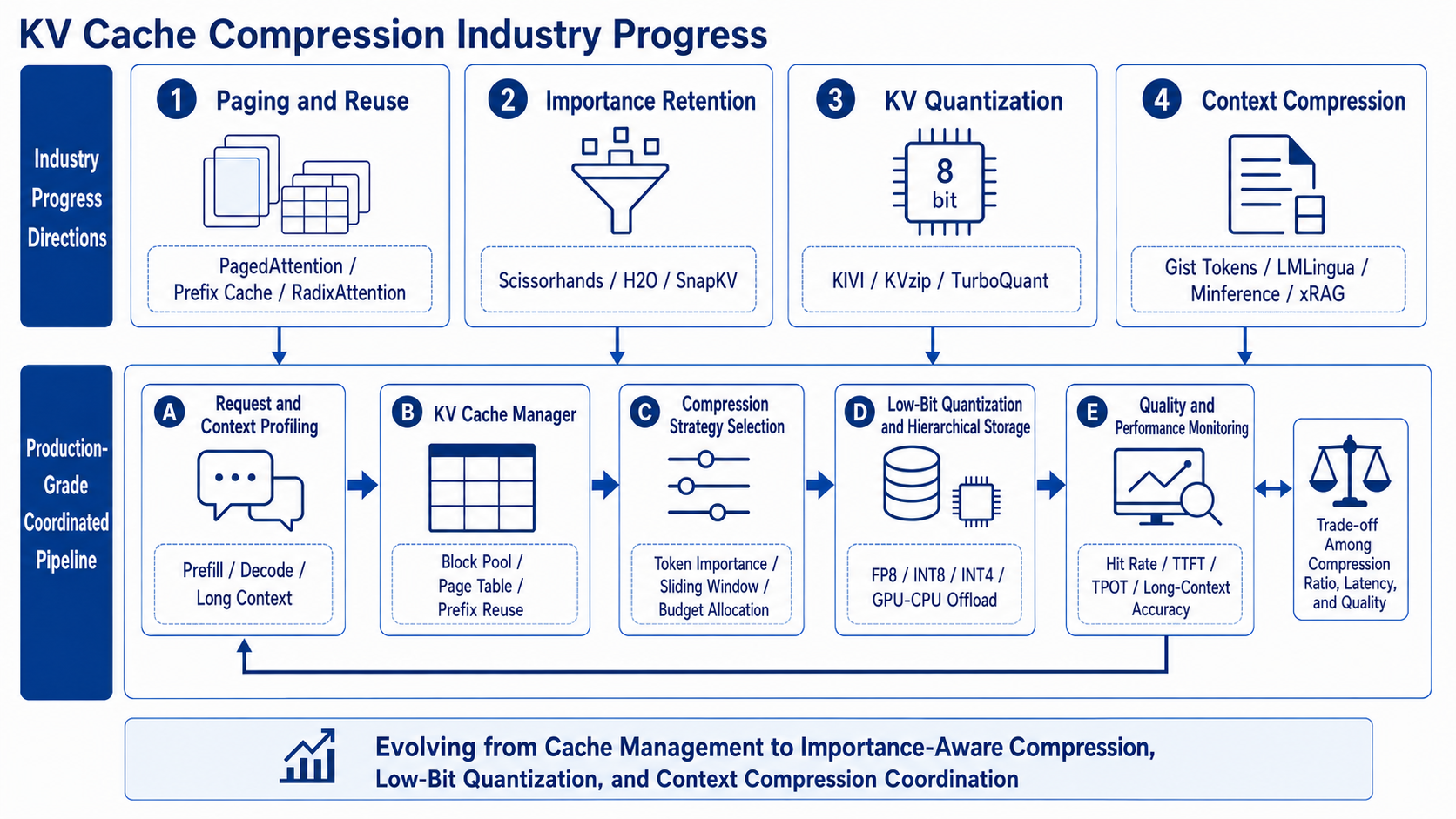}
\caption{Industry progress in KV cache compression.}
\label{fig:2_12}
\end{figure}

On the engineering system level, PagedAttention in vLLM is a representative advancement in KV cache management. Inspired by virtual memory in operating systems, it splits the KV cache into fixed-size blocks and uses a mapping from logical blocks to physical blocks to support non-contiguous storage, thereby reducing internal and external fragmentation caused by traditional contiguous pre-allocation\cite{ref048}. On this basis, vLLM's automatic prefix caching reuses shared KV caches via hashing\cite{ref107}, while SGLang's RadixAttention organizes KV caches using a radix tree (prefix tree) to reuse shared prefixes for structured generation, multi-branch inference, and multi-turn tasks\cite{ref108}. These mechanisms primarily address cache reuse under dynamic requests, shared prefixes, and multi-turn interactions.

Academic research has also proposed various KV cache compression methods based on importance and low-bit representations. Scissorhands, H\_2O, and SnapKV each retain more critical historical states from the perspectives of importance persistence, Heavy Hitter Tokens, and pre-generated attention patterns, compressing the KV cache under a fixed budget \cite{ref109,ref110,ref111}. KIVI, KVZip, and TurboQuant further reduce the cost of cache representation via KV cache quantization or context reconstruction \cite{ref112,ref113,ref114}. A common challenge for these approaches is balancing compression ratio, long-context quality, decoding latency, and kernel support.

Complementing KV cache compression, context compression techniques are becoming important supplements for reducing long-context inference costs. Methods such as Gist Tokens, AutoCompressor, and ICAE attempt to compress the original context into a few soft tokens, summary vectors, or memory slots, thereby reducing the context length that the model needs to process directly \cite{ref115,ref116,ref117}. LLMLingua and LongLLMLingua explicitly compress prompts and select tokens to reduce the number of input tokens, which is suitable for one-shot QA, summarization, and query-specific RAG scenarios \cite{ref118,ref119}. These methods act primarily on the input side and prefilling stage, complementing runtime KV cache compression.

Recent work has also begun to focus on computation-side and retrieval-side context compression. MInference 1.0 reduces the actual attention connections that need to be computed during the long-context prefilling stage through dynamic sparse attention, which is beneficial for scenarios with high prefilling costs such as long prompts, long code contexts, and long RAG inputs\cite{ref120}. xRAG, targeting RAG scenarios, fuses dense embeddings of retrieved documents into the language model's representation space to reduce or even replace raw document token inputs\cite{ref121}. Overall, production systems typically need to combine input-side compression, prefix reuse, paged management, importance-based eviction, KV quantization, and necessary quality fallback mechanisms, while simultaneously monitoring cache hit rate, GPU memory usage, TTFT, TPOT (time-per-output-token), throughput, and long-context accuracy. Therefore, KV cache compression should be regarded as a runtime core capability within the long-context compression ecosystem, rather than an isolated GPU memory optimization technique.

\subsection{Speculative Decoding}

\subsubsection{Technical Definition}

Speculative Decoding, also known as Speculative Sampling, is an inference acceleration technique designed for autoregressive large language models. Its core idea is to avoid generating tokens strictly one by one with the target large model. Instead, a lightweight ``drafter'' --- such as a draft model, draft head, early-exit layer, or n-gram proposer --- first proposes multiple candidate tokens in a single step. The target model then verifies and corrects these candidates through one parallel forward pass, thereby producing the same number of valid output tokens with fewer target-model invocations \cite{ref033}\cite{ref122}. This approach is particularly well aligned with token-oriented operations: it shifts the cost-optimization target from the request level down to the token level, directly improving accepted tokens per target forward pass while reducing TPOT, P95 latency, and unit token cost.

A typical workflow is as follows. Given a context, the drafter generates a sequence of K candidate tokens. The target model then computes the probabilities of these candidate positions in a single pass. The system accepts the candidates one by one according to a predefined acceptance criterion. If a candidate token is rejected, the system resamples from the target-model distribution and truncates the subsequent candidates. Under a strict accept-reject rule, classical speculative sampling preserves the same output distribution as the target model and can therefore be regarded as a form of lossless acceleration. However, some engineering implementations adopt approximate acceptance, typical acceptance, or greedy-specialized strategies, which require trade-offs among generation quality, throughput, and distributional consistency \cite{ref033}\cite{ref123}.

From the perspective of token-oriented operations, speculative decoding is not merely a standalone algorithm, but rather a system-level inference capability consisting of draft generation, parallel verification, dynamic scheduling, and metric-driven feedback loops. Key metrics include the draft acceptance rate, the average number of accepted tokens per decoding round, drafter overhead, target-model verification overhead, throughput variation under batching, gains in long-output scenarios, and fallback strategies across different traffic patterns. In general, speculative decoding is better suited to scenarios where the decode phase accounts for a large proportion of inference time, outputs are relatively long, the target model is constrained by memory bandwidth, and the drafter distribution is close to that of the target model. In contrast, its benefits may diminish significantly when the prefill phase dominates, the batch size is already large, the draft acceptance rate is low, or the drafter itself introduces substantial overhead.

\subsubsection{Industry Trends and Developments}

\begin{figure}[htbp]
\centering
\includegraphics[width=0.95\linewidth]{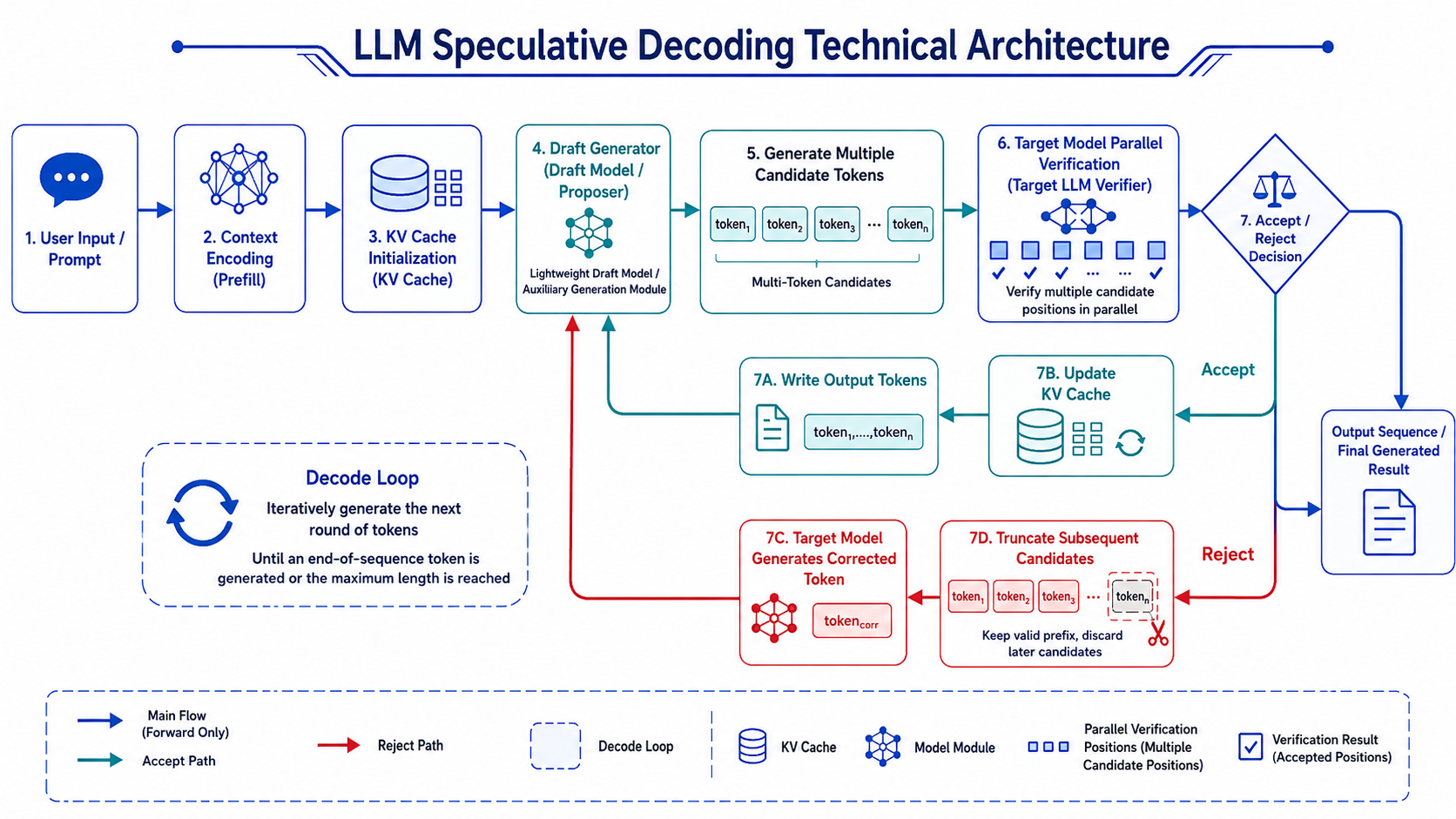}
\caption{Technical architecture of LLM speculative decoding.}
\label{fig:2_13}
\end{figure}

Early representative studies were primarily based on the ``two-model speculative sampling'' paradigm. Leviathan et al. proposed using a smaller model to generate candidate tokens and the target model to verify them in parallel, thereby reducing the number of serial decoding steps without changing the output distribution of the target model. Chen et al. further systematized this paradigm from the perspective of speculative sampling \cite{ref033}\cite{ref122}. The main bottleneck at this stage was that each target model required a high-quality draft model: if the draft model was too weak, the acceptance rate would be low; if it was too strong, its own computational overhead would offset the acceleration gains.

Subsequent research shifted from ``linear drafting'' to ``tree-structured drafting and batched verification.'' SpecInfer organizes multiple candidate sequences into a token tree and leverages the target LLM for tree-based parallel verification, reporting 1.5--3.5x gains in distributed and offloading-based inference scenarios \cite{ref124}. This line of work advanced speculative decoding from an algorithmic technique to a serving-system design problem, where KV cache management, tree-structured attention masks, concurrent scheduling, GPU memory consumption, and batching patterns all become critical factors affecting end-to-end performance.

After 2024, the field began to reduce its reliance on standalone small draft models. Medusa adds multiple decoding heads to the target model to directly predict several future tokens and verifies candidates simultaneously through tree attention. The paper reports that Medusa-1 achieves approximately 2.2x acceleration while preserving generation quality, with Medusa-2 further improving performance through joint training \cite{ref125}. LayerSkip, in contrast, introduces early-exit loss and layer dropout so that earlier layers acquire predictive capability. During inference, shallow layers perform ``self-speculation,'' while deeper layers conduct verification, reducing the additional memory footprint and deployment complexity associated with two-model approaches \cite{ref126}.

Another important direction is feature-level drafting. EAGLE argues that token-level autoregression is less stable than feature-level prediction, and instead predicts the features of the penultimate layer of the target model before generating candidate tokens. It reports 2.7--3.5x latency acceleration on models such as LLaMA2, Vicuna, and Mixtral \cite{ref127}. EAGLE-2 further introduces a context-aware dynamic draft tree, dynamically adjusting the draft tree according to contextual confidence. It reports 3.05--4.26x acceleration and an additional 20\%--40\% improvement over EAGLE-1 \cite{ref128}. In 2025, EAGLE-3 moved from feature prediction to direct token prediction, while integrating multi-layer features and a training-time test mechanism. Covering both chat and reasoning models, it reports up to 6.5x acceleration, making the EAGLE family one of the most important speculative decoding routes in current open-source inference engines \cite{ref129}.

In terms of engineering and productization, inference frameworks such as vLLM, SGLang, and TensorRT-LLM have incorporated multiple speculative decoding capabilities. The vLLM documentation lists several categories of proposers, including EAGLE, MTP, draft models, PARD, MLP, n-gram, and suffix decoding. SGLang supports EAGLE-2/EAGLE-3, MTP, classical draft models, and n-gram variants. NVIDIA TensorRT-LLM also provides EAGLE, Medusa, and draft-model-based approaches, and reports up to 3.6x improvement in token throughput in its official blog \cite{ref130}\cite{ref131}\cite{ref132}. This indicates that speculative decoding has moved from research prototypes into production inference stacks, with the central question shifting from ``whether it can accelerate inference'' to ``under which traffic profiles, batching regimes, hardware configurations, and SLA constraints it can deliver stable economic gains.''

Recent representative work continues to focus on stronger drafters, lower drafting overhead, and better operational deployability. ReDrafter uses an RNN-based drafter, beam search, and dynamic tree attention, and validates inference benefits on both H100 and Apple Silicon environments \cite{ref133}. SpecForge / SpecBundle further addresses the scarcity of high-quality EAGLE-3 draft weights and the lack of training infrastructure by providing production-oriented training frameworks and draft weights for mainstream open-source models \cite{ref134}. In 2026, P-EAGLE replaces EAGLE's autoregressive drafting with parallel drafting, predicting multiple draft tokens in a single forward pass and integrating the approach with vLLM. At the same time, SPEED-Bench emphasizes evaluating speculative decoding across multiple semantic domains and realistic serving regimes, helping avoid overestimating benefits based only on low-batch or synthetic-data settings \cite{ref135}\cite{ref136}.

Overall, the evolution of LLM speculative decoding can be summarized as a progression from a general lossless algorithm based on ``small-model drafting plus large-model verification'' toward a compositional system optimization paradigm involving multi-head prediction, self-speculation, feature-level drafting, dynamic trees, and parallel drafting. For token-oriented operations teams, the key to deployment is not to blindly pursue the highest acceleration ratio reported in papers, but to establish a proposer-selection strategy based on business traffic profiles. For code generation, summarization, RAG, and multi-turn dialogue, n-gram, EAGLE-3, and P-EAGLE can be prioritized for testing. For scenarios requiring strong distributional consistency, strict acceptance criteria should be preferred. For high-concurrency throughput scenarios, speculative decoding should be jointly optimized with continuous batching, KV cache management, quantization, and routing strategies, with ``effective token cost'' and ``throughput under SLA constraints'' serving as the final deployment criteria.

\subsection{Model Quantization}

\subsubsection{Technical Definition}

Model quantization refers to converting the weights, activations, or KV cache in large language models from high-precision representations such as FP16/BF16 to lower-precision representations such as INT8, INT4, FP8, or FP4. Through calibration, scaling, outlier handling, and high-performance kernel design, it aims to preserve model quality as much as possible while reducing memory footprint, memory bandwidth, and computational cost. From the perspective of intervention stage, quantization can be categorised into PTQ (Post-Training Quantization), QAT (Quantization-Aware Training), and low-bit fine-tuning. From the perspective of the target, it can be divided into weight quantization, activation quantization, weight-activation joint quantization, and KV cache quantization.

For token-oriented operations, the core value of quantization lies in lowering the deployment barrier of models and reducing the per-token resource cost. Weight quantization reduces the persistent memory footprint of a model, allowing the same hardware to accommodate a larger model or more instances. Activation and KV cache quantization affect the runtime memory and bandwidth under high concurrency and long-context scenarios. It should be noted that lower bit width does not automatically guarantee faster end-to-end performance; the actual benefit depends on hardware generation, whether the compute kernel fuses dequantization with matrix multiplication, batch size, context length, and the tolerance for task quality degradation.

\begin{figure}[htbp]
\centering
\includegraphics[width=0.95\linewidth]{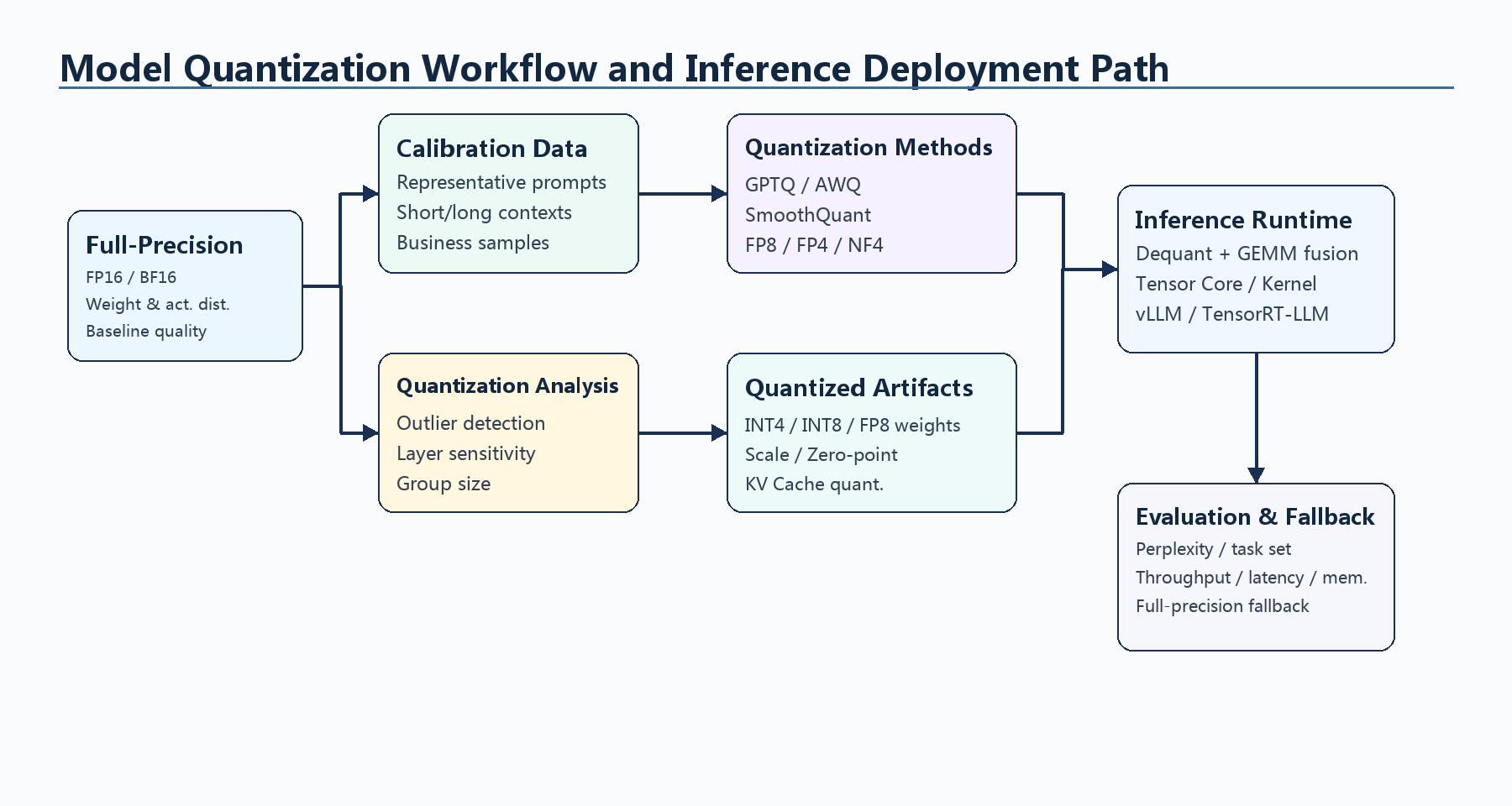}
\caption{Model quantization workflow and inference deployment pipeline.}
\label{fig:2_14}
\end{figure}

\subsubsection{Industry Trends and Developments}

A 2025 survey on low-bit large language models reviews LLM quantization from three aspects: basic numerical formats, system support, and algorithmic strategies. It points out that low-bit quantization now covers weights, activations, gradients, and inference runtime, and that PTQ, QAT, mixed precision, and hardware adaptation must be considered jointly \cite{ref137}. This indicates that model quantization is not a single compression algorithm but a deployment system co-determined by model format, calibration data, compute kernels, inference frameworks, and target hardware.

\begin{figure}[htbp]
\centering
\includegraphics[width=0.95\linewidth]{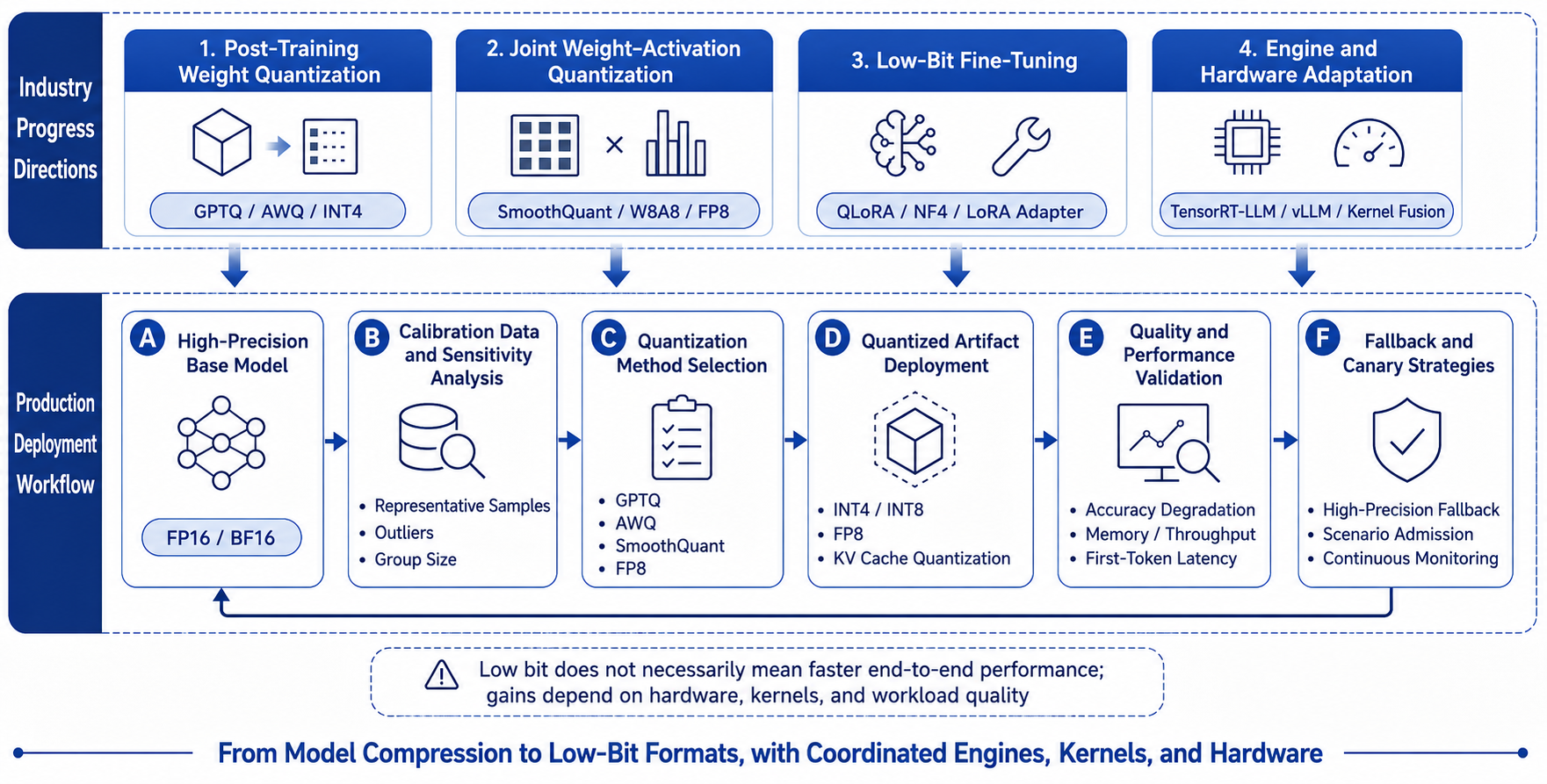}
\caption{Industry progress in model quantization.}
\label{fig:2_15}
\end{figure}

The mainstream practice of LLM quantization first focuses on post-training weight quantization. GPTQ uses approximate second-order information to perform layer-wise and block-wise quantization, enabling models with tens to hundreds of billions of parameters to retain low accuracy loss at 3-bit or 4-bit weights \cite{ref138}. AWQ identifies weight channels that are more critical to model output from an activation-aware perspective, and improves the generalisation stability of low-bit quantisation by protecting a small set of salient weights \cite{ref139}. These methods have driven the development of the 4-bit inference ecosystem for open-source models and made it possible to deploy larger models on a single machine or with fewer GPUs.

Weight-activation joint quantization focuses more on hardware acceleration. SmoothQuant observes that activation outliers in LLMs make activation quantization difficult, therefore it shifts the quantization difficulty from activations to weights, achieving W8A8 post-training quantization and facilitating the use of INT8 matrix multiplication on hardware \cite{ref140}. On new-generation GPUs, FP8 has also become an important low-precision format. Frameworks such as TensorRT-LLM and vLLM provide inference support around FP8 W8A8, FP8 KV cache, INT4 W4A16, AWQ, GPTQ, and other approaches \cite{ref141}\cite{ref142}. This shows that model quantization has moved from algorithmic papers to engineering recipes tailored to specific hardware and inference engines.

Low-bit fine-tuning expands the application boundaries of quantization. QLoRA combines a frozen 4-bit base model with LoRA adapters, and significantly reduces the memory requirement for fine-tuning through NF4 (NormalFloat4), double quantization, and paged optimisers \cite{ref143}. A systematic evaluation of quantization techniques published in 2025 also reminds that differences among low-bit methods stem not only from bit width but also from calibration data, grouping granularity, outlier handling, dequantisation kernels, and task evaluation setups \cite{ref144}. In enterprise scenarios, a base model can be deployed in low-bit form, and different business capabilities can be managed through lightweight adapters. However, at runtime, issues such as adapter combination, quantisation format compatibility, accuracy fallback, and multi-tenant resource isolation must also be addressed.

From a production perspective, model quantisation requires a complete validation pipeline. First, representative calibration data should be selected, covering short Q\&A, long context, code, structured output, and high-risk tasks. Second, outliers in each layer, sensitive layers, and grouping granularities should be analysed to choose between GPTQ, AWQ, SmoothQuant, FP8, or other routes. Then, end-to-end metrics must be verified on the target inference engine and hardware, including memory usage, throughput, time to first token (TTFT), time per token (TPT), long-context stability, and task accuracy degradation. Only when the quantisation format, compute kernels, engine, and hardware form a closed loop can quantisation be reliably translated into token-cost optimisation, rather than merely reducing the model file size.

\subsection{Model Distillation}

\subsubsection{Technical Definition}

Model Distillation, in the context of the Large Language Model era, refers to a collection of techniques and engineering practices aimed at transferring the knowledge, reasoning capabilities, alignment preferences, and complex task-solving skills of a large, computationally expensive Teacher Model into a smaller, more efficient Student Model. While traditional knowledge distillation in deep learning was largely limited to matching output probability distributions (logits) for classification tasks, the concept has been significantly generalized in the LLM era. The objective is no longer merely to make a smaller model produce the same outputs as a larger one, but rather to achieve efficient knowledge compression, faithful reproduction of reasoning processes, and deep internalization of complex agentic skills under constrained model capacity.

From the perspective of technical motivation, model distillation primarily addresses deployment latency on edge devices, cloud-service costs (e.g., improving throughput through frameworks such as vLLM), and the optimization of capabilities for specific vertical domains. Finding a Pareto-optimal trade-off among accuracy, memory footprint, inference speed, and training costs (including supervised fine-tuning and reinforcement learning budgets) has become the central theme driving the evolution of distillation technologies.

From an implementation perspective, contemporary LLM distillation can be categorized into five major technical paradigms.

White-box distillation / feature-level alignment is the most classical form of distillation and requires access to the weights of both the teacher and student models. The student model is trained to approximate the teacher by minimizing the KL divergence between their output probability distributions or by aligning intermediate hidden states and attention matrices. Because this approach provides high-bandwidth information transfer, it is theoretically efficient. However, with the widespread adoption of closed-source frontier-model APIs, white-box distillation is now primarily used for scaling within the same model family, such as distilling a 70B-parameter model into an 8B-parameter variant.

Black-box distillation / prompt-based synthetic data generation applies when teacher-model parameters are inaccessible and powerful proprietary models (e.g., GPT-4 or Claude 3.5) are treated as data generators. Carefully designed prompts are used to induce the teacher model to generate high-quality question-answer pairs, code samples, or multi-turn conversations. These synthetic datasets are then used to supervise the fine-tuning (SFT) of student models. The primary challenges of this paradigm include maintaining data diversity, filtering noise, and mitigating distribution shift between synthetic and real-world data.

Reasoning and chain-of-thought (CoT) distillation targets tasks such as mathematical reasoning, code generation, and complex logical inference. Training a student model solely on final answers often results in superficial pattern matching rather than genuine problem-solving capability, leading to increased hallucinations. Reasoning distillation addresses this issue by extracting high-quality chains of thought, multi-step solution trajectories, and verification or reflection processes generated by the teacher model. These are used to construct training triples of the form (instruction, reasoning process, final answer). Such supervision enables the student model to implicitly learn task decomposition strategies and step-by-step reasoning patterns, while also internalizing part of the optimal policies discovered through reinforcement learning.

Agent skill internalization and memory compression extend the goal of distillation beyond text generation toward interactive decision-making. This paradigm focuses on extracting successful trajectories accumulated by teacher models through trial-and-error exploration, tool use, and multi-step planning in complex environments such as ALFWorld and WebShop. By compressing these global memories and interaction experiences into high-quality supervised fine-tuning datasets, student models can internalize specific action policies and decision flows without requiring massive parameter counts. Consequently, relatively small models can acquire capabilities for task decomposition, state transitions, and environment interaction that resemble those of much larger systems.

On-policy distillation and distribution-shift correction address the limitations of traditional instruction distillation based on supervised fine-tuning, which is fundamentally an off-policy learning process: the student passively learns from static trajectories generated by the teacher. This approach suffers from severe distribution shift, often referred to as exposure bias. During training, the student predicts the next token under a perfect context generated by the teacher; during inference, however, it must continue generation based on its own outputs, which may contain small errors. These errors can accumulate over time, causing performance degradation. On-policy distillation addresses this issue by requiring the student model to actively generate outputs during training according to its own policy. These student-generated trajectories are subsequently evaluated, corrected, or ranked by a powerful teacher model, a verifier, or a reward model trained on teacher-generated data. This process forces the student to explore within its own probability distribution and learn how to recover from its characteristic failure modes. From a mathematical perspective, on-policy distillation represents an important transition from minimizing Forward KL divergence---which encourages the student to cover the entire teacher distribution---to minimizing Reverse KL divergence, which encourages the student to identify high-confidence solutions within its own capability region. This shift significantly improves generation quality and robustness in real-world deployment scenarios.

In summary, model distillation seeks to answer a central question: how can high-quality knowledge be efficiently acquired, cleaned, compressed, and injected into smaller models, and how can alignment algorithms enable these models to inherit the generalization and reasoning capabilities of much larger systems? In engineering practice, successful distillation pipelines often integrate distributed training frameworks such as Megatron-LM, efficient post-inference processing mechanisms, and dynamic sampling strategies to establish a complete closed-loop workflow spanning synthetic data generation, supervised fine-tuning, evaluation, and iterative optimization.

\subsubsection{Industry Trends and Developments}

The evolution of large language model distillation has been closely intertwined with the broader development of natural language processing paradigms. From early feature alignment methods, to the explosion of instruction-tuning datasets, and more recently to the transfer of reinforcement learning policies and advanced reasoning capabilities, the field has undergone several distinct stages.

\begin{figure}[htbp]
\centering
\includegraphics[width=0.95\linewidth]{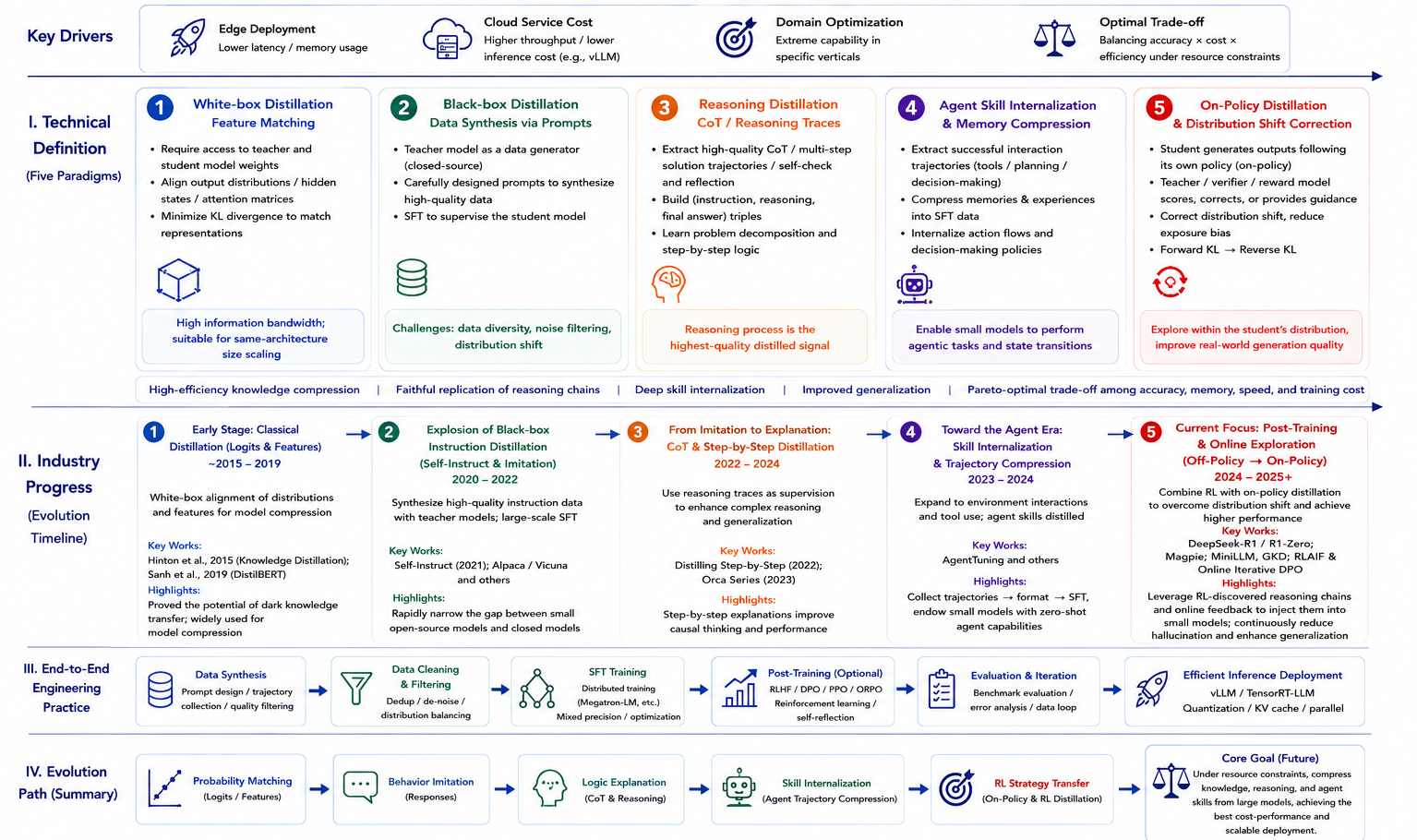}
\caption{Industry progress in model distillation.}
\label{fig:2_16}
\end{figure}

\paragraph{Early Stage: Classical Distillation Based on Logits and Feature Alignment}

During the era of pretrained language models such as BERT and RoBERTa, distillation primarily took the form of white-box knowledge transfer. The concept of knowledge distillation, formally introduced by Hinton et al. in 2015, demonstrated the potential of transferring "dark knowledge" by softening output probability distributions through a temperature parameter \cite{ref145}. Subsequently, numerous distillation methods tailored to Transformer architectures emerged. DistilBERT aligned output distributions during pretraining and compressed BERT by approximately 40\% while retaining 97\% of its language understanding capability \cite{ref146}. MiniLM further proposed aligning deep self-attention relations, including value-relation and query-key relation matrices, enabling more effective structural compression \cite{ref147}. At this stage, distillation was primarily concerned with model compression and relied heavily on homogeneous model architectures.

\paragraph{The Rise of Black-Box Instruction Distillation: Self-Instruct and Imitation Learning}

With the release of large-scale instruction-tuned models such as ChatGPT, direct access to high-dimensional logits became increasingly unavailable. Industry efforts therefore shifted rapidly toward generation-based data synthesis for black-box distillation. The Self-Instruct framework became a foundational work of this period: it showed how a powerful teacher model can autonomously generate large-scale and diverse instruction-following data from a small number of seed prompts, which can then be used to fine-tune high-performing student models \cite{ref148}. Building on this paradigm, models such as Stanford Alpaca and Vicuna narrowed the gap between open-source small models and proprietary frontier models in everyday conversational experience.

However, researchers soon identified the limitations of pure black-box imitation. The study The False Promise of Imitating Proprietary LLMs rigorously demonstrated that supervised fine-tuning solely on teacher-generated responses often teaches students only the teacher's style and formatting patterns rather than its underlying reasoning abilities or factual knowledge. In some out-of-distribution tasks, such students even perform worse than expected \cite{ref149}. This finding motivated the search for deeper forms of knowledge transfer.

\paragraph{From Imitation to Explanation: Chain-of-Thought and Reasoning Distillation}

To overcome the limitation of learning answers without understanding the underlying reasoning process, researchers increasingly shifted the distillation target from outputs to reasoning trajectories themselves. Google's Distilling Step-by-Step framework introduced teacher-generated rationales as additional supervision signals, significantly improving the performance of smaller models on natural language reasoning tasks while requiring substantially less training data \cite{ref150}.

Microsoft's Orca series further advanced this direction through explanation tuning. Rather than merely providing answers, the teacher model was required to explicitly generate detailed reasoning traces, system-instruction-following logic, and intermediate decision-making processes. By learning from these causally rich reasoning trajectories, Orca-13B achieved performance on complex tasks approaching that of substantially larger foundation models \cite{ref151}. This stage established a widely accepted principle: reasoning processes are among the most valuable forms of distillation data.

\paragraph{Entering the Agent Era: Skill Internalization and Trajectory Compression}

As language models acquired tool-use and multi-step environment interaction capabilities, distilling agent skills into smaller models became a major research frontier. Works such as AgentTuning demonstrated how powerful teacher models can explore virtual environments and external APIs, collect successful interaction trajectories, and convert them into supervised fine-tuning datasets, thereby enabling relatively small open-source models to acquire zero-shot agent capabilities \cite{ref152}.

A central challenge in this area is the signal-to-noise ratio of memory compression and skill distillation. During interaction with complex environments, teacher models often generate substantial trial-and-error behavior, redundant reflections, and backtracking operations. Blindly distilling these trajectories risks teaching student models to hesitate or overthink. Consequently, increasing attention has been devoted to techniques such as skill abstraction and trajectory pruning, which aim to extract core skill primitives from successful trajectories and allow smaller models to internalize specialized capabilities, such as code execution or multi-step question answering, under limited parameter budgets.

\paragraph{The Current Frontier: Post-Training, Reinforcement Learning, and Distillation of Advanced Reasoning Models}

Between late 2024 and 2025, the emergence of systems such as OpenAI's o1 series and DeepSeek-R1 shifted attention toward test-time computation and reinforcement learning, leading to another major leap in distillation methodologies. The release of DeepSeek-R1 demonstrated not only that large-scale reinforcement learning can elicit powerful long-horizon reasoning capabilities, but also that these capabilities can be efficiently transferred to smaller models through supervised distillation. Open-source distilled variants of R1-Zero based on architectures such as Qwen and Llama showed that high-quality reasoning trajectories discovered through reinforcement learning can be directly injected into models with only a few billion parameters, enabling them to achieve unexpectedly strong performance on mathematics and coding benchmarks \cite{ref076}.

Meanwhile, frameworks such as Magpie further streamlined the process of automated alignment-data generation. By exploiting the autoregressive properties of pretrained language models, Magpie can extract large-scale conversational and reasoning instructions directly from teacher models without requiring manually designed prompts, substantially reducing the cost of constructing distillation datasets \cite{ref153}.

For distribution-shift correction, the industry has increasingly recognized that purely offline imitation based on static datasets can lead small models to develop overconfident hallucinations and poor robustness under out-of-distribution prompts. On-policy distillation has therefore become an important pathway for capability transfer. MiniLLM highlighted that forcing a small model to mimic the broad and relatively flat output distribution of a large model can degrade generation quality, and introduced reverse-KL optimization on trajectories generated by the student itself, allowing the student to learn high-confidence expressions that remain within its own capability region while staying aligned with the teacher's logic \cite{ref154}. Generalized distillation further lets the student first generate trajectories and then receives probability-distribution guidance from the teacher on those student trajectories, reducing the gap between training and inference distributions \cite{ref155}. At the preference-alignment stage, on-policy distillation further evolves into paradigms such as RLAIF and online iterative DPO, where student models generate candidate responses online, stronger models score them to construct preference pairs, and the resulting feedback is used for further optimization \cite{ref156}.

However, many reasoning-distillation methods still rely on rejection sampling: a teacher model generates multiple reasoning trajectories for a given problem, and only correct trajectories are retained for student training. In this process, the teacher effectively acts as a static filter. For difficult boundary problems that the teacher cannot solve independently, all sampled trajectories may be marked unsolved and discarded, leaving the student trained mainly on easy and medium-difficulty examples. This creates an artificial teacher ceiling, in which the student's upper bound is constrained by the teacher's independent problem-solving ability. Empirical probing by China Unicom found that even when the sampling budget was increased to $N=64$, a Qwen3-32B teacher still failed to generate any valid trajectory for approximately 13\% of the most difficult AIME problems.

To address this limitation, China Unicom proposed HEAL (Hindsight Entropy-Assisted Learning) \cite{ref157}, a reinforcement-learning-free reasoning distillation framework inspired by the educational concept of the Zone of Proximal Development (ZPD). Its core insight is that a teacher's failure to solve a difficult problem independently does not necessarily mean that it lacks the relevant latent reasoning capability; it may only need a hint to enter the correct search space. HEAL reconstructs the distillation workflow along three dimensions: active synthesis, quality filtering, and progressive curriculum learning.

The GEAR (Guided Entropy-Assisted Repair) module performs entropy-driven reasoning repair. It simulates an expert problem-solving process in which the teacher first attempts a solution independently and consults the answer only when genuinely stuck. GEAR monitors entropy-gradient dynamics throughout the reasoning process and locates the step with the largest entropy surge as a critical reasoning breakpoint, namely the boundary at which the model transitions from certainty to uncertainty. At this breakpoint, the ground-truth answer is injected as a localized hint, allowing the teacher to resume reasoning from the last stable logical state and repair the broken trajectory. The search range is constrained to the first third of the sequence, because early logical errors in mathematical reasoning often irreversibly corrupt subsequent steps; GEAR therefore focuses on repairing foundational structural errors.

The PURE (Perplexity-Uncertainty Ratio Estimator) module addresses the risk of shortcut reasoning introduced by answer hints. A model may generate trajectories such as "Since the reference answer is 36, the final answer is therefore 36," which are syntactically coherent but logically vacuous. PURE identifies such shortcuts using a suspicion ratio, $\mathrm{PPL}(s_t) / (\mathrm{NLL}(a \mid s_{\le t}) + \epsilon)$. If a reasoning step exhibits high perplexity but unusually low answer uncertainty, it is likely to represent an imposed logical shortcut. PURE uses the global peak of this suspicion ratio as the trajectory-level anomaly score and removes the top 20\% most suspicious trajectories, preserving the logical integrity of the distillation data.

The PACE (Progressive Answer-Guided Curriculum Evolution) module organizes distillation into a three-stage curriculum. The first stage trains only on independently solved trajectories ($D_{\mathrm{base}}$). The second stage introduces globally answer-guided trajectories ($D_{\mathrm{hint}}$) and mixes them with the base data. The third stage incorporates GEAR-repaired trajectories ($D_{\mathrm{repair}}$) and oversamples them, allowing the student to focus on the most difficult reasoning breakpoints. This easy-to-hard curriculum improves training stability and reduces catastrophic forgetting.

HEAL shows that actively intervening in the teacher's reasoning process, repairing cognitive breakpoints, filtering shortcut reasoning, and progressively transferring knowledge can enable student models to break through the teacher ceiling imposed by traditional rejection-sampling-based distillation. In token-oriented operations, China Unicom has adopted HEAL as a core technology for high-quality knowledge refinement, compressing the capabilities of large reasoning models into lightweight models deployable on enterprise-grade servers and substantially reducing MaaS platform inference costs.

Overall, LLM distillation is evolving along a trajectory from probability matching to behavioral imitation, logical reasoning transfer, skill internalization, and reinforcement-learning policy transfer. The active-repair distillation paradigm represented by HEAL is pushing this trajectory from static knowledge transfer toward dynamic cognitive elicitation: rather than simply selecting problems already solved by the teacher, it helps the teacher cross its independent reasoning boundary and converts previously discarded difficult examples into high-quality distillation signals. Looking ahead, key challenges include how to compress the knowledge of large-scale agent systems and long-horizon reasoning structures into edge-deployable models through more efficient distributed training and data-cleaning pipelines, and how to extend entropy-guided repair mechanisms such as HEAL to self-distillation and online reinforcement-learning paradigms.

\section{Compute-Model Fusion}

Compute-model fusion is a key technical direction in large model inference performance optimization. Its core lies in performing collaborative optimization around model structure characteristics, operator execution methods, GPU memory access paths, and inference engine scheduling mechanisms, to achieve efficient matching between model algorithms and underlying computing resources. This section focuses on analysis across five aspects: operator fusion, GPU memory access optimization, basic operator acceleration, inference engine parameter tuning, and model architecture adaptation. Among them, model architecture adaptation is the prerequisite for compute-model fusion, determining how the inference system identifies the attention mechanisms, MoE expert layers, positional encodings, quantization formats, and multimodal components in different models; operator fusion and basic operator acceleration are the core means of improving per-computation efficiency, with the former reducing intermediate tensor reads and writes and lowering scheduling overhead, and the latter improving the execution efficiency of high-frequency operators such as GEMM, attention, Softmax, and normalization; GPU memory access optimization targets memory bottlenecks caused by KV cache, long context, and high concurrency in large model inference, improving resource utilization through cache management, data layout, hierarchical storage, and access path optimization; inference engine parameter tuning further optimizes parameters such as parallelization strategies, batching, prefill and decoding scheduling, and KV cache quotas at the service runtime layer, in order to balance time to first token, throughput, stability, and cost. Overall, these five categories of technologies together constitute a complete technical chain from ``model structure identification---operator and memory optimization---inference engine scheduling---service-oriented deployment evaluation,'' and serve as an important foundation for supporting the high-performance, low-cost, and stable production deployment of large models in heterogeneous computing environments.

\subsection{Operator Fusion}

\subsubsection{Technical Definition}

Operator fusion refers to identifying adjacent operators or subgraphs that can be merged in a deep learning computation graph and compiling them into a fused operator or fewer execution units, so as to reduce intermediate tensor reads and writes, Kernel Launch overhead, synchronization overhead, and scheduling overhead. oneDNN Graph defines operator fusion as follows: it accepts a complete computation graph, performs execution-engine-oriented graph partitioning, compiles candidate subgraphs for fusion, and executes them as fused operations; TVM also lists high-level operator fusion, hardware mapping, and memory latency hiding as core optimization problems for deep learning compilers.

\subsubsection{Industry Trends and Developments}

Operator fusion has evolved from pattern-based graph fusion in early CNN scenarios, such as Conv+BatchNorm+ReLU, Conv+Add, GELU, LayerNorm, and SkipLayerNorm, into a multi-layer optimization system spanning graph compilers, operator libraries, inference engines, and serving runtimes. ONNX Runtime divides graph optimizations into levels such as Basic, Extended, and Layout, where Extended optimizations include fusions such as GELU, LayerNorm, Attention, SkipLayerNorm, and BiasGELU; oneDNN Graph, by contrast, identifies candidate subgraphs through engine-aware graph partitioning and compiles and executes fused partitions\cite{ref158}.

\begin{figure}[htbp]
\centering
\includegraphics[width=0.95\linewidth]{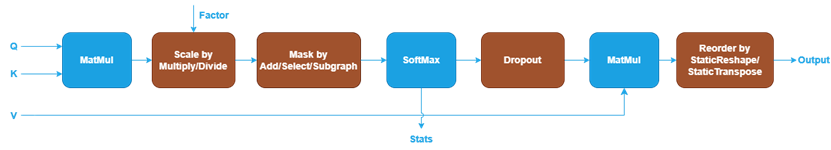}
\caption{Diagram of the Transformer Attention subgraph fusion pattern, showing how nodes such as MatMul, Scale, Mask, SoftMax, Dropout, MatMul, and Reorder constitute the SDPA DAG; adapted from Reference~\cite{ref160}.}
\label{fig:3_1}
\end{figure}

In Transformer and large model inference scenarios, Attention has become the core target of fusion optimization. TensorRT supports triggering MHA fusion either through the IAttention API or by constructing an attention graph from MatMul, Softmax, and other operations, and it supports GQA/MQA scenarios. Its documentation explicitly states that, for long sequences, MHA fusion can reduce the memory footprint of attention intermediate results from O(S\textasciicircum{}2) to O(S), while also reducing memory traffic, kernel launches, and synchronization overhead. TensorRT-LLM further distinguishes between the context phase and the generation phase for autoregressive GPT-like models: context FMHA can execute MHA/MQA with a single kernel, while long sequences adopt the ideas of FlashAttention/FlashAttention-2; the generation phase uses a masked MHA kernel and supports decoding-stage-oriented optimizations such as XQA, Paged KV cache, and INT8/FP8 KV cache.

\begin{figure}[htbp]
\centering
\includegraphics[width=0.95\linewidth]{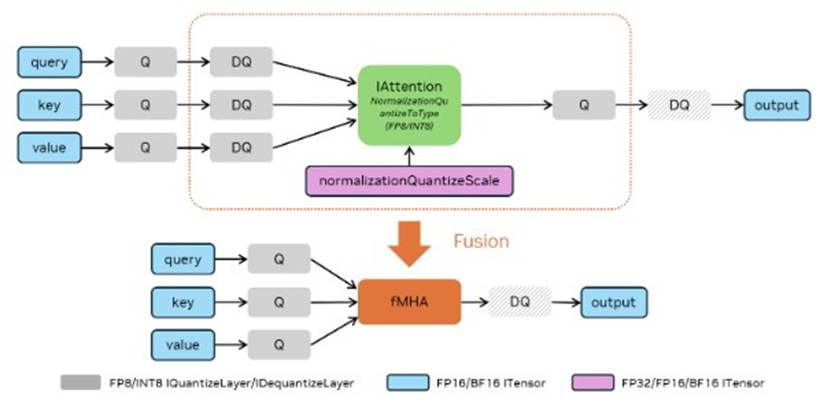}
\caption{The process of constructing an attention graph through the IAttention API to trigger MHA fusion; adapted from Reference~\cite{ref160}.}
\label{fig:3_2}
\end{figure}

Attention fusion has developed into a dedicated kernel system that is IO-aware, hardware-aware, and precision-aware. FlashAttention reduces reads and writes between HBM and on-chip SRAM through tiling \cite{ref044}; FlashAttention-2 improves parallelism and work partitioning \cite{ref045}; FlashAttention-3 introduces asynchronous execution, warp specialization, and FP8\cite{ref159} for Hopper.

NVIDIA cuDNN Frontend also treats SDPA/FlashAttention-style kernels as a key capability\cite{ref160}. Its SDPA operation uses the FlashAttention-2 algorithm, supports both training and inference, and covers paths such as FP16/BF16 forward/backward and FP8 forward/backward; the cuDNN Frontend README also lists high-performance kernel directions such as SDPA/Flash Attention, MoE Grouped GEMM fusions, and fused normalization + activation.

MoE and MLP subgraphs have also become hotspots for fusion. In traditional Transformer MLPs, the common Linear + Bias + GELU/SwiGLU + Linear pattern can reduce intermediate tensor reads and writes through methods such as epilogue fusion, GEMM+activation fusion, and GEMM+SwiGLU fusion. In MoE scenarios, the key lies in the coordinated optimization of token routing, permute/unpermute, expert grouping, Grouped GEMM, and activation functions. cuDNN Frontend already provides Grouped GEMM + SwiGLU fusion for Blackwell SM100+\cite{ref161}.

\subsection{GPU Memory Access Optimization}

\subsubsection{Technical Definition}

Memory access optimization refers to optimizing data layout, access order, access granularity, cache reuse, and data movement paths around hierarchical storage resources such as GPU global memory/HBM, L2/L1 caches, shared memory, registers, and, when necessary, CPU DRAM, SSDs, network, KV cache, etc., in order to reduce unnecessary memory transactions, repeated reads and writes, GPU memory fragmentation, cross-hierarchy copies, and sudden OOMs. Typical methods include contiguous and aligned access, warp-level coalesced access, tensor layout rearrangement, tiling/blocking, kernel fusion, shared memory, asynchronous copy and compute overlap, block-based KV cache management, prefix cache reuse, GPU memory pooling, and hierarchical cache management.

\subsubsection{Industry Trends and Developments}

In large model inference, the importance of optimizing GPU memory access continues to increase. The reason is that inference for decoder-only LLMs can be divided into two types of workloads: prefill and decode. Prefill processes the full input sequence and is more compute-intensive; decode generates tokens one by one, and each step needs to read the historical KV cache. As context length, the number of concurrent requests, and batch size grow, the capacity footprint and read bandwidth of the KV cache rapidly become bottlenecks. Research on KV cache quantization shows that KV cache not only significantly increases memory requirements, but its loading process may also leave compute cores idle, thereby limiting inference speed \cite{ref112}.

At the KV cache management level, PagedAttention draws on the paging concept from operating systems, dividing each request's KV cache into fixed-size KV blocks. This allows logically contiguous KV cache to be stored in non-contiguous physical GPU memory, and supports KV cache sharing within and across requests. The vLLM paper reports that its goal is to achieve near-zero KV cache memory waste and flexible sharing, and that at the same latency level it delivers a 2--4x throughput improvement over systems such as FasterTransformer and Orca \cite{ref048}.

\begin{figure}[htbp]
\centering
\includegraphics[width=0.95\linewidth]{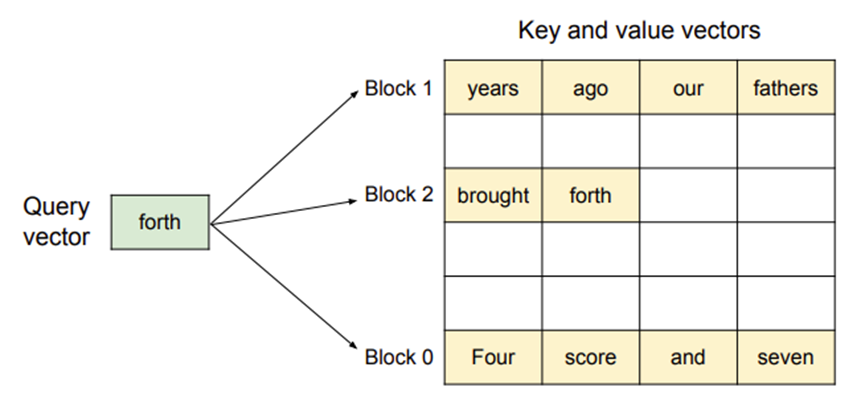}
\caption{Schematic diagram of the PagedAttention algorithm; adapted from Reference~\cite{ref048}.}
\label{fig:3_3}
\end{figure}

SGLang's RadixAttention targets scenarios such as complex programs, multi-round calls, few-shot, RAG, and multi-turn conversations. It organizes reusable prefix KV cache to avoid repeated prefill computation; its paper states that runtime KV cache reuse is performed through RadixAttention, achieving up to 6.4x throughput improvement over existing inference systems across multiple task types \cite{ref108}.

FlashInfer is an Attention engine for LLM inference services. To address the heterogeneity of KV cache storage, it introduces block-sparse and composable formats to optimize memory access and reduce redundancy, and adapts to different Attention variants through JIT templates. Meanwhile, its load-balancing scheduling algorithm takes into account both dynamic requests and the static configuration requirements of CUDAGraph. The paper reports that FlashInfer has been integrated into systems such as SGLang and vLLM, achieving a 29\%--69\% reduction in inter-token latency under different inference scenarios and a 28\%--30\% latency reduction for long-context inference\cite{ref162}.

Mooncake designs the service architecture as a KV-cache-centric disaggregated architecture, separating prefill and decode clusters, and uses underutilized CPU, DRAM, SSD, and other resources in GPU clusters to build a disaggregated KV cache. The paper reports that, in long-context scenarios, it can improve throughput by up to 525\% over the baseline, and enables Kimi to handle 75\% more requests under real workloads \cite{ref163}.

Low-precision and compressed KV cache is another main direction for long-context and high-concurrency inference. KIVI analyzes the element distribution of KV cache and proposes a 2-bit asymmetric scheme in which the key cache is quantized per channel and the value cache is quantized per token; the paper reports that on Llama, Falcon, and Mistral, it can reduce peak memory by 2.6x while largely preserving quality, increase batch size by up to 4x, and deliver a 2.35--3.47x throughput improvement \cite{ref112}. TurboQuant further formulates KV cache compression as an online vector quantization problem, reducing inner-product estimation bias through random rotation, scalar quantization, and 1-bit Quantized Johnson-Lindenstrauss residual correction. It reports that in KV cache quantization, 3.5 bits/channel can achieve quality neutrality, while 2.5 bits/channel causes only slight quality degradation\cite{ref114}.

\subsection{Fundamental Operators Acceleration}

\subsubsection{Technical Definition}

Foundational operator acceleration refers to optimizing high-frequency, general-purpose, and performance-sensitive basic computational units in deep learning models, such as GEMM/MatMul, convolution, Attention, Softmax, normalization, Pooling, activation functions, and tensor transformations. NVIDIA cuDNN is defined as a GPU-accelerated library of primitives for deep neural networks, providing highly optimized implementations of standard routines such as convolution, Attention, MatMul, Pooling, and Normalization; cuBLAS, meanwhile, is a GPU-accelerated BLAS/GEMM library for AI and HPC, supporting batched operations, execution across multiple GPUs, mixed- and low-precision execution, and fusion.

\subsubsection{Industry Trends and Developments}

Industry acceleration of fundamental operators has evolved from ``calling a single high-performance library'' into a layered system of ``standard libraries, compilers/inference engines, model-architecture-aware Kernels, and coordinated optimization of communication operators.'' The lower layer still centers on cuBLAS, cuDNN, CUTLASS, and similar libraries: cuDNN covers DNN primitives/operators such as Attention, convolution, MatMul, Normalization, Softmax, Pooling, and Pointwise; cuBLAS primarily serves GEMM and MatMul, supporting batched, multi-GPU, mixed-precision, low-precision, and fused GEMM; CUTLASS provides customizable CUDA GEMM/convolution/tensor-computation building blocks\cite{ref164}\cite{ref165}\cite{ref166}.

Large-model inference has further changed the focus of operator optimization: the Prefill stage has long sequences and large matrices, placing greater emphasis on GEMM/Attention throughput and HBM bandwidth utilization; the Decode stage has small batches and generates tokens one by one, placing greater emphasis on GEMM, low-latency scheduling, CUDA Graph, and memory access efficiency. MoE models introduce new critical paths: token routing, Grouped GEMM, expert load balancing, and so on. The DeepSeek-V3 technical report shows that DeepSeek-V3 adopts a large-scale MoE architecture and combines FP8 mixed-precision training, DualPipe, cross-node All-to-All communication optimization, and compute-communication overlap to reduce training and inference costs. This indicates that ``fundamental operator acceleration'' in MoE scenarios also includes communication Kernels, load balancing, and low-precision data movement \cite{ref042}.

One item especially worth attention in the past two years is DeepSeek DeepGEMM \cite{ref167}. DeepGEMM was initially open-sourced as a CUDA Kernel library for FP8 GEMM, used to support dense GEMM and MoE Grouped GEMM in DeepSeek V3/R1 training and inference. By April 2026, its README described it as a unified high-performance Tensor Core Kernel library covering key LLM computation primitives such as FP8, FP4, and BF16 GEMMs, MoE, MQA scoring, and HyperConnection, and using lightweight JIT to compile Kernels at runtime.

The core value of DeepGEMM lies in this: it is a model-architecture-aware Kernel for LLM/MoE forms. Its Grouped GEMM design groups only along the M dimension while keeping N/K fixed, making it suitable for MoE scenarios where multiple experts share the same weight shape but each expert receives a different number of tokens. Prefill/training can use a contiguous memory layout, concatenating tokens from different experts into a contiguous tensor. In the Decode stage, under CUDA Graph scenarios, it can use a masked layout, with masks representing valid token intervals.

DeepGEMM also has clear boundaries: it is currently mainly oriented toward NVIDIA SM90/SM100 architectures, and users need to handle FP8 type conversion, input transposition, and partial pre-fusion themselves. The contiguous memory layout of Grouped GEMM requires M-dimension computation-block alignment for each expert segment, which can introduce wasted computation under small batches or imbalanced loads. Around this pain point, the 2025 TMA adaptive FP8 Grouped GEMM proposed using a TMA descriptor pool and dual-phase load-store operations to eliminate the padding overhead caused by fixed 128 alignment. In experiments, compared with the ``padding + DeepGEMM'' baseline, it achieved a speedup of 1.7\%--20.4\% and memory savings of up to 23.8\%. This shows that DeepGEMM has become an important baseline for subsequent MoE Grouped GEMM research\cite{ref168}.

\subsection{Engine Parameter Tuning}

\subsubsection{Technical Definition}

Engine parameter tuning refers to the systematic optimization of an inference engine's parallelism parameters and scheduling strategies for large-model inference services, under the constraints of a given task type, compute resources, and quality-of-service objectives, so that the model service achieves the optimal balance among time-to-first-token latency, throughput, stability, and resource utilization. Its core tuning targets include parameters such as TP, PP, DP, PD disaggregation, prefill chunk size, and KV cache quota.

\subsubsection{Industry Trends and Developments}

Mainstream inference frameworks in the industry have already upgraded ``engine parameter tuning'' from static deployment parameter configuration to a system-level optimization problem. vLLM is one of the representatives of this direction. The vLLM V1 documentation incorporates chunked prefill, TP/PP/DP/EP, NUMA binding, attention backends, and parallel input processing into its tuning system. Among these, chunked prefill is enabled by default in applicable scenarios, and uses `max\_num\_batched\_tokens` to control the mixing ratio between Prefill and Decode: smaller values are more favorable to ITL, larger values are more favorable to TTFT and throughput, while excessively large values may sacrifice end-to-end latency\cite{ref169}.

At the scheduling algorithm level, Sarathi-Serve proposes splitting the Prefill of large prompts into multiple chunks and combining them with Decode requests into more balanced batches, thereby reducing Prefill's blocking effect on Decode and achieving a better balance between throughput and tail latency\cite{ref170}. DistServe, from the perspective of system architecture, proposes PD disaggregation, allocating compute-intensive Prefill and memory-access-sensitive Decode to different GPU pools, and separately optimizing resource allocation, parallelization strategies, and cross-stage placement for TTFT and TPOT, thereby improving goodput under SLO constraints\cite{ref171}.

SGLang also extends high-performance runtime capabilities to RadixAttention, a zero-overhead CPU scheduler, PD disaggregation, speculative decoding, TP, PP, EP, DP, chunked prefill, and more. In its PD disaggregation documentation, SGLang explicitly defines Prefill as a compute-intensive stage and Decode as a memory/KV-cache-intensive stage, and supports KV transfer backends such as Mooncake and NIXL. In addition, SGLang's tuning documentation already provides parameter recommendations for `--mem-fraction-static`, `--chunked-prefill-size`, `--max-running-requests`, `--cuda-graph-max-bs`, `--dp-size`, `--tp-size`, and others, indicating that its tuning focus already covers queue depth, KV cache utilization, CUDA Graph, parallelism modes, and prefix cache hit rate\cite{ref172}.

TensorRT-LLM supports in-flight batching, chunked context/prefill, paged KV cache, `max\_batch\_size`, `max\_num\_tokens`, KV cache block manager, KV cache reuse/offloading, quantization, and multiple parallelization strategies, and explicitly states that `max\_num\_tokens` affects the balance among available GPU memory for KV cache, throughput, TTFT, and TPOT\cite{ref173}. NVIDIA Dynamo further incorporates request routing above the inference engine, PD disaggregation, KV-aware routing, and GPU-to-GPU KV cache transfer into its distributed inference framework. Its official design documentation divides a single disaggregated request into three steps: Prefill computes KV, KV transfer, and Decode computation, and emphasizes that the key to high-performance disaggregated inference lies in efficient KV transfer and routing orchestration\cite{ref174}. Such systems show that, for large-scale inference clusters, engine parameter tuning has become a cluster-level optimization problem spanning instance grouping, GPU pool ratios, network interconnects, KV transfer backends, and cache routing strategies.

\begin{figure}[htbp]
\centering
\includegraphics[width=0.95\linewidth]{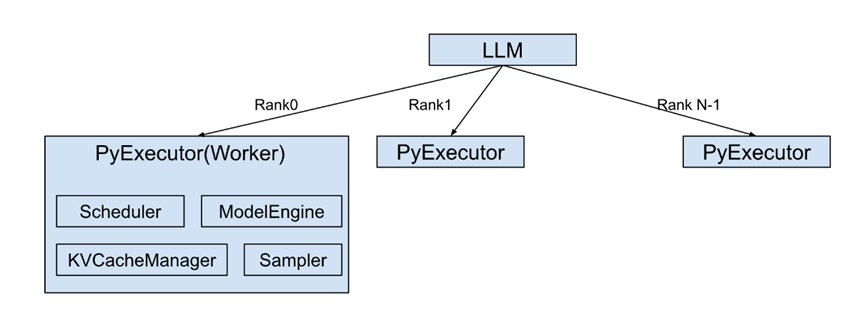}
\caption{Internal Structure Diagram of the TensorRT-LLM Inference Engine; adapted from Reference~\cite{ref173}.}
\label{fig:3_4}
\end{figure}

TGI has production capabilities such as continuous batching, SSE streaming output, Flash Attention, Paged Attention, KV cache, quantization, Tensor Parallelism, and an OpenAI-compatible API, but the official documentation also states that TGI has entered maintenance mode and recommends using inference engines such as vLLM or SGLang for newly built inference services\cite{ref175}. Therefore, in terms of the technical roadmap, TGI should be positioned as an object for ``legacy compatibility and benchmark comparison,'' rather than as the focus of subsequent capability development and tuning.

In the direction of domestic computing power, Ascend MindIE LLM provides large-model inference capabilities based on Ascend hardware and supports acceleration features such as multi-concurrent request scheduling, Continuous Batching, PageAttention, and FlashDecoding. Its LLM Manager is responsible for state management, task scheduling, unified memory pool management of the KV cache, and status monitoring interfaces, while the Modeling layer supports built-in models, Tensor partitioning, and multiple quantization methods\cite{ref176}. This indicates that domestic inference stacks are likewise carrying out systematic optimization around scheduling, caching, device memory pools, and runtime parameters.

\subsection{Model Architecture Adaptation}

\subsubsection{Technical Definition}

Model architecture adaptation refers to adapting the inference framework, compiler, operator library, and service scheduling logic to a specific model structure without changing the model's core semantics, such as MHA/MQA/GQA/MLA Attention, RoPE or ALiBi positional encoding, MoE expert layers, multimodal encoders, LoRA Adapter, quantization formats, and custom layers. The Hugging Face Transformers documentation explains that model code and configuration can be customized, and that users can add layers or optimize the Attention mechanism by modifying model components; the TensorRT-LLM documentation explicitly supports implementations of MHA, MQA, and GQA Attention for GPT-like models.

\subsubsection{Industry Trends and Developments}

As Llama, Qwen, DeepSeek, Gemma, multimodal models, and MoE models iterate rapidly, inference frameworks must identify a model's Attention type, positional encoding, expert-layer structure, vision/speech encoder, Adapter injection points, quantization format, context length, and chat template, and map these structural features to executable operators, cache layouts, parallel partitioning, and scheduling strategies.

Hugging Face Transformers enables new models to be integrated through configuration and model code via PreTrainedConfig, custom model classes, AutoClass, and AttentionInterface\cite{ref177}; vLLM further provides a Transformers modeling backend, allowing some models not natively supported by vLLM to enter the inference serving system through Transformers implementations. For MoE models, vLLM requires sparse MoE blocks to expose the expert structure, and requires Attention layers to invoke backends through a unified Attention function interface\cite{ref178}. For MLA introduced by the DeepSeek series, adaptation involves the compressed representation of KV cache, latent projection, different computation modes in the prefill/decode phases, and the selection of Kernel backends such as FlashMLA, FlashInfer, Triton, and CUTLASS\cite{ref057}\cite{ref179}.

MoE models drive inference systems to expand from tensor parallelism to expert parallelism and load balancing. The deployment bottlenecks of models such as DeepSeek-V3/R1, Mixtral, and Qwen-MoE lie not only in parameter scale, but also in expert-weight distribution, token-to-expert routing, All-to-All communication, and Grouped GEMM. vLLM has already incorporated expert parallelism into its distributed inference capabilities\cite{ref178}; SGLang explicitly uses expert parallelism to distribute MoE expert weights across multiple devices, and combines All-to-All communication with Grouped GEMM to reduce latency and improve throughput\cite{ref180}.

At the same time, inference frameworks must also handle vision encoders, audio encoders, projection layers, connectors, special tokens, image/video preprocessing, prompt templates, and multimodal KV cache. TensorRT-LLM's support matrix already covers language models, multimodal models, and vision generation models; vLLM is also expanding in areas such as LoRA, multimodal LoRA, prompt embeddings, and multimodal data processing\cite{ref181}.

Overall, the industry is forming a relatively clear model-architecture adaptation path: model structure identification $\rightarrow$ configuration and weight conversion $\rightarrow$ Tokenizer/Processor/Chat Template adaptation $\rightarrow$ Attention and KV cache strategy selection $\rightarrow$ MoE/multimodal/LoRA/quantization compatibility verification $\rightarrow$ Kernel and parallel-strategy mapping $\rightarrow$ service scheduling and performance stress testing $\rightarrow$ gray release and stability monitoring. Its core objective is to transform structural differences among models into maintainable, reusable, and evaluable engineering rules, thereby shortening the integration cycle for new models and maintaining stable throughput and latency across different GPU, NPU, or hybrid-computing environments.

\begin{figure}[htbp]
\centering
\includegraphics[width=0.95\linewidth]{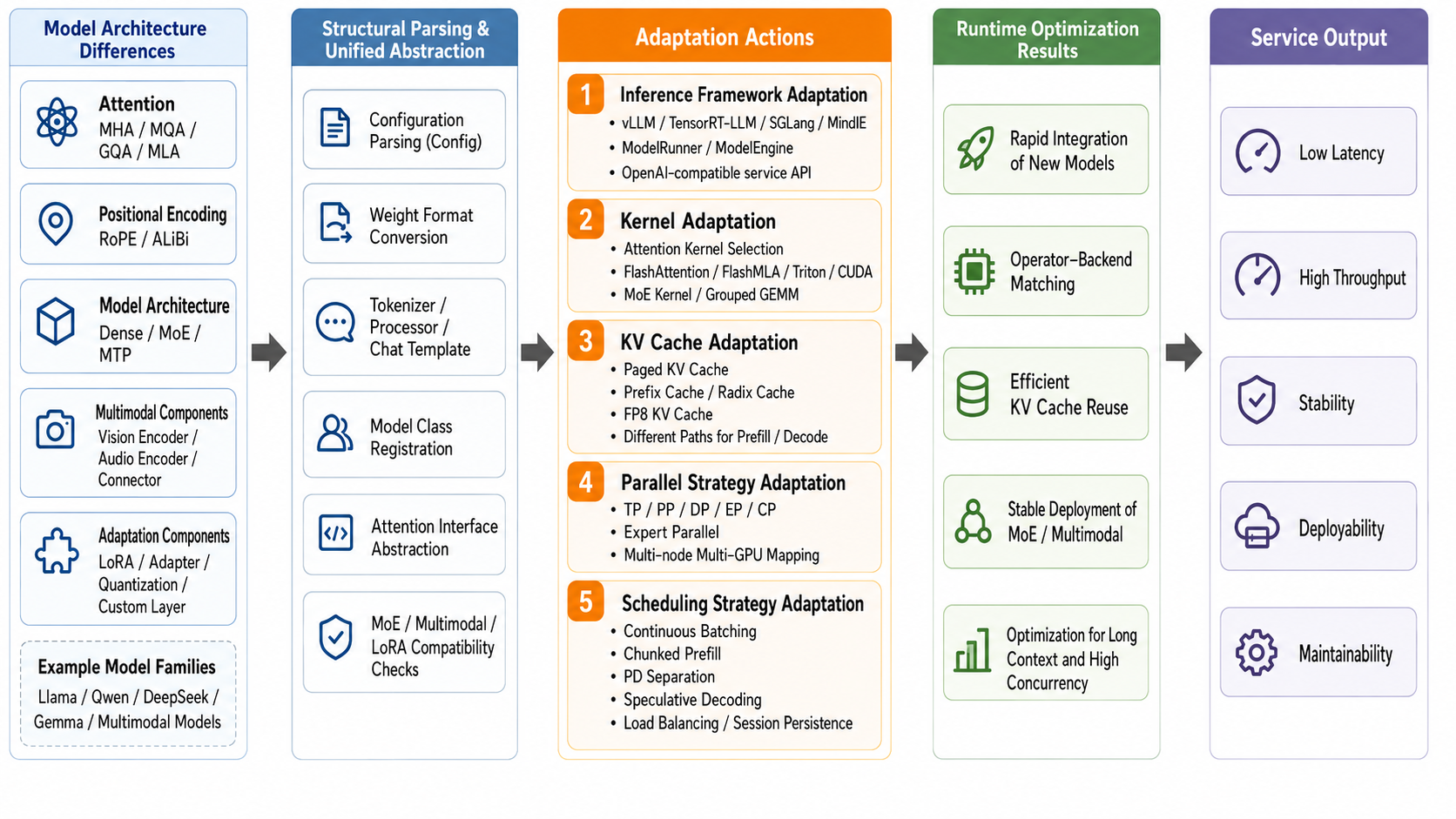}
\caption{Schematic Diagram of the Technical Process for Model Architecture Adaptation.}
\label{fig:3_5}
\end{figure}

\section{Compute-Network-Model Fusion}

Compute-network-model fusion is a key technical direction for large models as they move from single-node inference optimization toward clustered, platform-based, and service-oriented deployment. Its core lies in establishing a collaboratively optimized relationship among model inference requirements, computing resource organization, network communication capabilities, and gateway scheduling mechanisms. Unlike compute-model fusion, which emphasizes single-operator and single-instance execution efficiency, compute-network-model fusion focuses more on how large models can run stably, efficiently, and at low cost in multi-node, multi-GPU, multi-instance, multi-tenant, and high-concurrency scenarios.

This section focuses on eight aspects: multi-node inference parallelism, cluster-level KV cache scheduling, gateway sticky session routing, gateway semantic cache reuse, dynamic batching, high-performance communication libraries, load balancing, and rate limiting and degradation/fallback. Among them, multi-node inference parallelism is the foundation for scaling computing power, breaking through single-GPU memory and compute limits through tensor parallelism, pipeline parallelism, data parallelism, expert parallelism, context parallelism, and other methods. High-performance communication libraries provide data-plane support for multi-node parallelism and cache migration, and are responsible for efficiently completing collective communication, point-to-point communication, and KV cache transfer among GPUs, nodes, and resource pools.

Cluster-level KV cache scheduling, gateway sticky session routing, and semantic cache reuse jointly center on ``maximizing cache value,'' reducing repeated computation overhead at three levels: system-level cache management, request routing binding, and semantic-level result reuse. Dynamic batching further improves hardware utilization at the request scheduling level, striking a balance among throughput, time to first token, and tail latency. Load balancing, rate limiting, and degradation/fallback constitute service governance capabilities, ensuring that model services can maintain stable operation under traffic fluctuations, resource constraints, and local failures.

Overall, compute-network-model fusion forms a complete technical chain from ``model parallel partitioning --- network communication and transmission --- KV cache reuse --- request scheduling and orchestration --- gateway governance and protection,'' and serves as an important foundation for supporting large-model inference services as they move from single-point deployment to large-scale production operation.

\subsection{Multi-node Inference Parallelism}

\subsubsection{Technical Definition}

Multi-node inference parallelism refers to distributing large-model inference computation across multiple GPUs and multiple compute nodes, using tensor parallelism, pipeline parallelism, data parallelism, expert parallelism, context parallelism, and other methods to address insufficient single-GPU memory, insufficient single-node compute capacity, or insufficient throughput. TensorRT-LLM summarizes parallelization strategies as TP, PP, DP, EP, CP, and Wide-EP, where TP partitions model weights, PP distributes the model by layers, DP replicates the model to process different requests, EP is used for MoE expert distribution, and CP is used for long-context processing.

\subsubsection{Industry Trends and Developments}

DeepSeek's EPLB (Expert Parallelism Load Balancer) is currently one of the most noteworthy advances in multi-node MoE inference parallelism. Its goal is to generate an expert replication and placement plan based on expert load statistics during EP deployment: high-load experts are redundantly replicated, logical experts are mapped to multiple physical expert replicas, and the replicas are then packed onto GPUs to balance the number of received tokens and the expert computation load across GPUs. The core inputs of EPLB are the load statistics of each logical expert in each layer, and its outputs are mappings such as physical-to-logical, logical-to-physical, and the number of replicas for each logical expert\cite{ref182}.

EPLB's key design includes two types of strategies. The first is hierarchical load balancing: when the number of expert groups is divisible by the number of nodes, DeepSeek-V3's group-limited routing mechanism is used to first evenly pack expert groups onto different nodes, and then replicate experts within each node and place them onto individual GPUs, thereby minimizing cross-node communication traffic as much as possible; this strategy is more suitable for scenarios in the prefill stage where the expert parallelism scale is relatively small. The second is global load balancing: expert groups are no longer used as a constraint, and experts are instead replicated and placed globally, which is more suitable for scenarios in the decode stage where the expert parallelism scale is relatively large\cite{ref182}.

From the perspective of the open-source ecosystem, EPLB has already been incorporated into multiple inference systems. vLLM's expert-parallel deployment documentation already provides the `--enable-eplb` configuration option and explains that vLLM collects expert load statistics during each forward pass and periodically rebalances expert distribution; its configuration options include `window\_size`, `step\_interval`, `num\_redundant\_experts`, `use\_async`, and the communication backend, among others\cite{ref183}. SGLang has also integrated DeepSeek EPLB and recommends increasing batch size to stabilize expert activation statistics and using periodic rebalancing to adapt to changes in business traffic\cite{ref184}.

On the communication side, DeepSeek's open-sourced DeepEP further completes the underlying capabilities required for large-scale expert parallelism. DeepEP provides high-throughput, low-latency All-to-All GPU kernels for MoE dispatch/combine operations, supports low-precision communication such as FP8, and in V2 unifies the high-throughput and low-latency interfaces under ElasticBuffer, supporting larger-scale scale-up / scale-out expert-parallel domains. Its significance lies in the fact that the bottleneck of MoE inference is not only uneven expert placement, but also the irregular All-to-All communication caused by dynamic token routing across GPUs; EPLB is responsible for ``how experts are placed,'' while DeepEP is responsible for ``how tokens are efficiently delivered to experts''\cite{ref185}.

When SGLang reproduced a DeepSeek-style large-scale inference system in 2025, it used 12 nodes with 8 H100 GPUs per node, combining PD disaggregation, large-scale EP, DeepEP, DeepGEMM, and EPLB. Under 2000-token input sequences, it reported throughput of 52.3k input tokens/s and 22.3k output tokens/s per node; SGLang also noted that the effectiveness of EPLB depends on how well the input distribution matches real serving workloads, and that larger batch sizes and periodic rebalancing help mitigate random fluctuations in expert load\cite{ref184}.

\begin{figure}[htbp]
\centering
\includegraphics[width=0.95\linewidth]{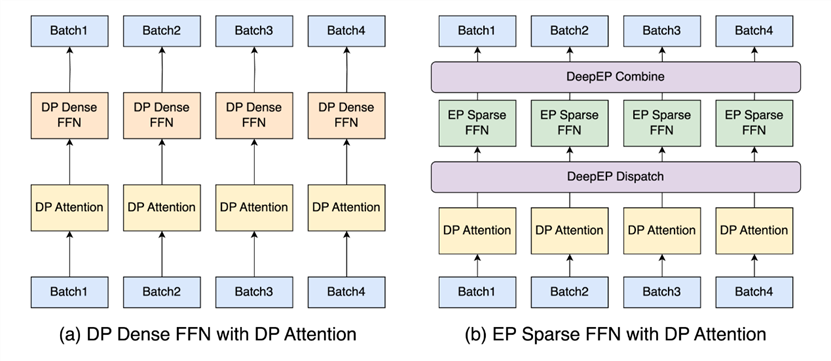}
\caption{Differences in parallelism strategies between dense models and MoE models in SGLang; the figure on the right shows the DeepEP implementation; adapted from Reference~\cite{ref184}.}
\label{fig:4_1}
\end{figure}

After EPLB, DeepSeek further open-sourced LPLB (Linear-Programming-Based Load Balancer) as a research-stage extension. LPLB uses linear programming to dynamically optimize per-batch token assignments: EPLB is more oriented toward handling long-term or static imbalance based on historical load statistics, while LPLB attempts to address transient dynamic imbalance caused by small-batch randomness. However, DeepSeek also clearly labels LPLB as still being in an early research stage, and lists limitations such as solver overhead, balancing only token count without modeling the nonlinear time cost of GroupGEMM, and potentially underperforming EPLB under extreme global imbalance\cite{ref186}.

Therefore, the current industry trend for multi-node inference parallelism can be summarized as follows: dense models are still based on combinations of TP/PP/DP/CP; MoE models rely more on DP Attention + EP MoE + PD disaggregation + dedicated All-to-All communication + EPLB load balancing.

\subsection{KV Cache Cluster-Level Scheduling}

\subsubsection{Technical Definition}

Cluster-level KV cache scheduling refers to elevating the KV cache, in multi-GPU, multi-node, multi-inference-instance, or PD-disaggregated clusters, from an intermediate GPU-memory state of a single request into a system-level resource that is manageable, indexable, migratable, and reusable. It performs block-based allocation, index matching, hit evaluation, cache-aware routing, cross-instance transfer, tiered storage, priority-based eviction, tenant isolation, and security control, in order to reduce TTFT, TPOT, and inference cost, while improving GPU memory utilization and system throughput.

\subsubsection{Industry Trends and Developments}

At the level of single-machine GPU memory management, vLLM's PagedAttention is a representative work. This approach divides each request's KV cache into fixed-size blocks/pages and avoids contiguous large GPU memory allocations through a mapping from logical blocks to physical blocks, thereby reducing GPU memory fragmentation and redundant copying, while supporting block-level sharing within the same request or across requests. The vLLM paper reports that, at the same latency level, it can achieve a 2--4x throughput improvement compared with systems such as FasterTransformer and Orca, with more pronounced benefits in long-sequence, large-model, and complex decoding scenarios \cite{ref048}. Meanwhile, vAttention uses CUDA virtual memory management APIs to decouple virtual addresses from physical GPU memory allocation, keeping the KV cache contiguous in virtual address space while allocating physical GPU memory on demand, so as to reduce the complexity of the attention-kernel and framework-memory-management modifications required by PagedAttention\cite{ref187}.

At the level of prefix caching and cache-aware scheduling, vLLM's automatic prefix caching identifies identical prefixes through hash-based block keys and incorporates the parent block hash, block tokens, LoRA ID, multimodal input hash, cache salt, and other factors into the cache key, showing that production systems have begun to focus on the combined issues of cache hits, cache isolation, and multi-tenant security. SGLang's RadixAttention further maintains the KV cache of all requests in a radix tree and, combined with LRU eviction and cache-aware scheduling, enables automatic KV cache reuse in complex LLM programs such as multi-turn dialogue, few-shot, agent, and multi-branch reasoning \cite{ref108}. TensorRT-LLM also uses a radix search tree to store reusable blocks and supports mechanisms such as priority-based LRU, retention policies, and cache salting.

At the level of hierarchical storage and cross-engine sharing, LMCache represents the direction of an ``independent KV cache layer.'' Its core idea is to extract the KV cache from the GPU HBM of a single inference engine, extend it to the GPU, CPU, storage, and network layers, and share it among engines such as vLLM and SGLang. Its paper emphasizes cache orchestration through batched data movement, computation and I/O pipelining, modular connectors, and control APIs\cite{ref188}.

At the level of PD disaggregated scheduling, Mooncake proposes a KV-cache-centric architecture for Kimi's production service, separating prefill/decode clusters and organizing CPU, DRAM, SSD, and NIC/RDMA resources into a distributed KV cache, with a global cache and scheduler making trade-offs between throughput and latency SLOs \cite{ref163}.

\begin{figure}[htbp]
\centering
\includegraphics[width=0.95\linewidth]{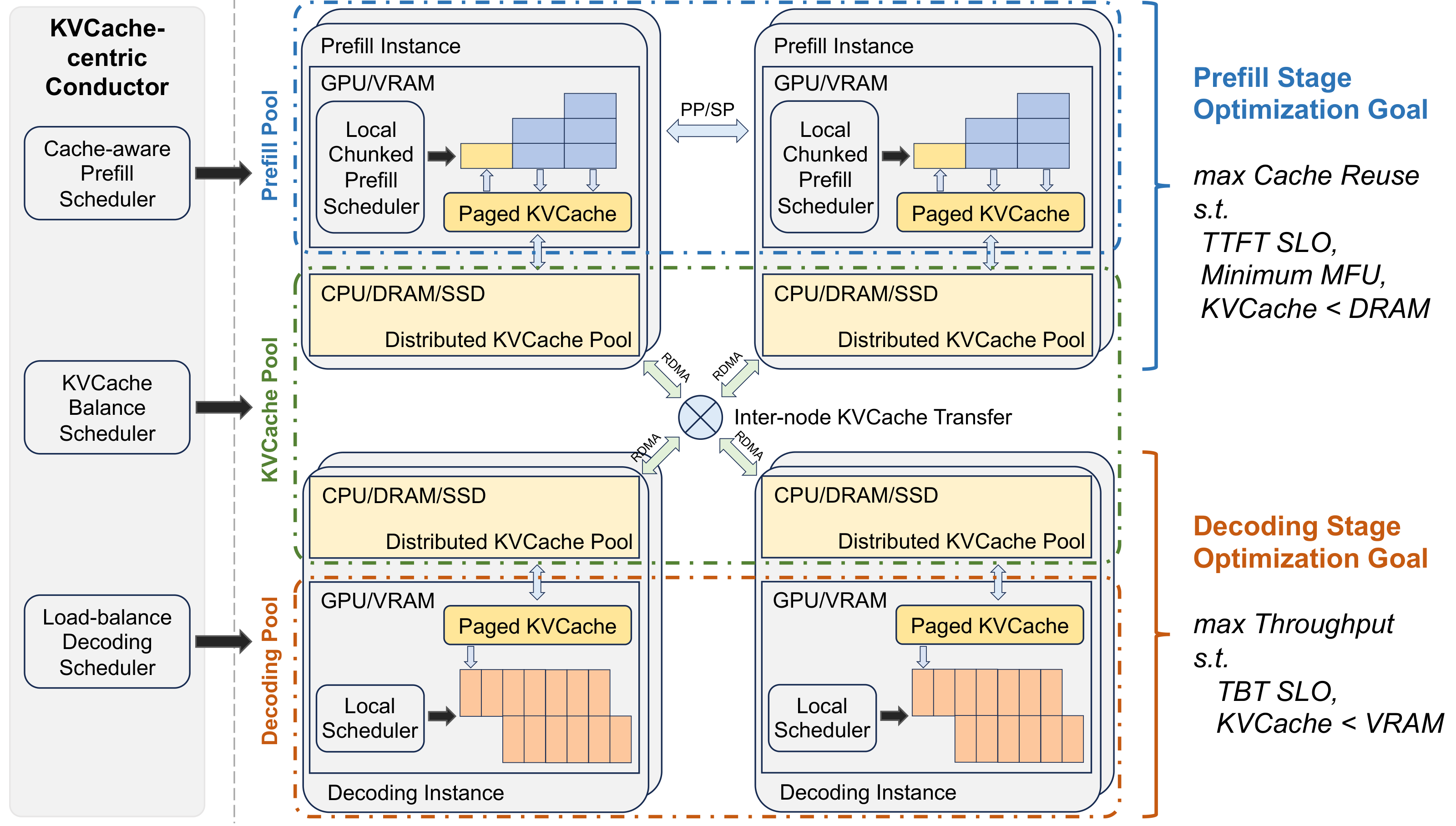}
\caption{Mooncake: a KV Cache-centric disaggregated architecture; adapted from Reference~\cite{ref163}.}
\label{fig:4_2}
\end{figure}

At the level of platformization and production deployment, TensorRT-LLM and NVIDIA Dynamo incorporate KV cache transfer, routing, offloading, and disaggregated serving into inference platform capabilities. TensorRT-LLM's disaggregated serving supports multiple access methods, including trtllm-serve, Dynamo, and Triton. Its KV cache exchange module is responsible for sending and receiving the KV cache, freeing cache space, and performing cache layout conversion. It supports communication backends such as MPI, UCX, and NIXL, and also reduces communication overhead by overlapping transfer and computation across multiple requests\cite{ref189}. NVIDIA Dynamo combines disaggregated serving, KV cache-aware routing, and KV cache offloading into data-center-level inference capabilities, and uses NIXL as a low-latency data transfer layer to support KV cache movement between nodes\cite{ref190}.

\subsection{Gateway Sticky Session Routing}

\subsubsection{Technical Definition}

Gateway sticky session routing refers to a mechanism at the AI gateway layer that consistently directs requests sharing identical contextual features to the same backend inference replica (Replica/Worker) utilizing specific hash algorithms or state mapping logic.

In traditional Web development, sticky sessions are primarily employed to maintain server-side in-memory states, such as preserving items in a user's shopping cart. However, in large language model (LLM) inference, their core purpose shifts to maximizing Key-Value (KV) cache reuse. Since LLM inference is an autoregressive generation process, each round of dialogue accumulates large intermediate computation states in video memory (VRAM). For a long context extending up to 32k tokens, recomputing the prefix for every incoming dialogue would not only introduce latency spikes of several seconds but also waste expensive computing resources.

To address this bottleneck, the AI gateway must implement coordinated request-processing logic. The process starts with feature extraction: upon receiving a request, the gateway rapidly parses the session ID or user ID from the request headers, or directly computes a prefix hash for the prompt. It then uses consistent hashing algorithms or dynamic lookup tables to continuously route requests carrying identical features to the same backend inference replica. Throughout this process, advanced gateways maintain real-time state awareness of backend nodes and monitor each machine's VRAM watermark to prevent out-of-memory (OOM) errors caused by over-prioritizing session stickiness at the expense of balanced GPU utilization.

This routing architecture is typically supported by three core components. The prefix identifier segments and hashes system prompts or dialogue histories to identify reusable text spans. The routing controller executes stickiness policies and balances two objectives: maximizing cache hit rates and maintaining backend load balance. The global cache manager serves as a metadata center, recording which context fragments are cached in the VRAM or RAM of specific nodes and providing a global view for the scheduling system.

\subsubsection{Industry Trends and Developments}

Industry scheduling strategies are shifting from simple hashing to cache-aware scheduling. Early routing methods relied mainly on user IDs or IP addresses for fixed binding, whereas recent systems increasingly support fine-grained and content-aware routing. A prominent development is the upward migration of radix-tree cache management. Inference frameworks such as vLLM and LightLLM originally managed caches internally within nodes using radix trees; contemporary gateway solutions, including Envoy-based AI extensions, attempt to decouple and elevate this tree-like logic to the gateway layer, enabling global prefix matching at the traffic ingress point \cite{ref048}. Hierarchical caching strategies are also co-evolving, with leading inference platforms implementing three-tier caching across VRAM, RAM, and disk. In this scenario, sticky routing can cooperate with SSD-assisted cache recovery: even if a request migrates to another node due to load conditions, the backend can rapidly restore context state from local disks and avoid severe performance degradation.

Taking vLLM, a widely used inference engine, as an example, its native server-side support for Automatic Prefix Caching (APC) requires the gateway layer to cooperate with cache-aware routing. In production environments, developers often deploy specific hash configurations in Nginx or write custom Lua scripts to steer requests sharing identical prompt prefixes to designated inference replicas. Practical implementations show that this hardware-software co-optimization can reduce Time-to-First-Token (TTFT) by more than 60\%.

In more advanced prefill/decode (PD) disaggregation architectures, sticky routing faces greater complexity. Under this paradigm, the gateway must maintain session affinity during decoding while also assuming a higher-level orchestration role. Once a prefill node completes the computation of a large context and generates the KV cache, the gateway must coordinate with the underlying communication mechanism to efficiently push or pull the cache data to designated decode nodes and route subsequent dialogue turns to the receiving node.

To support this gateway responsibility above inference engines, major open-source projects are evolving rapidly. For instance, the Envoy Gateway community is developing AI gateway extensions that incorporate dedicated KV-cache routing plug-ins. Domestic open-source gateways such as Higress and Kong have also begun integrating LLM plug-ins, supporting both token-count-based dynamic rate limiting and semantic-vector-based load balancing. These developments move sticky routing from simple identity-based binding toward content-aware binding. China Unicom uses a self-developed dynamic load balancer on its MaaS platform to implement sticky session routing through session identifiers, routing identical prefixes to the same node whenever possible and thereby improving KV-cache hit rates and reuse efficiency.

\begin{figure}[htbp]
\centering
\includegraphics[width=0.95\linewidth]{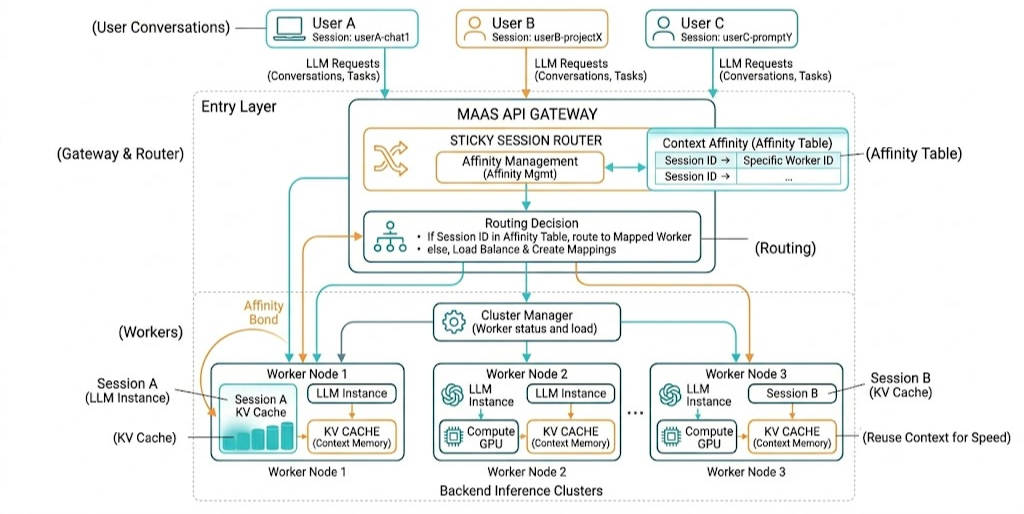}
\caption{Technical architecture of MaaS sticky session routing.}
\label{fig:4_3}
\end{figure}

\subsection{Gateway Semantic Cache Reuse}

\subsubsection{Technical Definition}

During the engineering deployment of large language models (LLMs), traditional caching schemes based on exact string matching are often insufficient because model inference has high computational cost and significant response latency. As the first filtering layer for incoming traffic, semantic caching has emerged as a critical component that uses representation learning and vector similarity retrieval to reuse responses for unstructured natural-language requests.

Traditional caching systems such as Redis rely on exact key-value matching. For instance, if a user sequentially inputs "What is the weather like in Beijing?" and "Please inform me about today's climate in Beijing," a traditional cache treats them as two independent strings and cannot reuse the cached response. Semantic caching instead maps text into a high-dimensional continuous vector space. In this embedding space, texts with different surface forms but similar semantics are close under vector-distance metrics such as Euclidean distance or cosine similarity, enabling the system to identify semantic equivalence rather than only literal equality.

In production environments, semantic caching at the gateway layer operates via a standardized and highly efficient workflow. When a new request arrives at the gateway, the vector encoding layer is first invoked to transform the raw user request into a dense vector in real time using a lightweight text embedding model. Subsequently, this vector is submitted to a high-performance vector database (such as Milvus or Vespa) for approximate nearest neighbor (ANN) search, rapidly identifying the closest historical requests within the vector space.

The retrieved candidate results are not directly returned to the user; instead, they undergo rigorous verification by the evaluation layer. This layer determines whether the retrieved historical cache content is logically equivalent to the current request, making the final semantic decision based on similarity scores. If a cache hit is determined, the gateway immediately intercepts the traffic and returns the cached response, instantaneously reducing a response latency that would otherwise take seconds down to the millisecond level. If a cache miss occurs, the request is forwarded normally to the backend LLM cluster, while the system asynchronously writes the response generated by the large model into the vector database, preparing for the next potential semantic hit.

\begin{figure}[htbp]
\centering
\includegraphics[width=0.95\linewidth]{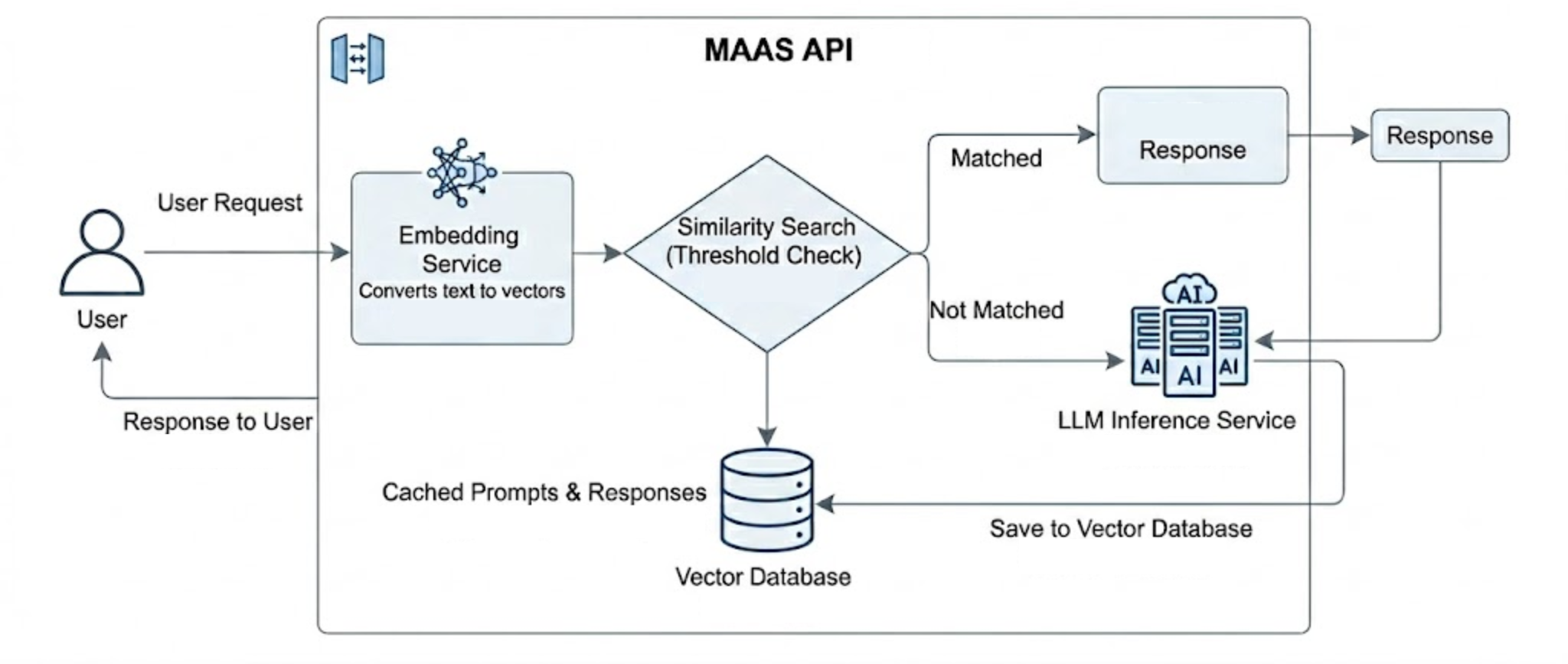}
\caption{MaaS API gateway.}
\label{fig:4_4}
\end{figure}

The performance of semantic caching is typically measured by the following three key metrics:

Hit Rate: In real-world business scenarios, due to the diversity of querying methods, semantic caching typically increases the effective hit rate by 3 to 5 times compared to traditional caching.

Search Latency: It must be ensured that the time consumed by Embedding + Vector Search is significantly less than the LLM generation time (typically requiring a ratio of less than 1:10).

Semantic Drift: This refers to the misidentification of requests as cache hits when they possess high similarity but distinct logical meanings. Resolving this issue relies on the precise design of the evaluation layer (Evaluator).

\subsubsection{Industry Trends and Developments}

Semantic caching technology has rapidly evolved from its early exploration stage in 2023 toward more intelligent and edge-oriented deployment in 2026. Its development trajectory reflects a shift from exact matching toward semantic alignment.

As the industry's first mature open-source semantic caching framework, the emergence of GPTCache marked the standardization of this technology. It proposed a five-layer modular architecture.

The storage layer supports elastic scaling from local disks to distributed vector databases.

The embedding layer supports HuggingFace, OpenAI, and various proprietary models.

The evaluation layer (Evaluator) introduces similarity threshold configurations, initially resolving reuse logic challenges.

The routing layer (Router) determines data flow and supports multi-model mixing.

The management layer (Manager) manages cache Time-To-Live (TTL) and eviction policies.

Leveraging its technical accumulation in the RAG domain, China Unicom implemented a semantic caching architecture similar to GPTCache, utilizing a self-developed small verifier model. Research data indicates that in specific vertical domains (such as financial consulting and government Q\&A), properly configuring GPTCache can help enterprises reduce repetitive inference costs by over 80\% while significantly enhancing user experience \cite{ref191}.

As application scenarios grew in complexity, evaluation strategies relying solely on vector distance (distance-based) exposed flaws related to plausible but incorrect hits \cite{ref192}. For example, "How do I activate a membership?" and "How do I cancel a membership?" may be extremely close in vector space, yet their answers are completely opposite. Consequently, the industry introduced the verifier model: Inala et al. proposed integrating miniature Transformer models as "validators" at the gateway layer. Before returning a cached response, the small model performs a rapid cross-validation between the Query and the Cache Key. This combination of "large model generation, small model verification" substantially improves result accuracy while preserving acceptable response latencies. China Unicom has also implemented this paradigm, training distinct small verifier models tailored to various scenarios of upper-layer applications.

Entering 2026, the Cortex architecture proposed by Wang et al. addressed data access challenges within ultra-large-scale distributed environments. By establishing semantic-aware mapping, Cortex achieved intelligent data prefetching at the edge. The system can predict a user's subsequent semantic path, proactively pushing potential cache hits to nodes closest to the user \cite{ref193}.

The current industry trend is progressing toward multimodal semantic caching. Beyond text, semantic caching for image generation (Stable Diffusion/Midjourney) and audio/video generation has begun entering production environments. By caching intermediate features within the latent space, the cost of multimodal inference is being further optimized.

\subsection{Dynamic Scheduling and Batching}

\subsubsection{Technical Definition}

Dynamic batching refers to the inference server dynamically combining multiple requests into batches for execution based on request arrival time, model instance state, input shape, queue waiting time, and the maximum batch size, in order to improve hardware utilization and throughput. The NVIDIA Triton documentation defines dynamic batching as the server-side capability to combine inference requests and dynamically create batches, which can typically improve throughput and can be configured through the maximum batch size, preferred batch sizes, maximum queue delay, and queue properties. For requests with different input shapes, Triton also provides Ragged Batching to reduce explicit padding.

\subsubsection{Industry Trends and Developments}

In model serving for traditional vision, speech, retrieval, classification, Embedding, and similar models, dynamic batching is mainly manifested as request-level batching: based on request arrival time, input shape, model instance availability, maximum queueing delay, queue priority, and timeout policy, the inference server merges multiple requests for the same model for execution, in order to improve GPU utilization and throughput per unit time. NVIDIA Triton is a typical representative of this type of capability; its dynamic batcher targets stateless models and supports scheduling based on batch size, queueing delay, and priority. It also supports some scenarios with variable-length inputs through Ragged Batching, provided that the model backend can process the concatenated ragged input and the corresponding intra-batch metadata\cite{ref194}.

In generative large model inference, the focus of dynamic batching shifts to token-level continuous scheduling. LLM requests usually include two stages: Prefill and Decode. The Prefill stage processes the complete prompt, is compute-intensive, and affects time-to-first-token latency; the Decode stage generates tokens one by one, with a smaller amount of computation per step but uncertain duration and output length. Static batches are slowed down by requests with long outputs, causing completed requests within the batch to be unable to exit promptly and new requests to be unable to enter promptly. ORCA proposed iteration-level scheduling and selective batching at OSDI 2022, lowering the scheduling granularity to the model-iteration level, so that the scheduler executes only one generation iteration each time and allows requests to dynamically enter or exit the batch during generation. Its paper reported, in GPT-3 175B evaluation, a 36.9x throughput improvement over FasterTransformer at the same latency level\cite{ref195}.

An important direction in recent years is staged optimization of Prefill/Decode. Sarathi-Serve argues that although the Prefill stage has high latency, it can easily saturate the GPU, while although the Decode stage has low latency, GPU utilization is insufficient. It therefore proposes chunked-prefill and stall-free scheduling, splitting the Prefill of long prompts into approximately equal-length chunks and inserting new requests without interrupting existing Decode operations, thereby achieving a better balance between large-batch throughput and low latency. Its OSDI 2024 paper reported that, on models such as Mistral-7B, Yi-34B, and Falcon-180B, service capacity can be improved relative to vLLM\cite{ref170}.

\begin{figure}[htbp]
\centering
\includegraphics[width=0.95\linewidth]{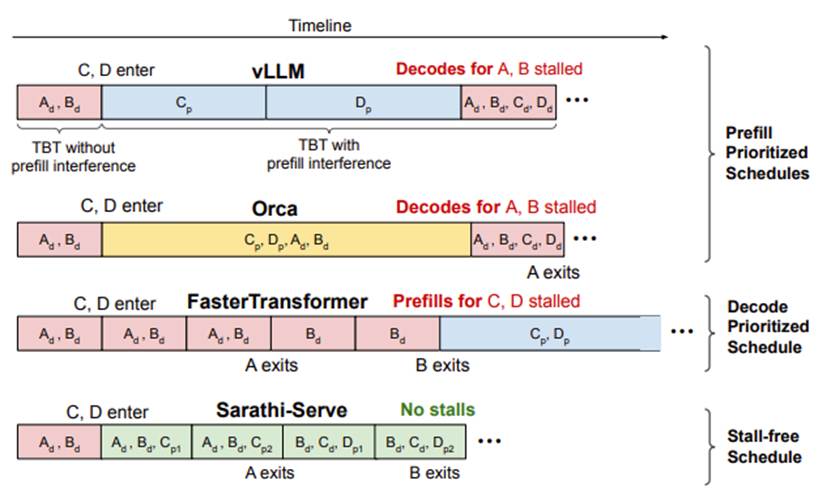}
\caption{Sarathi-Serve advances dynamic batching into Prefill/Decode phase-aware hybrid batching; adapted from Reference~\cite{ref170}.}
\label{fig:4_5}
\end{figure}

In multi-instance, multi-tenant, and mixed-workload scenarios, the industry has begun to combine dynamic batching with global queue management, request migration, priority isolation, and SLO-aware scheduling. Llumnix focuses on runtime rescheduling across multiple model instances, improving load balancing, mitigating resource fragmentation, strengthening priority isolation, and reducing tail latency through live migration of requests and in-memory states; its OSDI 2024 paper reports that it can improve tail latency by an order of magnitude and bring up to 1.5x acceleration for high-priority requests\cite{ref196}. QLM targets scenarios involving multiple models, multiple SLOs, and the co-location of online interactive requests with offline batch requests. It uses a waiting-time estimator and a global scheduler to orchestrate operations such as request pulling, eviction, load balancing, and model swapping, improving the SLO attainment rate by 40\%--90\% and throughput by 20\%--400\%\cite{ref197}.

Research after 2025 further extends from single-instance continuous batching to intra-device pipelining, cluster-level pooling, and token-level resource scheduling. NanoFlow argues that end-to-end LLM serving suffers utilization losses under many common workloads due to the serialization of heterogeneous operations such as computation, memory, and networking. It therefore proposes nano-batch scheduling and intra-device resource-overlap scheduling, achieving a 1.91x throughput improvement over existing systems under real-world workloads\cite{ref198}. Aegaeon, oriented toward concurrent multi-model serving, proposes GPU pooling and token-level auto-scaling. It adopts different scheduling strategies in the Prefill and Decode phases to improve, at token granularity, the resource pooling efficiency and SLO attainment capability of multi-model serving\cite{ref199}.

\subsection{High-Performance Communication Library}

\subsubsection{Technical Definition}

High-performance communication libraries refer to communication software stacks for multi-GPU, multi-node parallel computing and inference services. They provide collective communication and point-to-point communication capabilities, such as AllReduce, AllGather, ReduceScatter, All-to-All, Broadcast, and Send/Receive, and are optimized for interconnects such as NVLink, NVSwitch, PCIe, InfiniBand, RoCE, and RDMA. NCCL is officially defined as a library of fundamental multi-GPU and multi-node communication primitives optimized for NVIDIA GPUs and networking, providing topology-aware inter-GPU communication primitives and focusing on accelerating inter-GPU communication.

\subsubsection{Industry Trends and Developments}

High-performance communication libraries now cover the underlying data-plane software stack for training parallelism, MoE expert parallelism, inference service disaggregation, KV cache migration, and cross-resource-pool scheduling. On the training side, data parallelism, tensor parallelism, pipeline parallelism, and expert parallelism rely respectively on communication operators such as AllReduce, AllGather, ReduceScatter, All-to-All, Broadcast, and Send/Receive; NCCL natively supports these communication operators and optimizes for channels such as PCIe, NVLink, InfiniBand Verbs, and IP sockets\cite{ref200}. UCX, by contrast, more often serves as the underlying communication framework, providing transport abstractions for RDMA, TCP, GPU, shared memory, and network atomic operations, and supporting MPI, AI frameworks, and inference transport components in building unified communication paths\cite{ref201}.

In recent years, the optimization focus of communication libraries has shifted toward topology awareness, algorithmic adaptivity, and large-scale initialization efficiency. In NCCL 2.23, NVIDIA introduced the PAT algorithm to improve the scaling efficiency of AllGather and ReduceScatter in large-scale scenarios, while also enhancing initialization APIs, user buffer registration, and profiler plugin capabilities\cite{ref202}. Subsequently, NCCL further advanced cross-data-center communication capabilities. Through mechanisms such as fabricId and ncclNet, it incorporated network topology into the selection process for collective communication algorithms such as rings and trees, enabling cross-data-center training or cross-resource-pool training to minimize the use of expensive cross-domain links as much as possible\cite{ref203}.

For ultra-large-scale training, communication libraries have also begun to strengthen reliability, observability, and fault diagnosis capabilities. Since version 2.24, NCCL has introduced the RAS subsystem to query the health status of NCCL jobs, assist in locating crash issues, identify abnormal communication processes, and provide low-overhead state observability; NCCL 2.26 further enhanced GPU kernel and network profilers, network plugin quality-of-service control, and RAS capabilities. Meanwhile, tools such as NCCL Inspector have begun to provide always-on collective-communication performance and metadata logging capabilities, covering metrics such as algorithm bandwidth, bus bandwidth, execution time, message size, and collective communication type, thereby supporting the localization of communication bottlenecks in large-scale clusters\cite{ref204}.

In terms of communication-computation fusion, NCCL 2.28 and later provide the Device API, enabling CUDA kernels to directly call device-side communication primitives, and provide modules such as LSA, Multimem, and GIN for building reduction, broadcast, and communication-computation fusion kernels. Such capabilities are important for low-latency inference, MoE dispatch and combine, fine-grained expert parallelism, and custom collective communication, because communication can be brought closer to the GPU kernel scheduling path, reducing CPU intervention and synchronization overhead.

MoE and expert parallelism are driving communication libraries toward specialized communication libraries oriented to model structures. DeepEP represents this direction: it is positioned as an expert-parallel communication library for modern training and inference scenarios, with a focus on optimizing MoE dispatch and combine, low-latency All-to-All, and low-precision data transfer paths such as FP8, so as to reduce interference with compute kernels\cite{ref185}.

On the inference service side, the importance of high-performance communication libraries is further increasing. NIXL is oriented toward this category of inference data transfer scenarios and is positioned as an open, vendor-neutral data movement library. It supports CPU/GPU memory as well as different resources such as file, block, and object storage, and adapts to transport mechanisms such as UCX through pluggable backends; its companion tools NIXLBench and KVBench also show that inference communication is paying attention simultaneously to KV transfer latency, throughput, and end-to-end service metrics\cite{ref205}.

\begin{figure}[htbp]
\centering
\includegraphics[width=0.95\linewidth]{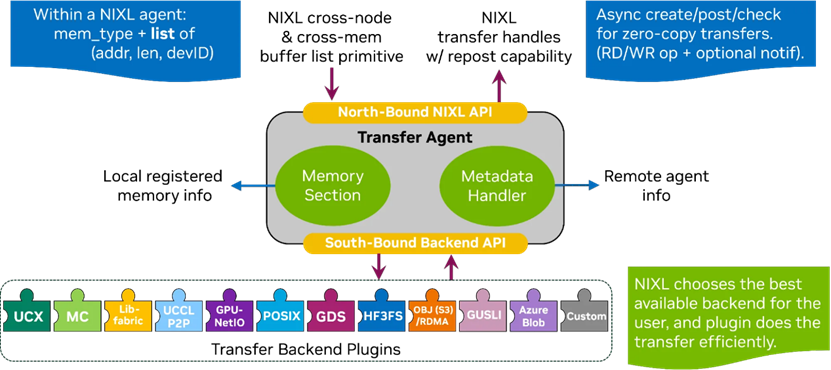}
\caption{NIXL High-Performance Inference Data Movement Library Architecture: abstracting high-performance data transfer among heterogeneous resources such as GPU/CPU/Storage through a unified API, Transfer Agent, metadata management, and pluggable backends; adapted from Reference~\cite{ref205}.}
\label{fig:4_6}
\end{figure}

Cross-vendor interoperability has also become an important direction. AMD RCCL provides collective communication and point-to-point communication interfaces similar to NCCL, oneCCL provides scale-up and scale-out collective communication capabilities for the oneAPI ecosystem, and UCC attempts to provide a unified collective communication layer for MPI, PGAS, and deep learning frameworks, and supports backends such as CUDA.

\subsection{Load Balancing}

\subsubsection{Technical Definition}

Positioned between clients and backend server arrays, a load balancer functions as a core reverse proxy and traffic distributor. Within modern distributed architectures, its value is manifested in three primary dimensions: horizontally scaling, which allows linear enhancement of overall system processing capacity by adding server instances; ensuring high availability, which dynamically monitors backend nodes through continuous health checks and automatically cordons failed servers to achieve smooth traffic migration; and optimizing performance, which allocates tasks based on the actual load of backend nodes to achieve globally optimal resource utilization across the cluster \cite{ref206}.

From a technical implementation standpoint based on the OSI seven-layer model, load balancing operates primarily across two dimensions. The first is Layer 4 (L4) load balancing, which functions at the TCP/UDP protocol layers and forwards traffic based on low-level network information such as source IP, destination IP, and port numbers. Because this process avoids packet decapsulation for complex application-layer protocols, its forwarding efficiency is exceptionally high, with LVS and F5 hardware devices serving as typical representatives in this domain.

The second is Layer 7 (L7) load balancing, which penetrates into application-layer protocols such as HTTP, HTTPS, and gRPC. Capable of inspecting specific contents like URLs, Cookies, and HTTP headers, L7 load balancing possesses robust "content-aware" capabilities. This enables support for more granular scheduling strategies, such as routing requests to specific microservices based on request paths, with standard tools including Nginx, HAProxy, and Envoy.

In specific scenario scheduling, the algorithmic model dictates traffic destination. The most fundamental is the round-robin algorithm, which distributes tasks sequentially and is ideally suited for scenarios where backend server configurations are completely identical; if backend hardware performance varies, weighted round-robin can be deployed to dynamically assign weights according to server hardware specifications.

For scenarios involving long-lived connections or highly variable processing times, the least-connections algorithm is a superior choice, preferentially dispatching new requests to nodes with the fewest active connections. To ensure that requests originating from the same source stably land on the same server, systems implement consistent hashing algorithms to maximize routing stability when nodes fluctuate.

\subsubsection{Industry Trends and Developments}

As cloud computing matures and demand for large-model inference grows rapidly, load balancing technology is evolving from passive routing toward active adaptation.

At the architectural level, the industry is accelerating the convergence of load balancing, traffic gateways, and security gateways. Serving as the standard data plane for Service Mesh, Envoy has become a common choice for modern internet enterprises to replace traditional Nginx clusters, owing to its dynamic configuration capabilities (xDS APIs) and plug-in extensibility \cite{ref207}. In recent practices, open-source projects such as Higress---which evolved based on Envoy \cite{ref208}---have achieved unified orchestration linking L4 to L7 load balancing with Ingress controllers. This architectural convergence reduces the need for traffic scheduling to hop repeatedly across multiple components, substantially simplifying enterprise operations and maintenance pipelines. China Unicom developed its own cloud-native gateway based on Envoy+Istio to replace traditional Nginx load balancing schemes, combining model probing technologies to dynamically perceive actual inference instance workloads and achieve intelligent, adaptive scheduling.

Given the unique characteristics of large-model inference, traditional load balancing strategies are increasingly insufficient. In LLM inference scenarios, the KV caches generated during each dialogue process have high reuse value, motivating semantic-aware load balancing. Upon receiving a request, the gateway no longer distributes it only according to connection-level metrics; instead, by rapidly analyzing the semantic similarity of the prompt, it routes the request to GPU nodes that have already cached the relevant context. This scheduling paradigm allows backend nodes to increase reuse of existing intermediate computation results, thereby improving overall inference throughput.

Concurrently, addressing the dynamic nature of large model generation tasks, load balancing is evolving toward predictive rate limiting and distribution. Traditional algorithms often rely on connection or request counts to evaluate backend pressure; however, in inference scenarios, under the same connection, generating 10 tokens and generating 2,000 tokens consume substantially different amounts of GPU compute. Consequently, the next generation of adaptive load balancers integrates real-time model token generation rates to dynamically estimate the actual load pressure of each node, enabling more rational traffic distribution decisions and moving beyond the coarse-grained scheduling methods of earlier systems.

\subsection{Gateway Rate Limiting and Circuit Breaking}

\subsubsection{Technical Definition}

Gateway rate limiting \cite{ref209} refers to restricting the number of incoming requests within a specific time window to prevent exceeding the system's maximum processing capacity. Serving as the central hub for traffic distribution, the gateway executes rate-limiting policies to protect expensive backend computing resources from being overwhelmed by transient peak floods.

Core algorithmic models include the following common approaches.

The fixed-window algorithm divides time into fixed periods and permits a predetermined quantity of requests per period. Its advantage lies in simplicity, whereas its disadvantage is the potential for double-traffic spikes at the window boundaries.

The sliding-window algorithm resolves the boundary limitations of the fixed-window algorithm and provides a smoother traffic curve.

The token-bucket algorithm generates tokens at a constant rate, and requests must acquire a token before being processed. It accommodates a certain degree of burst traffic.

The leaky-bucket algorithm lets requests enter the bucket like water and leak out for processing at a constant rate. It emphasizes an absolutely smooth output rate.

Gateway circuit breaking and degradation protect critical service functions by selectively shedding load or suspending calls to unhealthy backends. When the gateway detects slow responses, elevated error rates, or resource depletion in backend services, it proactively suspends calls to the affected service and executes predefined fallback logic to prevent fault propagation from causing cascading failures across the distributed system.

The core logical phases are as follows.

In the open state, when the error rate reaches a threshold (e.g., 50\% of requests failing within the last 30 seconds), the gateway directly intercepts subsequent requests.

In the half-open state, after a specified duration (the sleep window), the gateway permits a small fraction of traffic to attempt access. If successful, the system recovers; if it fails, it reverts to the open state.

Service fallback returns user-friendly static error pages, default values (cached data), or empty responses, ensuring that the client side does not hang.

\subsubsection{Industry Trends and Developments}

Entering 2026, rate limiting, circuit breaking, and degradation at the gateway layer have progressively evolved from static configurations into an operational model characterized by intelligent self-adaptation and full-link collaboration.

Traditional rate-limiting models rely heavily on manual estimation of QPS thresholds. In cloud environments where dynamic scaling has become standard, this approach is highly prone to configuration failures and imprecise protection, making it increasingly difficult to adapt to current complex system operation demands. Modern gateways such as Envoy have begun introducing adaptive rate-limiting schemes based on Little's Law. By constructing dynamic feedback mechanisms using CPU utilization, response time (RT), and concurrent request counts, the system can automatically adjust rate-limiting thresholds based on the current health index, reducing dependence on manual intervention. Alibaba's open-source Sentinel 2.0 \cite{ref210} further strengthens this adaptive traffic control capability; by precisely probing the system "queue length," it can identify system load inflection points and manage traffic proactively.

Simultaneously, rate limiting and circuit breaking capabilities are gradually shifting down to the infrastructure layer. The industry currently widely adopts Envoy-based gateway solutions. With the assistance of WebAssembly (Wasm) plug-ins, developers can dynamically deploy complex rate-limiting logic---such as dynamic quota adjustments for specific VIP users---without restarting the gateway, improving operational efficiency and system flexibility \cite{ref211}. China Unicom's self-developed model gateway implements various rate-limiting algorithms including QPS, TPM, and concurrency based on cloud-native Envoy+Istio and self-developed WASM plug-ins. It supports rate-limiting policy configurations across multiple dimensions, such as tenants, users, applications, and models, satisfying diverse practical application scenarios.

Regarding service degradation and fallback strategies, the industry continues to evolve toward intelligence and granularity. Incorporating the unique characteristics of LLM gateways, when backend inference models become overloaded, gateway degradation no longer simply returns a "system busy" prompt. Instead, it can fall back to small language models (SLMs) with fewer parameters, achieving a resilient service that provides degraded performance without service interruption and thereby minimizing the impact on user experience \cite{ref212}. Tailored to the characteristics of large model services, China Unicom has implemented adaptive model degradation; when a model is overloaded, the gateway proactively downgrades to a smaller-parameter model to provide resilient service.

\section{Future Outlook}

As token operations become a core mode of large model service delivery, inference optimization will increasingly determine the scalability, cost efficiency, and reliability of MaaS platforms. Future competition will not depend only on model capability or computing capacity. It will depend more on whether a platform can continuously deliver high-quality token services under real production constraints, including fluctuating traffic, heterogeneous workloads, long-context interaction, multi-agent execution, and strict business SLAs.

The next stage of token inference optimization will move from isolated acceleration techniques toward system-level collaborative optimization. Model selection, decoding acceleration, cache reuse, operator optimization, resource scheduling, and gateway governance are unlikely to remain isolated modules. Instead, these capabilities will be jointly optimized as part of a unified serving system. Such a system should connect capability profiling, model scheduling, token generation, cache management, low-level execution optimization, and service governance into a closed loop. The key value of this closed loop is to shift inference optimization from local speedup to end-to-end efficiency improvement.

Looking ahead, several trends may become increasingly important. First, optimization objectives are expected to shift from a single performance metric to a multi-objective balance among quality, cost, latency, throughput, stability, and security. Second, the optimization scope may expand from single models and single instances to coordinated scheduling across multiple models, multiple instances, and multiple clusters. In this process, intelligent routing, model cascading, model ensembling, and fallback mechanisms are likely to become increasingly common capabilities in MaaS platforms. Third, optimization methods may gradually move from static configuration and experience-based tuning toward dynamic adaptation. Future inference systems are expected to adjust scheduling policies and resource allocation in real time according to traffic patterns, model capability profiles, cache states, computing load, and business SLAs.

These trends will become particularly important in long-context and agent workloads. As service pipelines become longer, mechanisms such as Prefill, Decode, KV cache management, PD disaggregation, session affinity, semantic caching, dynamic batching, and cross-node scheduling will play a more critical role in controlling latency and cost. In parallel, heterogeneous computing environments and domestic accelerator ecosystems in China are expected to make model architecture adaptation, operator optimization, quantization compatibility, and inference engine tuning long-term priorities for large model platforms.

Overall, inference optimization for token operations is not merely a matter of accelerating model inference. It is a system engineering problem across models, computing resources, networks, gateways, and applications. The central objective is to achieve the best overall balance among quality, cost, latency, throughput, and stability under real business constraints. As token operations continue to mature, inference optimization will become essential for reducing token production costs, improving service efficiency, strengthening service resilience, and enabling large model services to evolve from callable APIs into scalable intelligent infrastructure.